\documentclass[plaindraft,letter]{jpp-AAS}

\usepackage{graphicx,amsmath,mathtools}
\usepackage{natbib}
\usepackage[pdftex,colorlinks,citecolor=blue]{hyperref}
\usepackage{cleveref}

\def\ang{Ann.~Geophys.}

\def\apjs{Astrophys.~J.~Supp.~Ser.}
\def\apj{Astrophys.~J.}
\def\apjl{Astrophys.~J.~Lett.}
\def\aap{Astron.~Astrophys.}
\def\araa{Ann.~Rev.~Astron.~Astrophys.}
\def\mnras{Mon.~Not.~R.~Astron.~Soc.}

\def\prl{Phys.~Rev.~Lett.}
\def\jgr{J.~Geophys.~Res.}
\def\ssr{Space Science Rev.}

\def\grl{Geophys.~Res.~Lett.}
\def\pop{Phys.~Plasmas}
\def\pof{Phys.~Fluids}

\def\jopp{J.~Plasma Phys.}
\def\jpp{J.~Plasma Phys.}
\def\jfm{J.~Fluid Mech.}

\def\rmp{Rev.~Mod.~Phys.}

\ifCUPmtlplainloaded \else
  \checkfont{eurm10}
  \iffontfound
    \IfFileExists{upmath.sty}
      {\typeout{^^JFound AMS Euler Roman fonts on the system,
                   using the 'upmath' package.^^J}%
       \usepackage{upmath}}
      {\typeout{^^JFound AMS Euler Roman fonts on the system, but you
                   dont seem to have the}%
       \typeout{'upmath' package installed. JPP.cls can take advantage
                 of these fonts, if you use 'upmath' package.^^J}%
       \providecommand\upi{\pi}%
      }
  \else
    \providecommand\upi{\pi}%
  \fi
\fi

\ifCUPmtlplainloaded \else
  \checkfont{msam10}
  \iffontfound
    \IfFileExists{amssymb.sty}
      {\typeout{^^JFound AMS Symbol fonts on the system, using the
                'amssymb' package.^^J}%
       \usepackage{amssymb}%
         
       \let\ge=\geqslant  
      }{}
  \fi
\fi

\ifCUPmtlplainloaded \else
  \IfFileExists{amsbsy.sty}
    {\typeout{^^JFound the 'amsbsy' package on the system, using it.^^J}%
     \usepackage{amsbsy}}
    {\providecommand\boldsymbol[1]{\mbox{\boldmath $##1$}}}
\fi

\newcommand{\pD}[2]{\frac{\partial #2}{\partial #1}}
\newcommand{\D}[2]{\frac{{\rm d} #2}{{\rm d} #1}}

\newcommand\bs[1]{\boldsymbol{#1}}
\newcommand\bb[1]{\mbox{\boldmath{$#1$}}}
\newcommand\grad{\bb{\nabla}}
\renewcommand\bcdot{\,\bb{\cdot}\,}

\newcommand\btimes{\,\bb{\times}\,}
\newcommand{\mc}[1]{\mathcal{#1}}
\newcommand{\mf}[1]{\,\mathfrak{#1}\!}

\newcommand{\msb}[1]{\mathsfbi{#1}}

\def\underwrite#1#2{\underbrace{\vphantom{\Biggl(}#2}_{\hspace{-0.27em}{\hspace{0.285em}\textrm{#1}}}}
\def\sqrtexplained#1{%
  \begingroup
    \sbox0{$#1$}
    \def\underbrace##1_##2{##1}
    \sbox2{$#1$}
    \dimen0=\wd0 \advance\dimen0-\wd2
    \mathrlap{\sqrt{\phantom{\displaystyle#1}\kern\dimen0 }}
    \hphantom{\sqrt{\vphantom{\displaystyle#1}}}
  \endgroup
  #1}
  
\newcommand{\gas}{\!\bs{R}_s}
\newcommand{\gai}{\!\bs{R}_i}
\newcommand{\ggs}{{\negthickspace\!\bs{R}_s}}

\newcommand{\ggr}{{\negthickspace\!\bs{r}}}
\newcommand{\imag}{{\rm i}}
\newcommand{\rmd}{{\rm d}}
\newcommand{\eb}{\hat{\bb{b}}}
\newcommand{\ez}{\hat{\bb{z}}}
\newcommand{\ex}{\hat{\bb{x}}}
\newcommand{\ey}{\hat{\bb{y}}}
\newcommand{\valf}{v_{\rm A}}
\newcommand{\vasq}{v^2_{\rm A}}
\newcommand{\valfeff}{v_{{\rm A}\ast}}

\newcommand{\vthprl}[1]{v_{{\rm th}\parallel #1}}
\newcommand{\vthprp}[1]{v_{{\rm th}\perp #1}}

\newcommand{\betaprp}[1]{\beta_{\perp #1}}
\newcommand{\betaprl}[1]{\beta_{\parallel #1}}
\newcommand{\cell}[1]{C^\parallel_{\ell #1}}
\newcommand{\czero}[1]{C^\parallel_{0 #1}}
\newcommand{\cone}[1]{C^\parallel_{1 #1}}
\newcommand{\ctwo}[1]{C^\parallel_{2 #1}}
\newcommand{\cthree}[1]{C^\parallel_{3 #1}}
\newcommand{\kell}[1]{C^\perp_{\ell #1}}
\newcommand{\kzero}[1]{C^\perp_{0 #1}}
\newcommand{\kone}[1]{C^\perp_{1 #1}}
\newcommand{\ktwo}[1]{C^\perp_{2 #1}}

\newcommand{\dpprp}[1]{\delta p_{\perp #1}}
\newcommand{\pprp}[1]{p_{\perp 0 #1}}
\newcommand{\dpprl}[1]{\delta p_{\parallel #1}}
\newcommand{\pprl}[1]{p_{\parallel 0 #1}}

\newcommand{\dne}{\delta n_e}
\newcommand{\dni}{\delta n_i}
\newcommand{\nsp}{n_{0s}}
\newcommand{\nem}{n_{0e}}
\newcommand{\nip}{n_{0i}}

\newcommand{\dtprp}[1]{\delta T_{\perp #1}}
\newcommand{\tprl}[1]{T_{\parallel 0{#1}}}
\newcommand{\tprp}[1]{T_{\perp 0{#1}}}
\newcommand{\tauprl}[1]{\tau_{\parallel{#1}}}
\newcommand{\tauprp}[1]{\tau_{\perp{#1}}}

\newcommand{\dupar}{u'_{\parallel 0s}}
\newcommand{\duparsq}{u'^2_{\parallel 0s}}

\newcommand{\dBprl}{\delta B_\parallel}
\newcommand{\dBprp}{\delta \bb{B}_\perp}
\newcommand{\chik}{\chi_{\bs{k}}}

\newcommand{\dBprlk}{\delta B_{\parallel\bs{k}}}
\newcommand{\dBprpk}{\delta \bb{B}_{\perp\bs{k}}}
\newcommand{\varphik}{\varphi_{\bs{k}}}
\newcommand{\Aprlk}{A_{\parallel\bs{k}}}
\newcommand{\Phik}{\Phi_{\bs{k}}}
\newcommand{\Psik}{\Psi_{\bs{k}}}

\title[Kinetic turbulence in pressure-anisotropic plasmas]{Astrophysical gyrokinetics: \\ Turbulence in pressure-anisotropic plasmas at ion scales and beyond}

\author[M.~W.~Kunz and others]%
{M.~W.~Kunz\ls$^{1,2}$%
  \thanks{Email address for correspondence: mkunz@princeton.edu}, 
I.~G.~Abel$^{3,4}$, 
K.~G.~Klein$^{5,6}$,
and A.~A.~Schekochihin$^{7,8}$}

\affiliation{$^1$Department of Astrophysical Sciences, Princeton University, Peyton Hall, Princeton, NJ 08544, USA\\[\affilskip]
$^2$Princeton Plasma Physics Laboratory, PO Box 451, Princeton, NJ 08543, USA\\[\affilskip]
$^3$Princeton Center for Theoretical Science, Princeton University, Jadwin Hall, Princeton, NJ 08544, USA\\[\affilskip]
$^4$Chalmers University of Technology, 41296 Gothenburg, Sweden\\[\affilskip]
$^5$CLASP, University of Michigan, Space Research Building, Ann Arbor, MI 48109, USA\\[\affilskip]
$^6$Lunar and Planetary Laboratory, University of Arizona, Tucson, AZ 85721\\[\affilskip]
$^7$Rudolf Peierls Centre for Theoretical Physics, University of Oxford, 1 Keble Road, Oxford OX1 3NP, UK\\[\affilskip]
$^8$Merton College, Merton Street, Oxford OX1 4JD, UK}

\pubyear{2018}
\volume{}
\pagerange{}
\date{\today}
\begin{document}

\maketitle

\begin{abstract}
We present a theoretical framework for describing electromagnetic kinetic turbulence in a multi-species, magnetized, pressure-anisotropic plasma. The turbulent fluctuations are assumed to be small compared to the mean field, to be spatially anisotropic with respect to it, and to have frequencies small compared to the ion cyclotron frequency. At scales above the ion Larmor radius, the theory reduces to the pressure-anisotropic generalization of kinetic reduced magnetohydrodynamics (KRMHD) formulated by Kunz \etal, 2015, {\it J.~Plasma Phys.}, vol.~81, 325810501. At scales at and below the ion Larmor radius, three main objectives are achieved. First, we analyse the linear response of the pressure-anisotropic gyrokinetic system, and show it to be a generalisation of previously explored limits. The effects of pressure anisotropy on the stability and collisionless damping of Alfv\'{e}nic and compressive fluctuations are highlighted, with attention paid to the spectral location and width of the frequency jump that occurs as Alfv\'{e}n waves transition into kinetic Alfv\'{e}n waves. Secondly, we derive and discuss a very general gyrokinetic free-energy conservation law, which captures both the KRMHD free-energy conservation at long wavelengths and dual cascades of kinetic Alfv\'{e}n waves and ion entropy at sub-ion-Larmor scales. We show that non-Maxwellian features in the distribution function change the amount of phase mixing and the efficiency of magnetic stresses, and thus influence the partitioning of free energy amongst the cascade channels. Thirdly, a simple model is used to show that pressure anisotropy, even within the bounds imposed on it by firehose and mirror instabilities, can cause order-of-magnitude variations in the ion-to-electron heating ratio due to the dissipation of Alfv\'{e}nic turbulence. Our theory provides a foundation for determining how pressure anisotropy affects the turbulent fluctuation spectra, the differential heating of particle species, and the ratio of parallel and perpendicular phase mixing in space and astrophysical plasmas.
\end{abstract}

\section{Introduction}

In a previous paper \citep[][hereafter Paper I]{kunz15}, we presented a theoretical framework for low-frequency electromagnetic (drift-)kinetic turbulence valid at scales larger than the particles' Larmor radii (``long'' wavelengths) in a collisionless, multi-species plasma. That result generalised reduced magnetohydrodynamics (RMHD; \citealt{kp74,strauss76,strauss77,zm92a}) and kinetic RMHD \citep[][hereafter S09]{schekochihin09} to the case where the mean distribution function of the plasma is pressure-anisotropic and different ion species are allowed to drift with respect to each other -- a situation routinely encountered in the solar wind \citep[e.g.][]{hundhausen67,feldman73,marsch82a,marsch82b,marsch06} and presumably ubiquitous in hot dilute astrophysical plasmas such as the intracluster medium of galaxy clusters \citep[e.g.][]{schekochihin05,sc06} and low-luminosity black-hole accretion flows \citep[e.g.][]{quataert98,sharma06,yn14,kunz16}. This framework was obtained via two routes: one starting from Kulsrud's formulation of kinetic MHD \citep{kulsrud64,kulsrud83} and one starting from applying the nonlinear gyrokinetic reduction \citep[e.g.][]{fc82,howes06} of the Vlasov-Maxwell set of equations. The latter approach also enables the study of fluctuations at and below the ion Larmor scale, the subject of this Paper.

Before embarking on any quantitative analysis or even qualitative discussion of what the gyrokinetic framework entails, we catalogue the principal theoretical achievements and implications of Paper I.\footnote{All sections and equations in Paper I are referenced using the prefix ``I--''; e.g.~(I--C1) refers to equation (C1) of Paper I and \S I--4 refers to section 4 of Paper I.} First, we showed that the main physical feature of low-frequency, long-wavelength plasma turbulence survives the generalisation to non-Maxwellian equilibrium distribution functions: Alfv\'{e}nic and compressive fluctuations are energetically decoupled, with the latter passively advected by the former. The Alfv\'{e}nic cascade is fluid, satisfying RMHD equations (with the Alfv\'{e}n speed modified by pressure anisotropy and interspecies drifts; see \S\S I--2.4, I--3.3), whereas the compressive cascade is kinetic and subject to collisionless damping (\S\S I--4.1, I--4.4). For a single-ion-species, bi-Maxwellian plasma, the kinetic cascade splits into three independently cascading parts (\S\S I--4.3, I--4.5, I--4.6): two parts associated with density and magnetic-field-strength fluctuations and a purely kinetic part associated with the entropy of the perturbed distribution function. Secondly, the organising principle of this long-wavelength turbulence was elucidated in the form of a conservation law for the appropriately generalised kinetic free energy (\S I--5.1). Using this alongside linear theory, we showed that certain non-Maxwellian features in the distribution function can reduce the rate of collisionless damping and the efficacy of magnetic stresses, and that these changes influence the partitioning of free energy amongst the various cascade channels. As the firehose or mirror instability thresholds are approached, the dynamics of the plasma are modified so as to reduce the energetic cost of bending magnetic-field lines or of compressing them.

In this paper, we concentrate on the sub-Larmor-scale ``dissipation'' range. We investigate the linear properties of kinetic Alfv\'{e}n waves (KAWs) and their nonlinear phase-space cascade in a plasma whose mean particle distribution functions exhibit pressure anisotropy and interspecies drifts. We find that the stability conditions imposed on the KAWs by the anisotropy of the distribution functions are a combination of those experienced by Alfv\'{e}n waves and compressive fluctuations at long wavelengths. We further show that, similar to the dual Alfv\'{e}nic-kinetic cascade of free energy in the inertial range (Paper I), there are two sub-ion-Larmor-scale kinetic cascades: one of KAWs, which is governed by a set of fluid-like electron reduced magnetohydrodynamic equations, and a passive phase-space cascade of ion-entropy fluctuations. These cascades have already been considered for a single-ion-species plasma whose equilibrium distribution function is isotropic and Maxwellian (S09). Here we focus on whether and to what extent their results carry over to the more general case. Special attention is paid to the transition from the inertial range across the ion-Larmor scale to the kinetic range, and to the effect of pressure anisotropy on the spectral location of this transition and on the amount of Landau-damped energy that ultimately makes its way to collisional scales.

\section{Prerequisites}

\subsection{Basic equations and notation}

For completeness, we provide here the basic equations derived in Paper I from which this paper's results follow, as well as the notation introduced in Paper I by which this paper's results may be understood.\footnote{A glossary of frequently used symbols can be found in appendix E of Paper I.} This recapitulation starts with the Vlasov-Landau equation,
\begin{equation}\label{eqn:vlasov}
\dot{f}_s \doteq \pD{t}{f_s} + \bb{v}\bcdot\grad f_s + \frac{q_s}{m_s} \biggl( \bb{E} + \frac{\bb{v}\btimes\bb{B}}{c} \biggr) \bcdot \pD{\bb{v}}{f_s} = \biggl( \pD{t}{f_s} \biggr)_{\rm coll} ,
\end{equation}
governing the space-time evolution of the particle distribution function of species $s$, $f_s = f_s(t,\bb{v},\bb{r})$, where $\bb{v}$ is the velocity-space variable and $\bb{r}$ is the real-space variable. The charge and mass of species $s$ are denoted $q_s$ and $m_s$, respectively; $c$ is the speed of light. The electric field $\bb{E}$ and magnetic field $\bb{B}$ are expressed in terms of scalar and vector potentials:
\begin{equation}\label{eqn:EandB}
\bb{E} = -\grad\varphi - \frac{1}{c} \pD{t}{\bb{A}} \quad {\rm and} \quad \bb{B} = B_0 \ez + \grad\btimes\bb{A} ,
\end{equation}
where $B_0 \ez$ is the guide magnetic field, taken to lie along the $z$ axis, and $\grad\bcdot\bb{A} = 0$ (the Coulomb gauge). These fields satisfy the plasma quasineutrality constraint,
\begin{equation}\label{eqn:quasineutrality}
0 = \sum_s q_s n_s = \sum_s q_s \int\rmd^3\bb{v} \, f_s ,
\end{equation}
and the pre-Maxwell version of Amp\`{e}re's law,
\begin{equation}\label{eqn:ampere}
- \nabla^2 \bb{A} = \frac{4\upi}{c} \bb{j} = \frac{4\upi}{c} \sum_s q_s n_s \bb{u}_s = \frac{4\upi}{c} \sum_s q_s \int\rmd^3\bb{v} \, \bb{v} f_s ,
\end{equation}
where $n_s$ and $\bb{u}_s$ are the number density and mean velocity of species $s$ and $\bb{j}$ is the current density. 

The term on the right-hand side of (\ref{eqn:vlasov}) represents the effect of collisions on the distribution function; in this paper, collisions are assumed to be sub-dominant and thus its specific form will not be required (precisely what `sub-dominant' means will be stated in short order). The assumption of weak collisionality gives the pressure tensor
\begin{equation}\label{eqn:ptensor}
\msb{P}_s \doteq \int \rmd^3\bb{v} \, m_s ( \bb{v} - \bb{u}_s ) ( \bb{v} - \bb{u}_s ) f_s
\end{equation}
the freedom to be anisotropic, even in the mean (zeroth-order) background. An example of such a pressure tensor is that describing a gyrotropic plasma (see \S\ref{sec:gkreduction}),
\begin{equation}\label{eqn:pgyrotensor}
\msb{P}_s = p_{\perp s} \bigl( \msb{I} - \eb\eb \bigr) + p_{\parallel s} \eb\eb ,
\end{equation}
where $\msb{I}$ is the unit dyadic, $\eb \doteq \bb{B}/B$ is the unit vector in the direction of the magnetic field, the subscript $\perp$ ($\parallel$) denotes the component perpendicular (parallel) to $\eb$, and
\begin{gather}\label{eqn:pressures}
p_{\parallel s} \doteq n_s T_{\parallel s} = \int \rmd^3\bb{v} \, m_s ( v_\parallel - u_{\parallel s} )^2 f_s, \\*
p_{\perp s} \doteq n_s T_{\perp s} = \int \rmd^3\bb{v} \, \frac{m_s}{2} | \bb{v}_\perp - \bb{u}_{\perp s} |^2 f_s 
\end{gather}
are the parallel and perpendicular pressures, respectively, of species $s$. An oft-employed distribution function that exhibits such pressure anisotropy is the bi-Maxwellian
\begin{equation}\label{eqn:biMax}
f_{\textrm{bi-M},s}(v_\parallel,v_\perp) \doteq \frac{n_s}{\sqrt{\upi}\vthprl{s}} \exp\Biggl[ - \frac{(v_\parallel - u_{\parallel s})^2}{\vthprl{s}^2} \Biggr] \frac{1}{\upi\vthprp{s}^2} \exp\Biggl[ - \frac{|\bb{v}_\perp - \bb{u}_{\perp s}|^2}{\vthprp{s}^2}\Biggr] ,
\end{equation}
where
\begin{equation}\label{eqn:vthermals}
\vthprl{s} \doteq \sqrt{\frac{2 T_{\parallel s}}{m_s}} \quad {\rm and} \quad \vthprp{s} \doteq \sqrt{\frac{2 T_{\perp s}}{m_s}}
\end{equation}
are the parallel and perpendicular thermal speeds of species $s$. Pressure anisotropy is caused in a weakly collisional plasma by adiabatic invariance: conservation of the magnetic moment $\mu_s \doteq m_s | \bb{v}_\perp - \bb{u}_{\perp s} |^2 / 2 B$ implies that a slow change in magnetic-field strength must be accompanied by a proportional change in the perpendicular temperature of species $s$ \citep{cgl56}. While such velocity-space anisotropy is generically exhibited by the gyrokinetic fluctuations regardless of whether the mean distribution function is proved (or assumed) to be isotropic and Maxwellian, in what follows we also allow for the possibility of a background pressure anisotropy.

\subsection{Gyrokinetic ordering}\label{sec:gkordering}

Our aim is to reduce (\ref{eqn:vlasov})--(\ref{eqn:ampere}) so that they describe only those fields whose fluctuating parts are small compared to the mean field, are spatially anisotropic with respect to it, have frequencies $\omega$ small compared to the Larmor frequency $\Omega_s \doteq q_s B_0 / m_s c$, and have parallel length scales $k^{-1}_\parallel$ large compared to the Larmor radius $\rho_s \doteq \vthprp{s}/\Omega_s$. While such specifications may appear to be quite restrictive, modern theories \citep[e.g.][]{gs95} and numerical simulations \citep[e.g.][]{smm83,opm94,cv00,mg01} of magnetized turbulence provide a strong foundation for expecting such anisotropic low-frequency fluctuations to comprise much of the energy in the turbulent cascade. Such spatial anisotropy is also now routinely measured in the solar wind \citep[e.g.][]{bwm96,hfo08,podesta09,wicks10,chen11a,chen16} and suggested by observations of turbulent density fluctuations in the interstellar medium \citep[e.g.][]{armstrong90,rickett02}.

The reduction is carried out in detail in appendix C of Paper I; here we describe its primary ingredients and principal consequences. The fields are split into their mean parts (denoted with a subscript `0') and fluctuating parts (denoted with $\delta$), the former characterized by spatial homogeneity on the fluctuating scales of interest (i.e.~$kL \gg 1$ where $L$ is some representative macroscale). The latter are taken to satisfy the asymptotic ordering
\begin{equation}\label{eqn:ordering}
\frac{\delta f_{1s}}{f_{0s}} \sim \frac{\delta\bb{B}}{B_0} \sim \frac{\delta\bb{E}}{(\vthprl{s}/c)B_0} \sim \frac{k_\parallel}{k_\perp} \sim \frac{\omega}{\Omega_s} \doteq \epsilon \ll 1 , \quad k_\perp \rho_s \sim 1 ,
\end{equation}
where we have expanded the distribution function in powers of $\epsilon$:
\begin{equation}
f_s = f_{0s} + \delta f_s = f_{0s} + \delta f_{1s} + \delta f_{2s} + \dots .
\end{equation}
Note that the fluctuations are permitted to have perpendicular scales on the order of the Larmor radius. We further assume that the collision frequency $\nu_s \lesssim \epsilon^2 \Omega_s$, thereby allowing non-Maxwellian $f_{0s}$ (cf.~\S A2.2 of \citealt{howes06}).\footnote{With such weak collisionality, one might worry about the possible formation of sharp structures in velocity space and the consequent importance of the parallel nonlinearity in the gyrokinetic equation, $\delta E_\parallel (\partial/\partial v_\parallel) \delta f_{1s} $, which is rigorously ordered out of standard (collisional) astrophysical gyrokinetics by ordering $\nu_s \sim \epsilon \Omega_s$ (see \S2 and Appendix A of \citealt{howes06}). To see that this is not a problem here, we remind the worried reader that the parallel nonlinearity is of comparable size to the other terms in the gyrokinetic equation (\ref{eqn:gyrokinetic}) only if the gyrokinetic response $h_s$ (see \S\ref{sec:gyrokineticresponse}) satisfies $\partial h_s / \partial v_\parallel \sim h_s / (\vthprl{s} \epsilon)$. To estimate the velocity derivative, we follow the argument in \S 7.9 of S09. After one turbulence cascade time, we anticipate that structures are formed such that, at some collisional dissipation scale $\ell_\nu$ satisfying $(\partial/\partial t) h_s(\ell_\nu) \sim C[h_s(\ell_\nu)]$ with $C$ being the collision operator, we have $\partial / \partial v_\parallel \sim \omega/\nu_s$. If $\nu_s \lesssim \epsilon\omega$ but $\nu_s \gg \epsilon^2 \omega$, then this velocity-space structure obeys $\partial / \partial v_\parallel \ll 1/(\vthprl{s}\epsilon)$. Thus, the parallel nonlinearity is unimportant, even at the dissipation scale. To justify the absence of collisions in the gyrokinetic equation obtained from our orderings (equation (\ref{eqn:gyrokinetic})), we estimate from equation (252) of S09 that the dissipation range satisfies $k_\perp \rho_i \gg 1/\sqrt{\epsilon}$, which is beyond our $k_\perp \rho_i \sim 1$ ordering in (\ref{eqn:ordering}). Thus, we require neither collisions nor the parallel nonlinearity in our gyrokinetic equation. These constraints are broken only if one were to (erroneously) evolve our equations for a time asymptotically longer than a cascade time. For context, order-of-magnitude estimates for the scales in the $\beta \sim 1$ solar wind yield a collisional mean free path $\lambda_{\rm mfp} \sim 1~{\rm au}$, $\nu_i \sim 10^{-7}~{\rm Hz}$, $\rho_i \sim 10^{-7}~{\rm au}$, $\Omega_i \sim 1~{\rm Hz}$, and an inertial-range turbulent cascade starting at a wavelength $\lambda \sim 10^{-3}~{\rm au}$ and proceeding down to electron Larmor scales (see, e.g., \citealt{ht14,kiyani15,wilson18}).} In the solar wind, large-scale pressure anisotropy is driven as the plasma expands in a predominantly radial magnetic field \citep[e.g.][]{matteini12}. In low-luminosity accretion flows, such anisotropy is thought to be driven on large scales by kinetic magnetorotational turbulence \citep[e.g.][]{sharma06,kunz16}. In either case, the turbulence on ion-Larmor scales ($k_\perp \rho_i \sim 1$) evolves on time scales much faster than those on which the macroscopic (``background'') pressure anisotropy is driven. This motivates our assumption of a fixed background pressure anisotropy of the fluctuating time scales of interest.

The gyrokinetic ordering guarantees that (to lowest order) all species drift perpendicularly to the magnetic field with identical velocities, $\bb{u}_{\perp s} = \bb{u}_\perp = c\bb{E}\btimes\bb{B}/B^2$. It then follows that the mean drift of any species relative to the centre-of-mass velocity $\bb{u} \doteq \sum_s m_s n_s \bb{u}_s / \sum_s m_s n_s$ must be in the parallel direction, {\em viz.}, $\bb{u}_s = \bb{u} + u'_{\parallel s} \eb$, with
\begin{equation}
u'_{\parallel s} = \frac{1}{n_s} \int\rmd^3\bb{v} \, ( v_\parallel - u_\parallel ) f_s .
\end{equation}
(Note that $\sum_s \, m_s n_s u'_{\parallel s} = 0$ by definition.) Our collisionless ordering permits parallel interspecies drifts (denoted by $\dupar$) in the background state, and we formally order $\dupar \sim \vthprl{s}$ for all species $s$. We further assume that the Alfv\'{e}n speed
\begin{equation}
\valf \doteq \frac{B_0}{\sqrt{4\upi\rho_0}} \sim \vthprl{s} ,
\end{equation}
where $\rho_0$ is the mean mass density of the plasma. This implies that the parallel and perpendicular plasma beta parameters,
\begin{equation}
\betaprl{s} \doteq \frac{8\upi\pprl{s}}{B^2_0} \quad {\rm and} \quad \betaprp{s} \doteq \frac{8\upi\pprp{s}}{B^2_0}
\end{equation}
respectively, are considered to be of order unity in the gyrokinetic expansion. The other dimensionless parameters in the system -- namely, the electron-ion mass ratio $m_e / m_i$, the charge ratio $Z_i \doteq q_i / | q_e | = q_i / e$, the parallel and perpendicular temperature ratios
\begin{equation}
\tauprl{s} \doteq \frac{\tprl{s}}{\tprl{e}} \quad {\rm and} \quad \tauprp{s} \doteq \frac{\tprp{s}}{\tprp{e}} ,
\end{equation}
and the temperature anisotropy $\tprp{s} / \tprl{s}$ of species $s$ -- are all considered to be of order unity as well. Subsidiary expansions with respect to these parameters can (and will) be made after the gyrokinetic expansion is performed.

Since we have $\omega \sim k_\parallel \vthprl{s} \sim k_\parallel \valf$, fast magnetosonic fluctuations are ordered out of our equations. Such fast-wave fluctuations are rarely seen in the solar wind \citep{howes12}. Observations of turbulence in the solar wind confirm that it is primarily Alfv\'{e}nic \citep[e.g.][]{bd71,chen16} and that its compressive component is approximately pressure-balanced \citep{burlaga90,roberts90,mt93,mccomas95,bpb04,bc05}. A more serious limitation of our analysis is perhaps the exclusion of cyclotron resonances, which have been traditionally considered necessary to explain the strong perpendicular heating observed in the solar wind \citep{leamon98a,isenberg01,kasper13}. Larmor-scale fluctuations whose amplitudes are large enough to break adiabatic invariance and thus drive chaotic gyromotion and stochastic particle heating \citep{chandran10,chandran13} are also precluded. That being said, the gyrokinetic framework does capture much of the physics governing both the inertial and dissipative ranges of kinetic turbulence, and so it is a sensible step to incorporate realistic background distribution functions into the gyrokinetic description of weakly collisional astrophysical plasmas. It is with that goal in mind that we commence with a presentation of the gyrokinetic theory.

\subsection{Gyrokinetic reduction}\label{sec:gkreduction}

\subsubsection{Gyrotropy of the background distribution function}

Under the ordering (\ref{eqn:ordering}), the largest term in the Vlasov-Landau equation (\ref{eqn:vlasov}) corresponds to Larmor motion of the mean distribution about the uniform guide field:
\begin{equation}\label{eqn:gyrophase1}
-\Omega_s \ez \bcdot \biggl( \bb{v} \btimes \pD{\bb{v}}{f_{0s}} \biggr) = 0 .
\end{equation}
This directional bias allows us to set up a local Cartesian coordinate system and decompose the particle velocity in terms of the parallel velocity $v_\parallel$, the perpendicular velocity $v_\perp$, and the gyrophase angle $\vartheta$,
\begin{equation}
\bb{v} = v_\parallel \ez + v_\perp ( \cos\vartheta \,\ex + \sin\vartheta \,\ey ) .
\end{equation}
Equation (\ref{eqn:gyrophase1}) then takes on the simple form
\begin{equation}
-\Omega_s \pD{\vartheta}{f_{0s}} = 0 ,
\end{equation}
which states that the mean distribution function is gyrotropic (independent of the gyrophase):
\begin{equation}
f_{0s} = f_{0s}(v_\parallel, v_\perp, t).
\end{equation}
All velocity-space derivatives of $f_{0s}$ that enter (\ref{eqn:vlasov}) are thus with respect to $v_\parallel$ and $v_\perp$, {\em viz.}
\begin{equation}
\frac{q_s}{m_s} \pD{\bb{v}}{f_{0s}} = - ( v_\parallel - \dupar ) \ez \,\frac{q_s f^\parallel_{0s}}{\tprl{s}} - \bb{v}_\perp \,\frac{q_s f^\perp_{0s}}{\tprp{s}} ,
\end{equation}
where
\begin{equation}\label{eqn:fprlfprp}
f^\parallel_{0s} \doteq - \vthprl{s}^2 \,\pD{(v_\parallel - \dupar)^2}{f_{0s}} \quad {\rm and} \quad f^\perp_{0s} \doteq -\vthprp{s}^2 \, \pD{v^2_\perp}{f_{0s}}
\end{equation}
are dimensionless derivatives of a species' mean distribution function with respect to the square of the parallel velocity (peculiar to the species drift velocity) and the perpendicular velocity, respectively. Their weighted difference,
\begin{equation}
\mf{D}f_{0s} \doteq \frac{\tprp{s}}{\tprl{s}} f^\parallel_{0s} - f^\perp_{0s} ,
\end{equation}
measures the velocity-space anisotropy of the mean distribution function. For a bi-Maxwellian distribution (\ref{eqn:biMax}), $f^\parallel_{0s} = f^\perp_{0s} = f_{0s}$ and 
\begin{equation}
\mf{D}f_{0s} = \biggl( \frac{\tprp{s}}{\tprl{s}} - 1 \biggr) f_{0s} \doteq \Delta_s f_{0s} ,
\end{equation}
where $\Delta_s$ is the temperature (or, equivalently, pressure) anisotropy of the mean distribution function of species $s$.

\subsubsection{Boltzmann response}

At $\mc{O}(\epsilon \Omega_s f_{0s})$, we learn from (\ref{eqn:vlasov}) that the first-order distribution function $\delta f_{1s}$ may be split into two parts. The first of these is the so-called adiabatic (or `Boltzmann') response,
\begin{equation}\label{eqn:boltzmann}
\delta f_{1s,{\rm Boltz}} = - \frac{q_s \varphi'_s}{\tprl{s}}  f^\parallel_{0s} + \frac{q_s}{\tprp{s}} \biggl( \varphi - \frac{v_\parallel A_\parallel}{c} \biggr) \mf{D}f_{0s} ,
\end{equation}
where
\begin{equation}\label{eqn:varphiprime}
\varphi'_s \doteq \varphi - \frac{\dupar A_\parallel}{c}
\end{equation}
is the fluctuating electrostatic potential in the frame of the parallel-drifting species $s$. This part of $\delta f_{1s}$ represents the (leading-order) evolution of $f_{0s}$ under the influence of the perturbed electromagnetic fields. To see this, we first introduce the total particle energy in the parallel-drifting frame,
\begin{equation}\label{eqn:energy}
\overline{\varepsilon}_s = \varepsilon_{0s} + \varepsilon_{1s} \doteq \frac{1}{2} m_s \bigl| \bb{v} - \dupar \ez \bigr|^2 + q_s \varphi'_s
\end{equation}
and the (gyrophase-dependent part of the) first adiabatic invariant,
\begin{equation}\label{eqn:mu}
\overline{\mu}_s = \mu_{0s} + \mu_{1s} \doteq \frac{m_s v^2_\perp}{2B_0} + \frac{q_s}{B_0} \biggl( \varphi - \frac{v_\parallel A_\parallel}{c} \biggr) ,
\end{equation}
both written out to first order in the fluctuation amplitudes \citep[e.g.][]{kruskal58,hth67,taylor67,ctb81,parrathesis}. It is then straightforward to show by using 
\begin{equation}
f^\parallel_{0s} = -\tprl{s} \pD{\varepsilon_{0s}}{f_{0s}} \quad {\rm and} \quad f^\perp_{0s} = -\tprp{s} \biggl( \frac{1}{B_0} \pD{\mu_{0s}}{f_{0s}} + \pD{\varepsilon_{0s}}{f_{0s}} \biggr)
\end{equation}
(see \S I--C.4) that the sum of the mean distribution function and the Boltzmann response is simply
\begin{equation}
f_{0s} (v_\parallel, v_\perp) + \delta f_{1s,{\rm Boltz}} = f_{0s}(\overline{\varepsilon}_s,\overline{\mu}_s) + \mc{O}(\epsilon^2) .
\end{equation}
In other words, the Boltzmann response does not change the form of the mean distribution function if the latter is written as a function of the constants of the motion (calculated sufficiently accurately).

\subsubsection{Gyrokinetic response}\label{sec:gyrokineticresponse}

The second part of $\delta f_{1s}$, which we denote by $h_s$, represents the response of rings of charge to the fluctuating fields, and is thus referred to as the {\em gyrokinetic response}. It satisfies
\begin{equation}
 \bb{v}_\perp\bcdot\grad_\perp h_s - \Omega_s \pD{\vartheta}{h_s} \biggr|_{\bs{r}} = - \Omega_s \pD{\vartheta}{h_s} \biggr|_{\,\gas} = 0,
\end{equation}
where we have transformed the $\vartheta$ derivative taken at constant position $\bb{r}$ to one taken at constant guiding centre
\begin{equation}
\bb{R}_s = \bb{r} + \frac{\bb{v}\btimes\ez}{\Omega_s} .
\end{equation}
Thus, $h_s$ is independent of the gyrophase angle at constant guiding centre $\bb{R}_s$ (but not at constant position $\bb{r}$):
\begin{equation}
h_s = h_s(t,\bb{R}_s,v_\parallel,v_\perp) .
\end{equation}

\subsubsection{Gyrokinetic equation}

At $\mc{O}(\epsilon^2 \Omega_s f_{0s})$, we find from (\ref{eqn:vlasov}) that the gyrokinetic response evolves via the {\em gyrokinetic equation}
\begin{align}\label{eqn:gyrokinetic}
\pD{t}{h_s} + v_\parallel \pD{z}{h_s} + \frac{c}{B_0} \{ \langle \chi \rangle_{\gas} , h_s \} &= \frac{q_s f^\parallel_{0s}}{\tprl{s}} \biggl( \pD{t}{} + \dupar \pD{z}{} \biggr) \langle\chi\rangle_{\gas} \nonumber\\*
\mbox{} &- \frac{q_s\mf{D}f_{0s}}{\tprp{s}} \biggl( \pD{t}{} + v_\parallel \pD{z}{} \biggr) \langle\chi\rangle_{\gas} ,
\end{align}
where
\begin{equation}\label{eqn:chi}
\chi \doteq \varphi - \frac{v_\parallel A_\parallel}{c} - \frac{\bb{v}_\perp\bcdot\bb{A}_\perp}{c} 
\end{equation}
is the gyrokinetic potential and
\begin{equation}\label{eqn:ringaverage}
\langle \chi(t, \bb{r},\bb{v}) \rangle_{\gas} \doteq \frac{1}{2\upi} \oint \rmd\vartheta \, \chi \biggl( t, \bb{R}_s - \frac{\bb{v}\btimes\ez}{\Omega_s} , \bb{v} \biggr)
\end{equation}
denotes the ring average of $\chi$ at fixed guiding centre $\bb{R}_s$. The Poisson bracket
\begin{equation}\label{eqn:poisson}
\{ \langle\chi\rangle_{\gas} , h_s \} \doteq \ez \bcdot \biggl( \pD{\bb{R}_s}{\langle\chi\rangle_{\gas}} \btimes \pD{\bb{R}_s}{h_s} \biggr)
\end{equation}
represents the nonlinear interaction between the gyrocentre rings and the ring-averaged electromagnetic fields. 

The gyrokinetic equation (\ref{eqn:gyrokinetic}) can also be written in the following, perhaps more physically illuminating, form:
\begin{equation}\label{eqn:gyrokinetic2}
\pD{t}{h_s} + \langle \dot{\bb{R}}_s \rangle_{\gas} \bcdot \pD{\bb{R}_s}{h_s} = - \langle \dot{\overline{\varepsilon}}_s \rangle_{\gas} \pD{\overline{\varepsilon}_s}{f_{0s}} - \langle \dot{\overline{\mu}}_s \rangle_{\gas} \pD{\overline{\mu}_s}{f_{0s}} ,
\end{equation}
where
\begin{equation}
\langle\dot{\bb{R}}_s\rangle_{\gas} = v_\parallel \ez - \frac{c}{B_0} \pD{\bb{R}_s}{\langle\chi\rangle_{\gas}}\btimes\ez
\end{equation}
is the ring velocity,
\begin{equation}
\langle\dot{\overline{\varepsilon}}_s\rangle_{\gas} = q_s \biggl( \pD{t}{} + \dupar \pD{z}{} \biggr) \langle\chi\rangle_{\gas}
\end{equation}
is the ring-averaged rate of change of the particle energy (\ref{eqn:energy}), and 
\begin{equation}
\langle\dot{\overline{\mu}}_s\rangle_{\gas} = \frac{q_s}{B_0} \biggl( \pD{t}{} + v_\parallel \pD{z}{} \biggr) \langle\chi\rangle_{\gas}
\end{equation}
is the ring-averaged rate of change of the (gyrophase-dependent part of the) first adiabatic invariant (\ref{eqn:mu}). The right-hand side of (\ref{eqn:gyrokinetic2}) represents the effect of collisionless work done on the rings by the fields (the wave-ring interaction). Written in this way, (\ref{eqn:gyrokinetic}) is simply the ring-averaged Vlasov equation,
\begin{equation}
\langle \dot{f}_s(t,\bb{r},\overline{\varepsilon}_s,\overline{\mu}_s) \rangle_{\gas} = 0
\end{equation}
to lowest order in $\epsilon$.

It is a manifestly good idea in much of what follows to absorb the final term of (\ref{eqn:gyrokinetic}) (and, likewise, of (\ref{eqn:gyrokinetic2})), into $h_s$ by writing the latter in terms of the velocity-space coordinates $(v_\parallel, \mu_s)$, where
\begin{equation}\label{eqn:mus}
\mu_s \doteq \overline{\mu}_s - \frac{q_s}{B_0} \langle\chi\rangle_{\gas}
\end{equation}
is the full adiabatic invariant, {\em viz.}, $\dot{\mu}_s \sim \mc{O}(\epsilon^2 \omega \tprp{s} / B_0)$. At long wavelengths satisfying $k_\perp \rho_i \ll 1$,
\begin{equation}\label{eqn:mus_longwavelength}
\mu_s \simeq \frac{m_s | \bb{v} - \bb{u}_\perp - \bb{v}\bcdot\eb\eb |^2}{2B} \doteq \frac{m_s w^2_\perp}{2B}, 
\end{equation}
which is simply the magnetic moment of a particle in a magnetic field of strength $B$ drifting across said field at the $\bb{E}\btimes\bb{B}$ velocity,
\begin{equation}\label{eqn:uexb}
\bb{u}_\perp = \frac{c}{B_0} \ez\btimes\grad_\perp \varphi(\bb{r}) . 
\end{equation}
Then, introducing\footnote{Our $\widetilde{h}_s$ is equivalent to $\delta H_0$ of \citet{fc82} -- see their equation (42), with their $\delta G_0$ being our $h_s$.}
\begin{subequations}
\begin{align}\label{eqn:htilde}
\widetilde{h}_s (v_\parallel, \mu_s ) &\doteq h_s ( v_\parallel, v_\perp ) + \frac{q_s \mf{D}f_{0s} }{\tprp{s}} \langle\chi\rangle_{\gas} \\*
\mbox{} &= h_s(\varepsilon_{0s},\mu_{0s}) + \frac{q_s}{B_0} \pD{\mu_{0s}}{f_{0s}} \langle\chi\rangle_{\gas},
\end{align}
\end{subequations}
the gyrokinetic equation reads
\begin{equation}\label{eqn:gkequation2}
\pD{t}{\widetilde{h}_s} + \langle\dot{\bb{R}}_s\rangle_{\gas} \bcdot \pD{\bb{R}_s}{\widetilde{h}_s} =  \frac{q_s f^\parallel_{0s}}{\tprl{s}} \biggl( \pD{t}{} + \dupar \pD{z}{} \biggr) \langle\chi\rangle_{\gas} .
\end{equation}
This form of the gyrokinetic equation is particularly well suited for deriving the gyrokinetic invariants (\S\ref{sec:invariant}). Its right-hand side represents the collisionless work done on the rings by the fields in a frame comoving with the parallel drift velocity of species $s$. 

It will also prove useful in what follows to modify the energy variable $\overline{\varepsilon}_s$ to obtain
\begin{equation}
\varepsilon_s \doteq \overline{\varepsilon}_s - q_s \langle\varphi'_s\rangle_{\gas} ,
\end{equation}
which is the kinetic energy of the particle as measured in the frame moving with the $\dupar$ and $\bb{E}\btimes\bb{B}$ drifts \citep[e.g.][]{parrathesis}; indeed, 
\begin{equation}
\varepsilon_s \simeq \frac{1}{2} m_s (v_\parallel -\dupar)^2 + \frac{1}{2} m_s w^2_\perp
\end{equation}
at long wavelengths. If the mean distribution function is expressed in terms of these new velocity-space variables, {\em viz.}~$f_s = \widetilde{f}_{0s}(\varepsilon_s,\mu_s) + \delta\widetilde{f}_s$, then the perturbed distribution function $\delta\widetilde{f}_s$ becomes (see (I--C52))
\begin{subequations}\label{eqn:dftilde}
\begin{align}
\delta\widetilde{f}_s(\varepsilon_s,\mu_s) &= \delta f_{1s} - \Bigl( \delta f_{1s,{\rm Boltz}} - \langle \delta f_{1s,{\rm Boltz}} \rangle_{\gas} \Bigr) - \frac{q_s \mf{D} f_{0s} }{\tprp{s}} \biggl\langle \frac{\bb{v}_\perp\bcdot\bb{A}_\perp}{c}\biggr\rangle_{\ggs}  \\*
\mbox{} &= \widetilde{h}_s(v_\parallel,\mu_s) - \frac{q_s f^\parallel_{0s}}{\tprl{s}} \langle\varphi'_s\rangle_{\gas} .
\end{align}
\end{subequations}
This particular form of the perturbed distribution function is quite useful; it is the $k_\perp \rho_s \sim 1$ generalisation of the perturbed distribution function that prominently features in the generalised free energy of KRMHD (\S I--5.1), and thus is anticipated to appear in the generalised free energy of the gyrokinetic theory. The latter is derived in \S\ref{sec:invariant}.

\subsubsection{Field equations}

The equations governing the electromagnetic potentials are most easily obtained by substituting the decomposition
\begin{equation}
f_s = f_{0s} ( v_\parallel , v_\perp ) + \delta f_{1s,{\rm Boltz}} + h_s ( t, \bb{R}_s, v_\parallel, v_\perp ) + \dots
\end{equation}
into the leading-order expansions of the quasineutrality constraint (\ref{eqn:quasineutrality}) and Amp\`{e}re's law (\ref{eqn:ampere}). The result is (see \S I--C.3)
\begin{equation}\label{eqn:gkqn}
0 = \sum_s q_s \Biggl[ \int\rmd^3\bb{v} \, \langle h_s \rangle_{\bs{r}} - \frac{q_s \nsp}{\tprp{s}} \Biggl( \kzero{s} \varphi - \kone{s} \frac{\dupar A_\parallel}{c} \Biggr) \Biggr] ,
\end{equation}
\begin{equation}\label{eqn:gkprlamp}
\nabla^2_\perp A_\parallel = - \frac{4\upi}{c} \sum_s q_s \Biggl[ \int\rmd^3\bb{v} \, v_\parallel \langle h_s \rangle_{\bs{r}} - \frac{q_s\nsp\vthprl{s}}{2\tprp{s}} \Biggl( \kone{s} \varphi \frac{2\dupar}{\vthprl{s}} + \widetilde{\Delta}_s \frac{\vthprl{s} A_\parallel}{c} \Biggr) \Biggr] ,
\end{equation}
\begin{equation}\label{eqn:gkprpamp}
\nabla^2_\perp \dBprl = - \frac{4\upi}{c} \ez \bcdot \Biggl[ \grad_\perp \btimes \sum_s q_s \int\rmd^3\bb{v} \, \langle\bb{v} h_s \rangle_{\bs{r}} \Biggr] ,
\end{equation}
where
\begin{equation}
\widetilde{\Delta}_s \doteq \frac{\tprp{s}}{\tprl{s}} - \ktwo{s} \Biggl( 1 + \frac{2\duparsq}{\vthprl{s}^2} \Biggr)
\end{equation}
is the temperature anisotropy of species $s$ augmented by the parallel ram pressure from background parallel drifts, $C^\perp_{\ell s}$ are parallel moments of the perpendicular-differentiated mean distribution function (all of which equate to unity for a drifting bi-Maxwellian distribution; see appendix \ref{app:coeffs}), and 
\begin{equation}
\langle h_s (t, \bb{R}_s, v_\parallel, v_\perp ) \rangle_{\bs{r}} \doteq \frac{1}{2\upi} \oint\rmd\vartheta \, h_s \biggl( t, \bb{r} + \frac{\bb{v}\btimes\ez}{\Omega_s}, v_\parallel, v_\perp \biggr)
\end{equation}
denotes the gyro-average of $h_s$ at fixed $\bb{r}$. Together with the gyrokinetic equation (\ref{eqn:gyrokinetic}), the field equations (\ref{eqn:gkqn})--(\ref{eqn:gkprpamp}) constitute a closed system that describes the evolution of a gyrokinetic plasma with non-Maxwellian $f_{0s}$ and parallel interspecies drifts.

This completes our abbreviated review of the material derived in Paper I on the gyrokinetic framework for homogeneous, non-Maxwellian plasmas. We now proceed to analyse the linear and nonlinear behaviour of the perturbations governed by this system of equations.

\section{Linear gyrokinetic theory}\label{app:lineargk}

\subsection{From rings to gyrocentres}

The most straightforward way of making contact with the results of Paper I, while facilitating the extension of the theoretical framework into the kinetic range, is via the linear gyrokinetic theory \citep[pioneered by][]{rf68,th68,catto78,al80,ctb81}. This is obtained most easily by shifting the description of the plasma from one composed of extended rings of charge that move in a vacuum to one of a gas of point-particle-like gyrocentres moving in a polarizable medium. This transformation is enacted by working with the {\em gyrocentre distribution function}
\begin{subequations}\label{eqn:gyrocentredistribution}
\begin{align}
g_s &\doteq \widetilde{h}_s - \frac{q_s f^\parallel_{0s}}{\tprl{s}} \biggl\langle \varphi'_s - \frac{\bb{v}_\perp\bcdot\bb{A}_\perp}{c} \biggr\rangle_{\ggs} \\*
\mbox{} &= \delta\widetilde{f}_s + \frac{q_s f^\parallel_{0s}}{\tprl{s}} \biggl\langle \frac{\bb{v}_\perp\bcdot\bb{A}_\perp}{c} \biggr\rangle_{\ggs}  \\*
\mbox{} &= \langle \delta f_{1s} \rangle_{\gas} + \frac{q_s f^\perp_{0s}}{\tprp{s}} \biggl\langle \frac{\bb{v}_\perp\bcdot\bb{A}_\perp}{c}\biggr\rangle_{\ggs}  .
\end{align}
\end{subequations}
This new function not only helps simplify the algebra involved in deriving the linear theory, but also makes a good deal of physical sense. In the electrostatic limit, the use of $g_s$ (which, in this limit, equals $\langle \delta f_{1s} \rangle_{\gas}$) aids in the interpretation of polarization effects within gyrokinetics \citep{krommes12}, places the gyrokinetic equation in a numerically convenient characteristic form \citep{lee83}, and arises naturally from the Hamiltonian formulation of gyrokinetics \citep{dubin83,bh07}. In the electromagnetic case, introducing $g_s$ takes advantage of the fact that the Alfv\'{e}nic fluctuations have a gyrokinetic response that is approximately cancelled at long wavelengths by the Boltzmann response (see \S I--C.4), i.e.~$\delta f_{1s} \simeq g_s$ for long-wavelength Alfv\'{e}nic fluctuations.

Using (\ref{eqn:gyrocentredistribution}) to replace $\widetilde{h}_s$ in the gyrokinetic equation (\ref{eqn:gkequation2}), we find that $g_s$ evolves according to
\begin{align}\label{eqn:ggyrokinetic}
\pD{t}{g_s} + v_\parallel \pD{z}{g_s} + \frac{c}{B_0} \{ \langle \chi \rangle_{\gas} \, , g_s \} &= -\frac{q_s f^\parallel_{0s}}{\tprl{s}}\, \bigl( v_\parallel - \dupar \bigr)  \Biggl\langle \frac{1}{B_0} \{ A_\parallel , \varphi - \langle \varphi \rangle_{\gas} \} 
\nonumber\\*
\mbox{}  &+  \frac{1}{c} \pD{t}{A_\parallel} + \eb \bcdot \grad \varphi - \eb \bcdot \grad \biggl\langle \frac{\bb{v}_\perp\bcdot\bb{A}_\perp}{c} \biggr\rangle_{\ggs} \Biggr\rangle_{\ggs}  ,
\end{align}
where
\begin{equation}\label{eqn:gkbdotgrad}
\eb \bcdot \grad = \pD{z}{} + \frac{\dBprp}{B_0} \bcdot \grad_{\!\perp} = \pD{z}{} - \frac{1}{B_0} \bigl\{ A_\parallel \, , \dots \bigr\}
\end{equation}
is the spatial derivative along the perturbed magnetic field and
\begin{equation}\label{eqn:dBprp}
\dBprp = - \ez\btimes\grad_\perp A_\parallel(\bb{r}) .
\end{equation}
We have used compact notation in writing out the nonlinear terms: $\langle \{ A_\parallel \, , \varphi - \langle \varphi \rangle_{\gas} \} \rangle_{\gas} = \langle \{ A_\parallel (\bb{r}) \, , \varphi (\bb{r}) \} \rangle_{\gas} - \{ \langle A_\parallel \rangle_{\gas} \, , \langle \varphi \rangle_{\gas} \}$, where the first Poisson bracket involves derivatives with respect to $\bb{r}$ and the second with respect to $\bb{R}_s$. We now develop the linear theory.

\subsection{Linear gyrokinetic equation}\label{sec:gklin}

We begin by linearizing the gyrokinetic equation (\ref{eqn:ggyrokinetic}) in the fluctuations' amplitudes:
\begin{equation}\label{eqn:lineargk}
\pD{t}{g_s} + v_\parallel \pD{z}{g_s} = - \frac{q_s f^\parallel_{0s}}{\tprl{s}} \, \bigl( v_\parallel - \dupar \bigr) \biggl\langle \frac{1}{c} \pD{t}{A_\parallel} + \pD{z}{\varphi} - \pD{z}{} \frac{\bb{v}_\perp\bcdot\bb{A}_\perp}{c} \biggr\rangle_{\ggs} .
\end{equation}
Decomposing the perturbed distribution function $g_s$ and the fluctuating electromagnetic potentials $\varphi$ and $\bb{A}$ into plane-wave solutions,
\begin{gather*}
g_s(t,\bb{R}_s,v_\parallel,v_\perp) = \sum_{\bs{k}} g_{s\bs{k}}(v_\parallel,v_\perp) \, {\rm e}^{-\imag(\omega t - \bs{k\cdot R}_s)} , \\*
\varphi(t,\bb{r}) = \sum_{\bs{k}} \varphik \, {\rm e}^{-\imag(\omega t - \bs{k\cdot r})} , \quad \bb{A}(t,\bb{r}) = \sum_{\bs{k}} \bb{A}_{\bs{k}} \, {\rm e}^{-\imag(\omega t - \bs{k\cdot r})} ,
\end{gather*}
and substituting these expressions into (\ref{eqn:lineargk}), we find that
\begin{equation}\label{eqn:gsk}
g_{s\bs{k}} = -\Biggl[ {\rm J}_0 (a_s) \frac{q_s}{\tprl{s}} \biggl( \varphik - \frac{\omega\Aprlk}{k_\parallel c}  \biggr) +  \frac{2v^2_\perp}{\vthprl{s}^2} \frac{{\rm J}_1(a_s)}{a_s} \frac{\dBprlk}{B_0} \Biggr]  \frac{v_\parallel - \dupar}{v_\parallel - \omega / k_\parallel} f^\parallel_{0s} ,
\end{equation}
where ${\rm J}_0(a_s)$ and ${\rm J}_1(a_s)$ are, respectively, the zeroth- and first-order Bessel functions of $a_s \doteq k_\perp v_\perp / \Omega_s$ (cf.~equation I--B1). The Bessel functions arise from performing the ring averages in the Fourier space (see \S I--C6 for details). 

\subsection{Gyrokinetic field equations}\label{sec:gkfieldeqns}

Next, we insert (\ref{eqn:gsk}) into the field equations (\ref{eqn:gkqn})--(\ref{eqn:gkprpamp}). This procedure involves computing several $v_\parallel$-, $v_\perp$-, and Bessel-function--weighted Landau-like integrals over the mean distribution function. These integrals (denoted $\Gamma^\parallel_{\ell m}$ and $\Gamma^\perp_{\ell m}$ for integer $\ell$ and $m$) are defined in appendix \ref{app:coeffs} and evaluated to leading order in $\alpha_s \doteq (k_\perp \rho_s)^2 / 2$. Using these definitions, the quasineutrality constraint (\ref{eqn:gkqn}) and the parallel (\ref{eqn:gkprlamp}) and perpendicular (\ref{eqn:gkprpamp}) components of Amp\`{e}re's law may be written, respectively, as
\begin{align}\label{eqn:ggkqn2}
&\sum_s \frac{q^2_s \nsp \varphik}{\tprp{s}} \biggl[ \frac{\tprp{s}}{\tprl{s}} \Gamma^\parallel_{00}(\xi_s, \alpha_s) + \kzero{s} - \Gamma^\perp_{00}(\alpha_s) \biggr] 
\nonumber\\*
\mbox{} &\quad - \sum_s \frac{q^2_s \nsp \dupar \Aprlk}{c\tprp{s}} \biggl[ \frac{\tprp{s}}{\tprl{s}} \Gamma^\parallel_{01}(\xi_s,\alpha_s) + \kone{s} - \Gamma^\perp_{01}(\alpha_s) \biggr]
\nonumber\\*
\mbox{} &\quad + \sum_s q_s \nsp \biggl[ \frac{\tprp{s}}{\tprl{s}} \Gamma^\parallel_{10}(\xi_s,\alpha_s) - \Gamma^\perp_{10}(\alpha_s) \biggr] \frac{\dBprlk}{B_0}  = 0 , 
\end{align}
\begin{align}\label{eqn:ggkprlamp2}
&\sum_s \frac{q^2_s\nsp\dupar\varphik}{\tprp{s}} \biggl[ \frac{\tprp{s}}{\tprl{s}} \Gamma^\parallel_{01}(\xi_s,\alpha_s) + \kone{s} - \Gamma^\perp_{01}(\alpha_s) \biggr] + \Biggl\{ \frac{c^2k^2_\perp}{4\upi} + \sum_s \frac{q^2_s\nsp}{m_s} \nonumber\\*
\mbox{} &\quad- \sum_s \frac{q^2_s\nsp}{m_s} \frac{\tprl{s}}{\tprp{s}} \Biggl( 1 + \frac{2\duparsq}{\vthprl{s}^2} \Biggr) \biggl[ \frac{\tprp{s}}{\tprl{s}} \Gamma^\parallel_{02}(\xi_s,\alpha_s) + \ktwo{s} - \Gamma^\perp_{02}(\alpha_s) \biggr] \Biggr\} \frac{\Aprlk}{c} \nonumber\\*
\mbox{} &\quad + \sum_s q_s \nsp \dupar \biggl[ \frac{\tprp{s}}{\tprl{s}} \Gamma^\parallel_{11}(\xi_s,\alpha_s) - \Gamma^\perp_{11}(\alpha_s) \biggr] \frac{\dBprlk}{B_0} = 0,
\end{align}
and
\begin{align}\label{eqn:ggkprpamp2}
&\sum_s \betaprp{s} \frac{q_s \varphik}{\tprp{s}} \biggl[ \frac{\tprp{s}}{\tprl{s}} \Gamma^\parallel_{10}(\xi_s,\alpha_s) - \Gamma^\perp_{10}(\alpha_s) \biggr] 
\nonumber\\*
\mbox{} &\quad - \sum_s \betaprp{s} \frac{q_s\dupar \Aprlk}{c \tprp{s}} \biggl[ \frac{\tprp{s}}{\tprl{s}} \Gamma^\parallel_{11}(\xi_s,\alpha_s) - \Gamma^\perp_{11}(\alpha_s) \biggr]
\nonumber\\*
\mbox{} &\quad + \Biggl\{ \sum_s \betaprp{s} \biggl[ \frac{\tprp{s}}{\tprl{s}} \Gamma^\parallel_{20}(\xi_s,\alpha_s) - \Gamma^\perp_{20}(\alpha_s) \biggr] - 2 \Biggr\} \frac{\dBprlk}{B_0}  = 0,
\end{align}
where $\xi_s \doteq (\omega - k_\parallel \dupar)/k_\parallel \vthprl{s}$ is the dimensionless phase velocity of the fluctuations in the parallel-drifting frame. 
The first two of these equations -- quasineutrality (\ref{eqn:ggkqn2}) and the parallel component of Amp\`{e}re's law (\ref{eqn:ggkprlamp2}) -- can be combined as follows:
\begin{align}\label{eqn:lingkvorticity}
&\frac{\omega}{k_\parallel}  \sum_s \frac{q^2_s\nsp\varphik}{\tprp{s}} \Biggl\{ \kzero{s} - \Gamma^\perp_{00}(\alpha_s) - \frac{k_\parallel\dupar}{\omega} \bigl[ \kone{s} - \Gamma^\perp_{01}(\alpha_s) \bigr] \Biggr\} \nonumber\\*
\mbox{} &\quad - \frac{\omega}{k_\parallel} \sum_s \frac{q^2_s\nsp\dupar\Aprlk}{c\tprp{s}} \Biggl\{ \kone{s} - \Gamma^\perp_{01}(\alpha_s) - \frac{k_\parallel\dupar}{\omega} \bigl[ \ktwo{s} - \Gamma^\perp_{02}(\alpha_s) \bigr] \Biggr\} \nonumber\\*
\mbox{} &\quad = \sum_s \frac{q^2_s\nsp\Aprlk}{m_s c} \Biggl\{ \frac{2\alpha_s}{\beta_\perp} + 1 - \Gamma_{00}(\alpha_s) - \frac{\tprl{s}}{\tprp{s}} \bigl[ \ktwo{s} - \Gamma^\perp_{02}(\alpha_s) \bigr] \Biggr\} \nonumber\\*
\mbox{} &\quad + \frac{\omega}{k_\parallel} \sum_s q_s n_{0s} \frac{\dBprlk}{B_0} \Biggl[ \Gamma^\perp_{10}(\alpha_s) - \frac{k_\parallel\dupar}{\omega} \Gamma^\perp_{11}(\alpha_s) \Biggr] ,
\end{align}
where $\beta_\perp \doteq \sum_s\betaprp{s}$. Equation (\ref{eqn:lingkvorticity}) amounts to a statement of vorticity conservation. To see that this is the case, note that the lowest-order terms in this equation, which are first order in $(k_\perp\rho_s)^2 \ll 1$, give
\begin{equation}\label{eqn:linvorticity}
\frac{\omega}{k_\parallel} \rho_0 k^2_\perp \frac{c\varphik}{B_0} = \Biggl( \pprp{} - \pprl{} + \sum_s m_s \nsp \duparsq + \frac{B^2_0}{4\pi} \Biggr) k^2_\perp \frac{\Aprlk}{B_0} ,
\end{equation}
where $\rho_0 \doteq m_s \nsp$, $p_{\perp 0} \doteq \sum_s \pprp{s}$, and $p_{\parallel 0} \doteq \sum_s \pprl{s}$. (In obtaining (\ref{eqn:linvorticity}), we have used the constraints $\sum_s m_s \nsp \dupar = 0$, $\sum_s q_s \nsp = 0$, and $\sum_s q_s \nsp \dupar = 0$.) Using (\ref{eqn:uexb}) and (\ref{eqn:dBprp}) to relate the potentials $\varphi$ and $A_\parallel$ to the fluctuating fields $\bb{u}_\perp$ and $\delta\bb{B}_\perp$, respectively, we see that equation (\ref{eqn:linvorticity}) is just the Fourier transform of the linearized ``MHD'' vorticity equation,
\begin{equation}\label{eqn:vorticity}
\grad\btimes \Biggl[ \rho_0 \pD{t}{\bb{u}_\perp} = \Biggl( \pprp{} - \pprl{} - \sum_s m_s \nsp \duparsq + \frac{B^2_0}{4\upi} \Biggr) \nabla_\parallel \frac{\dBprp}{B_0} \Biggr] ,
\end{equation}
modified by the presence of background pressure anisotropy and interspecies drifts.

\subsection{Gyrokinetic dispersion relation for arbitrary $f_{0s}$}

The linear dispersion relation for pressure-anisotropic, multi-species gyrokinetics is obtained by combining (\ref{eqn:ggkqn2})--(\ref{eqn:ggkprpamp2}) and demanding non-zero solutions. We have found its general form to be neither physically illuminating nor particularly useful for our purposes. In lieu of numerically computing its general solution across an expansive parameter space, we opt to examine a number of illustrative asymptotic limits for which analytical solutions may be obtained. These limits are treated in the remainder of this section, and are supported by exact numerical solutions relevant to bi-Maxwellian $f_{0s}$ in a hydrogenic plasma presented in \S\ref{sec:numerical}.

Before proceeding with this programme, however, it is useful to examine the general dispersion relation for a non-Maxwellian plasma without interspecies drifts in the limit $\omega \rightarrow 0$. Doing so reveals that solutions are stable if and only if
\begin{equation}\label{eqn:generalfirehose}
1 + \sum_s \frac{\betaprl{s}}{2} \Biggl[ \frac{\tprp{s}}{\tprl{s}} \frac{1-\Gamma^\parallel_{02}(0,\alpha_s)}{\alpha_s} - \frac{\ktwo{s}-\Gamma^\perp_{0s}(\alpha_s)}{\alpha_s} \Biggr] \ge 0 
\end{equation}
and
\begin{equation}\label{eqn:generalmirror}
1 - \sum_s \betaprp{s} \Biggl[ \frac{\tprp{s}}{\tprl{s}} \frac{\Gamma^\parallel_{20}(0,\alpha_s)}{2} - \frac{\Gamma^\perp_{20}(\alpha_s)}{2} \Biggr] \ge - { {\displaystyle \Biggl\{ \sum_s c_s \biggl[ \frac{\tprp{s}}{\tprl{s}} \Gamma^\parallel_{10}(0,\alpha_s) - \Gamma^\perp_{10}(\alpha_s) \biggr] \Biggr\}^2} \over {\displaystyle \sum_s \frac{2c^2_s}{\betaprp{s}}  \biggl[ \frac{\tprp{s}}{\tprl{s}} \Gamma^\parallel_{00}(0,\alpha_s) + \kzero{s} - \Gamma^\perp_{00}(\alpha_s) \biggr]} } ,
\end{equation}
where $c_s \doteq q_s \nsp/e\nem$ is the charge-weighted ratio of number densities; note that $c_e = -1$ and $\sum_s c_s = 0$. While it may not be plainly evident at this stage, (\ref{eqn:generalfirehose}) and (\ref{eqn:generalmirror}) are, respectively, the {\em firehose} and {\em mirror} stability criteria. We will make repeated reference to these stability criteria throughout this paper as we take various limits of the general dispersion relation. Note that the instabilities themselves, whose growth rates peak at parallel scales set by finite-Larmor-radius effects \citep[e.g.][]{ks67,dv68,yoon93,hm00,hellinger07,rosin11,rincon15} and whose nonlinear saturation relies on particle trapping and/or pitch-angle scattering of particles \citep[e.g.][]{kss14,riquelme15}, fall outside of the gyrokinetic ordering employed here. (Though, see \citet{pj13} and \citet{pj17} for attempts to modify gyrokinetics to describe the oblique firehose and mirror instabilities.)

\subsection{Long-wavelength limit for arbitrary $f_{0s}$: Linear KRMHD}\label{sec:linearKRMHD}

We first examine the long-wavelength ($k_\perp \rho_s \ll 1$) limit of (\ref{eqn:ggkqn2})--(\ref{eqn:ggkprpamp2}), which is obtained by taking the leading-order expressions for the $\Gamma_{\ell m}(\xi_s, \alpha_s)$ factors given in (\ref{eqn:gammas}) and (\ref{eqn:gammaprl}) and by dropping the $k^2_\perp c^2 / 4\upi$ term in (\ref{eqn:ggkprlamp2}). In this approximation, the parallel Amp\`{e}re's law is equivalent to the quasineutrality constraint. The remaining field equations ({\em viz.}~quasineutrality and the perpendicular Amp\`{e}re's law) may be written in the following compact form:
\begin{equation}\label{eqn:longwavelength}
\left[
\begin{array}{cc}
\sum_s c^2_s \czero{s}(\xi_s) {\displaystyle \frac{2}{\betaprl{s}}} & \sum_s c_s \Delta_{1s}(\xi_s)\\* [3em]
\sum_s c_s \Delta_{1s}(\xi_s) & \sum_s \betaprp{s}\Delta_{2s}(\xi_s) - 1
\end{array}
\right] \negthickspace \left[
\begin{array}{c}
{\displaystyle \frac{4\upi e\nem}{B^2_0} \biggl( \varphik - \frac{\omega\Aprlk}{k_\parallel c} \biggr)} \\* [1.5em]
{\displaystyle \frac{\dBprlk}{B_0}}
\end{array}
\right] = 0 ,
\end{equation}
where
\begin{equation}\label{eqn:Delta_ells}
\Delta_{\ell s}(\xi_s) \doteq \cell{s}(\xi_s) \frac{\tprp{s}}{\tprl{s}} - 1
\end{equation}
for integer $\ell$.

There are two types of solutions to (\ref{eqn:longwavelength}). The first is straightforwardly obtained by setting the determinant of the matrix to zero, yielding the dispersion relation
\begin{equation}
\Biggl[ \sum_s c^2_s \czero{s}(\xi_s) \frac{2}{\betaprl{s}} \Biggr] \Biggl[ \sum_s \betaprp{s} \Delta_{2s}(\xi_s) - 1 \Biggr] = \Biggl[ \sum_s c_s \Delta_{1s}(\xi_s) \Biggr]^2 .
\end{equation}
This equation is identical to the KRMHD dispersion relation for the compressive fluctuations (cf.~I--B9).\footnote{There is a typographical error in (I--B8): the minus sign there should be a plus sign. This error does not affect any of the subsequent formulae or analysis in Paper I.} In the absence of interspecies drifts, its solutions are stable if
\begin{equation}
1 - \sum_s \betaprp{s} \biggl[ \frac{\tprp{s}}{\tprl{s}} \ctwo{s}(0) - 1 \biggr] \ge - { {\displaystyle \Biggl\{ \sum_s c_s \biggl[ \frac{\tprp{s}}{\tprl{s}} \cone{s}(0) - 1 \biggr] \Biggr\}^2 } \over {\displaystyle \sum_s \frac{2c^2_s}{\betaprp{s}} \frac{\tprp{s}}{\tprl{s}} \czero{s}(0) } } ,
\end{equation}
which is the long-wavelength limit of the general mirror stability criterion (\ref{eqn:generalmirror}) (see also (I--B14) and \citet{hellinger07}).

The other type of solution is obtained by stipulating that $\dBprlk = 0$ and thus requiring $\varphik = \omega\Aprlk / k_\parallel c$. If we write the potentials in terms of the perpendicular velocity and magnetic-field fluctuations via (\ref{eqn:uexb}) and (\ref{eqn:dBprp}), this gives $\bb{u}_{\perp\bs{k}} = - (\omega / k_\parallel) (\dBprpk / B_0)$, which we recognize as the eigenvector describing the Alfv\'{e}nic fluctuations. To obtain the corresponding eigenvalues, we use $\varphik = \omega\Aprlk / k_\parallel c$ in (\ref{eqn:ggkprlamp2}) and examine the terms that are leading order in $k_\perp \rho_s$. The result is equivalent to (\ref{eqn:linvorticity}), which yields the Alfv\'{e}n-wave eigenvalues (cf.~I--3.1),
\begin{equation}\label{eqn:alfvenwave}
\omega = \pm k_\parallel \valf \Biggl[ 1 + \sum_s \frac{\betaprl{s}}{2} \Biggl( \Delta_s - \frac{2\duparsq}{\vthprl{s}^2} \Biggr) \Biggr]^{1/2} \doteq \pm k_\parallel \valfeff,
\end{equation}
where we have defined the effective Alfv\'{e}n speed $\valfeff$. For $\pprp{} - \pprl{} - \sum_s m_s \nsp \duparsq < 0$, the speed at which deformations in the magnetic field are propagated is effectively reduced by the excess parallel pressure, which undermines the restoring force exerted by the tension of the magnetic-field lines. When
\begin{equation}\label{eqn:firehose}
1 + \sum_s \frac{\betaprl{s}}{2} \Biggl( \Delta_s - \frac{2\duparsq}{\vthprl{s}^2} \Biggr) < 0 ,
\end{equation}
the effective Alfv\'{e}n speed becomes imaginary and the firehose instability results. Neglecting interspecies drifts, equation (\ref{eqn:firehose}) is the same as the long-wavelength limit of the general firehose stability criterion (\ref{eqn:generalfirehose}).

Thus, at long wavelengths, the linear gyrokinetic theory correctly reduces to the linear theory of KRMHD (Paper I).

\subsection{Gyrokinetic dispersion relation for an electron-ion bi-Maxwellian plasma}\label{sec:lineargk_biMax}

As the ion gyroscale is approached, $k_\perp \rho_i \sim 1$, the Alfv\'{e}n waves are no longer decoupled from the compressive fluctuations and, therefore, can be collisionlessly damped. Nonlinearly, the fraction of the Alfv\'{e}n-wave energy that remains in the turbulent cascade is channeled to yet smaller scales, where the Alfv\'{e}n-wave cascade transitions into a cascade of dispersive KAWs. This cascade proceeds further to electron Larmor scales, $k_\perp \rho_e \sim 1$, at which point the KAWs are Landau-damped on the electrons. In this section, the linear theory of Maxwellian collisionless gyrokinetics that forms the basis of these statements (\citealt{howes06}, S09) is extended to a plasma consisting of a single ion species and electrons, each with a bi-Maxwellian equilibrium distribution function.\footnote{While this paper was in an advanced stage of preparation, a paper by \citet{verscharen17} appeared in which the linear gyrokinetic theory for a bi-Maxwellian ion-electron plasma was derived using the nonlinear gyrokinetic theory presented in Paper I and the long-wavelength limit was analyzed. Where there is overlap with the results presented in this section, agreement is found.}

For $f_{0s} = f_{\textrm{bi-M},s}(v_\parallel,v_\perp)$, the integrals over the perpendicular velocity space in the $\Gamma_{\ell m}(\xi_s,\alpha_s)$ coefficients (\ref{eqn:gammas}) and (\ref{eqn:gammaprl}) may be expressed in terms of the zeroth-order (${\rm I}_0$) and first-order (${\rm I}_1$) modified Bessel functions:
\begin{subequations}\label{eqn:gkG}
\begin{align}
\Gamma_0(\alpha_s) &\doteq \int^\infty_0 \frac{\rmd v^2_\perp}{\vthprp{s}^2} \, \bigl[ {\rm J}_0(a_s) \bigr]^2 \, {\rm e}^{-v^2_\perp / \vthprp{s}^2} = {\rm I}_0(\alpha_s) \, {\rm e}^{-\alpha_s} , \\*
\Gamma_1(\alpha_s) &\doteq \int^\infty_0 \frac{\rmd v^2_\perp}{\vthprp{s}^2} \, \frac{v^2_\perp}{\vthprp{s}^2} \frac{2{\rm J}_1(a_s){\rm J}_1(a_s)}{a_s}  \, {\rm e}^{-v^2_\perp / \vthprp{s}^2} = \bigl[ {\rm I}_0(\alpha_s) - {\rm I}_1(\alpha_s) \bigr] \, {\rm e}^{-\alpha_s} , \\*
\Gamma_2(\alpha_s) &\doteq \int^\infty_0 \frac{\rmd v^2_\perp}{\vthprp{s}^2} \biggl[ \frac{2v^2_\perp}{\vthprp{s}^2} \frac{{\rm J}_1(a_s)}{a_s} \biggr]^2 {\rm e}^{-v^2_\perp / \vthprp{s}^2} = 2 \Gamma_1(\alpha_s) .
\end{align}
\end{subequations}
In addition, we can express the integrals over the parallel velocity space in the $\Gamma^\parallel_{\ell m}(\xi_s,\alpha_s)$ coefficients in terms of the (Maxwellian) plasma dispersion function $Z_{\rm M}(\xi)$:
\begin{equation}\label{eqn:gkZ}
\frac{1}{\sqrt{\upi}} \int^\infty_{-\infty} \frac{\rmd v_\parallel}{\vthprl{s}} \, \frac{v_\parallel}{v_\parallel - \omega / k_\parallel} \, {\rm e}^{-v_\parallel^2 / \vthprl{s}^2} = 1 + \xi_s Z_{\rm M}(\xi_s) ,
\end{equation}
where $\xi_s \doteq \omega / k_\parallel \vthprl{s}$ is the dimensionless phase speed \citep{fc61}. Thus, combining (\ref{eqn:gkG}) and (\ref{eqn:gkZ}), we have
\begin{equation}\label{eqn:gkGp}
\Gamma^\parallel_{0\ell} (\alpha_s) = \Gamma_\ell(\alpha_s) \bigl[ 1 + \xi_s Z_{\rm M} (\xi_s) \bigr]
\end{equation}
for integer $\ell$. Similarly, $C^\perp_{\ell s} = 1$.

With these simplifications, equations (\ref{eqn:ggkqn2})--(\ref{eqn:ggkprpamp2}) may be written succinctly in matrix form:
\begin{equation}\label{eqn:gkmatrix}
\left[
\begin{array}{ccc}
\mc{A} & \mc{A} - \mc{B} & \mc{C} \\ [1.6em]
\mc{A} - \mc{B} & \mc{A} - \mc{B} - {\displaystyle \frac{\alpha_\ast}{\overline{\omega}^2} } & \mc{C} + \mc{E} \\ [1.6em]
\mc{C} & \mc{C} + \mc{E} & \mc{D} - {\displaystyle \frac{2}{\betaprp{i}} }
\end{array}
\right] \negthickspace \left[
\begin{array}{c}
\varphik \\ [1.5em]
- {\displaystyle \frac{\omega\Aprlk}{k_\parallel c}} \\ [1.5em]
{\displaystyle \frac{\tprp{i}}{q_i} \frac{\dBprlk}{B_0} }
\end{array}
\right]
= 0,
\end{equation}
where we have employed the shorthand notation (cf.~\S2.6 of \citealt{howes06})
\begin{subequations}\label{eqn:howescoeffs}
\begin{align}
\mc{A} &\doteq 1 + \Gamma_0(\alpha_i) \biggl[ \Delta_i  + \xi_i Z_{\rm M}(\xi_i) \frac{\tprp{i}}{\tprl{i}} \biggr] + \frac{\tauprp{i}}{Z_i} \biggl\{ 1 + \Gamma_0(\alpha_e) \biggl[ \Delta_e + \xi_e Z_{\rm M}(\xi_e) \frac{\tprp{e}}{\tprl{e}} \biggr] \biggr\} ,\\*
\mc{B} &\doteq 1 - \Gamma_0(\alpha_i) + \frac{\tauprp{i}}{Z_i} \bigl[ 1 - \Gamma_0(\alpha_e) \bigr] ,\\*
\mc{C} &\doteq \Gamma_1(\alpha_i) \biggl[ \Delta_i + \xi_i Z_{\rm M}(\xi_i) \frac{\tprp{i}}{\tprl{i}} \biggr] - \Gamma_1(\alpha_e) \biggl[ \Delta_e + \xi_e Z_{\rm M}(\xi_e) \frac{\tprp{e}}{\tprl{e}} \biggr]  ,\\*
\mc{D} &\doteq \Gamma_2(\alpha_i) \biggl[ \Delta_i + \xi_i Z_{\rm M}(\xi_i) \frac{\tprp{i}}{\tprl{i}} \biggr] + \Gamma_2(\alpha_e) \frac{Z_i}{\tauprp{i}} \biggl[ \Delta_e + \xi_e Z_{\rm M}(\xi_e) \frac{\tprp{e}}{\tprl{e}} \biggr] ,\\*
\mc{E} &\doteq \Gamma_1(\alpha_i) - \Gamma_1(\alpha_e) , \\*
\alpha_\ast &\doteq \alpha_i + \frac{\betaprl{i}}{2} \Delta_i \bigl[ 1 - \Gamma_0(\alpha_i) \bigr] + \frac{\betaprl{e}}{2} \Delta_e \bigl[ 1 - \Gamma_0(\alpha_e) \bigr] \frac{m_i}{Z_i m_e} \frac{\tauprp{i}}{Z_i} ,\label{eqn:alphatilde}
\end{align}
\end{subequations}
and $\overline{\omega} \doteq \omega / k_\parallel \valf$. 

Setting the determinant of the matrix in (\ref{eqn:gkmatrix}) equal to zero yields the gyrokinetic dispersion relation, which may be written in the following compact form after multiplying by $\mc{A}$ (cf.~eq.~41 of \citealt{howes06}):
\begin{equation}\label{eqn:gkdisprel}
\underwrite{Alfv\'{e}n}{\biggl( \frac{\alpha_\ast \mc{A}}{\overline{\omega}^2} - \mc{A}\mc{B} +  \mc{B}^2 \biggr)} \underwrite{slow}{\biggl( \frac{2\mc{A}}{\betaprp{i}} - \mc{A}\mc{D} + \mc{C}^2 \biggr)} = \underwrite{FLR coupling}{\bigl( \mc{A}\mc{E} + \mc{B}\mc{C} \bigr)^2} .
\end{equation}
We have labelled each factor in the dispersion relation (\ref{eqn:gkdisprel}) according to its physical meaning: the first term in parentheses corresponds to the Alfv\'{e}n-wave branch, the second corresponds to the slow-wave branch, and the right-hand side represents the finite-Larmor-radius (FLR) coupling between the two branches that occurs as $k_\perp \rho_s$ approaches and exceeds unity. For a hydrogenic plasma (i.e.~$Z_i = 1$, $m_i / m_e \approx 1836$), the complex eigenvalue solution $\overline{\omega}$ to (\ref{eqn:gkdisprel}) depends on five dimensionless parameters: the ratio of the ion Larmor radius to the perpendicular wavelength, $k_\perp \rho_i$; the ion plasma $\betaprl{i}$; the ion-electron perpendicular temperature ratio, $\tauprp{i}$; the ion pressure anisotropy, $\Delta_i$; and the electron pressure anisotropy, $\Delta_e$.\footnote{Alternatively, one may specify $\tauprp{i}$, $\tauprl{i}$, and $\Delta_i$, which, combined, implies a choice of $\Delta_e$.} In what follows, we vary these parameters to obtain asymptotic limits of the dispersion relation (\ref{eqn:gkdisprel}).

\subsubsection{KRMHD limit: $Z_i m_e / m_i$, $k_\perp \rho_i \rightarrow 0$}\label{sec:gkkrmhd}

In the limit where $k_\perp \rho_i$ and the electron-ion mass ratio $Z_i m_e / m_i$ are both asymptotically small, one should recover the linear theory for bi-Maxwellian KRMHD (cf.~\S I--4.4 and \S\ref{sec:linearKRMHD}). In this limit, $\mc{B} \simeq \alpha_i$, $\mc{E} \simeq -(3/2)\alpha_i$, and the dispersion relation (\ref{eqn:gkdisprel}) becomes
\begin{equation}\label{eqn:gkkrmhd}
\biggl( \frac{\alpha_\ast}{\overline{\omega}^2} - \alpha_i \biggr) \biggl( \frac{2\mc{A}}{\betaprp{i}} - \mc{A}\mc{D} + \mc{C}^2 \biggr) = 0 .
\end{equation}
%
%
%

Setting the first factor of (\ref{eqn:gkkrmhd}) to zero and simplifying $\alpha_\ast \simeq \alpha_i [ 1 + (\betaprl{i}/2) \Delta_i + (\betaprl{e}/2) \Delta_e]$ (see (\ref{eqn:alphatilde})), we obtain the dispersion relation for undamped Alfv\'{e}n waves modified by the ion and electron pressure anisotropies:
\begin{equation}\label{eqn:gkaw}
\omega = \pm k_\parallel \valf \sqrt{ 1 + \frac{\betaprl{i}}{2} \Delta_i + \frac{\betaprl{e}}{2} \Delta_e }.
\end{equation}
Again, when
\begin{equation}\label{eqn:iefirehose}
1 + \frac{\betaprl{i}}{2} \Delta_i + \frac{\betaprl{e}}{2} \Delta_e < 0 ,
\end{equation}
the effective Alfv\'{e}n speed becomes imaginary and the firehose instability results (cf.~(\ref{eqn:firehose})).

Setting the second factor of (\ref{eqn:gkkrmhd}) to zero, and using the leading-order expressions for $\mc{C} \simeq (\tprp{i} / \tprl{i}) [ 1 + \xi_i Z_{\rm M}(\xi_i) - \tauprl{i}/\tauprp{i} ]$ and $\mc{D} \simeq 2\mc{C} + 2 ( 1 + Z_i/\tauprp{i}) \Delta_e$, we obtain (after some straightforward but tedious algebra) the dispersion relation for the compressive fluctuations (i.e., those with density and magnetic-field-strength fluctuations),
\begin{equation}\label{eqn:compressive}
\bigl[ 1 + \xi_i Z_{\rm M}(\xi_i) - \Lambda^+ \bigr] \bigl[ 1 + \xi_i Z_{\rm M}(\xi_i) - \Lambda^- \bigr] = 0,
\end{equation}
where
\begin{equation}\label{eqn:lambdapm}
\Lambda^\pm = - \frac{\tauprl{i}}{Z_i} + \frac{\tprl{i}}{\tprp{i}}\frac{\varsigma_i}{\betaprp{i}} \pm \sqrt{ \biggl( \frac{\tauprl{i}}{\tauprp{i}} + \frac{\tauprl{i}}{Z_i} \biggr)^2 + \biggl( \frac{\tprl{i}}{\tprp{i}} \frac{\varsigma_i}{\betaprp{i}} \biggr)^2 } ,
\end{equation}
and
\begin{equation}\label{eqn:varsigma}
\varsigma_i \doteq 1 - \beta_\perp \Delta_e .
\end{equation}
Equation (\ref{eqn:compressive}) indicates two compressive branches of solutions, a ``+'' branch (corresponding to the compressive-wave eigenvector $G^+$) and a ``$-$'' branch (corresponding to the other compressive-wave eigenvector $G^-$), which are discussed in \S\S I--4.3 and I--4.4 (see (I--4.20{\it a,b}) in particular) and consist of linear combinations of density and magnetic-field-strength fluctuations. An important limit of (\ref{eqn:compressive}) is obtained for $\betaprp{s} \sim \betaprl{s} \sim 1/\Delta_s \gg 1$, in which the ``+'' compressive branch, consisting primarily of magnetic-field-strength fluctuations, is collisionlessly damped at a rate
\begin{equation}\label{eqn:slowmode}
\gamma \doteq -\imag\omega = - \frac{|k_\parallel|\valf}{\sqrt{\upi\betaprl{i}}} \frac{\tprl{i}^2}{\tprp{i}^2} \Biggl( 1 - \sum_s \betaprp{s} \Delta_s \Biggr) .
\end{equation}
(In this limit, the ``$-$'' branch consists mainly of density fluctuations and is strongly damped with ${\rm Im}(\xi_i) \sim 1$.) Equation (\ref{eqn:slowmode}) captures the effect of pressure anisotropy on the \citet{barnes66} damping of slow modes (in the limit $k_\parallel / k_\perp \ll 1$). The damping is due to Landau-resonant particles interacting with the mirror force associated with the magnetic compressions in the wave. When 
\begin{equation}\label{eqn:iemirror}
1 - \betaprp{i} \Delta_i - \betaprp{e} \Delta_e < 0,
\end{equation}
the proportional increase (for $\Delta_s > 0$) in the number of large-pitch-angle particles in the magnetic troughs ($\dBprl < 0$) of the slow mode results in more perpendicular pressure than can be stably balanced by the magnetic pressure. The result is the {\em mirror instability} \citep[see, e.g.,][]{sk93}. A more general criterion for the long-wavelength mirror instability in a single-ion-species plasma is given by (I--B14) -- see also (\ref{eqn:generalmirror}); for an electron-ion bi-Maxwellian plasma with arbitrary $\betaprp{s}$ and $\betaprl{s}$, it becomes
\begin{equation}\label{eqn:iemirror2}
1 - \betaprp{i} \Delta_i - \betaprp{e} \Delta_e < - \frac{\betaprl{e}}{2} \frac{ ( \Delta_i - \Delta_e)^2 }{1 + Z_i / \tauprl{i}} ,
\end{equation}
a more restrictive condition than (\ref{eqn:iemirror}). The difference between the right-hand sides of (\ref{eqn:iemirror}) and (\ref{eqn:iemirror2}) is due to the stabilizing effect of the parallel electric field, which is small when $\betaprp{s} \sim \betaprl{s} \sim 1/\Delta_s \gg 1$. We refer the reader to \S I--4.4.2 for further analysis and discussion.

\subsubsection{KAW limit: $k_\perp \rho_e \ll 1 \ll k_\perp \rho_i$, $\overline{\omega} \sim \mc{O}(\alpha^{1/2}_i)$}\label{sec:linKAW}

In the limit $k_\perp \rho_e \ll 1 \ll k_\perp \rho_i$, we have $\Gamma_0(\alpha_i)$, $\Gamma_1(\alpha_i) \rightarrow 0$ for the ions and $\Gamma_0(\alpha_e) \simeq \Gamma_1(\alpha_e) \simeq 1$ for the electrons, whence $\mc{B} \simeq 1$ and $\mc{E} \simeq -1$ in (\ref{eqn:gkdisprel}). We also drop the electron plasma dispersion functions to lowest order in $k_\perp \rho_e$, a simplification that will be justified {\it a posteriori}. The gyrokinetic dispersion relation (\ref{eqn:gkdisprel}) becomes
\begin{equation}
\biggl( \frac{\alpha_\ast \mc{A}}{\overline{\omega}^2} - \mc{A} + 1 \biggr) \biggl( \frac{2\mc{A}}{\betaprp{i}} - \mc{A}\mc{D} + \mc{C}^2 \biggr) = \bigl( -\mc{A} + \mc{C} \bigr)^2 .
\end{equation}
With $\betaprp{s} \sim \betaprl{s} \sim \Delta^{-1}_s \sim 1$ and $\overline{\omega} \sim \mc{O}(\alpha^{1/2}_i)$, we have $\mc{A} \simeq 1 + \tprp{i}/Z_i\tprl{e}$, $\mc{C} \simeq - \Delta_e$, and $\mc{D} \simeq 2(Z_i/\tauprp{i}) \Delta_e$.\footnote{The final term in the definition of $\alpha_\ast$ (see (\ref{eqn:alphatilde})) must be retained in this limit, despite its dependence on the higher-order term $1-\Gamma_0(\alpha_e)$. This is because its leading-order term is proportional to $\alpha_e (m_i/Z_i m_e) (\tauprp{i}/Z_i) = \alpha_i \gg 1$. Thus, $\alpha_\ast \simeq \alpha_i \, ( 1 + \betaprl{e} \Delta_e / 2)$. This correction to $\alpha_i$ stems from the difference between the Boltzmann response (\ref{eqn:boltzmann}) and its ring average, and modifies the effective Alfv\'{e}n speed in both Alfv\'{e}n-wave (see (\ref{eqn:gkaw})) and KAW (see (\ref{eqn:kaw})) dispersion relations.} The solutions are then
\begin{equation}\label{eqn:kaw}
\omega = \pm \frac{k_\parallel \valf k_\perp \rho_i}{ \sqrt{\betaprp{i} +  2 / (1 +  Z_i \tprl{e} / \tprp{i} ) - 2 \mc{K}} } \, \sqrt{ \bigl( 1 - \betaprp{e} \Delta_e + \mc{K} \bigr)  \biggl( 1 + \frac{\betaprl{e}}{2} \Delta_e \biggr) } ,
\end{equation}
where
\begin{equation}\label{eqn:Kcoeff}
\mc{K} \doteq \frac{\betaprl{e}}{2} \frac{\Delta^2_e}{1 + Z_i \tprl{e} / \tprp{i}} > 0 
\end{equation}
is a stabilizing term due to the parallel electric field at sub-ion-Larmor scales (cf.~the right-hand side of (\ref{eqn:iemirror2})). Equation (\ref{eqn:kaw}) is a generalisation of the standard KAW dispersion relation \citep[e.g.][]{kingsep90},
\[
\omega = \pm \frac{k_\parallel \valf k_\perp \rho_i}{\sqrt{\beta_i + 2/( 1 + Z_i T_{0e} / T_{0i} )}} ,
\]
for bi-Maxwellian plasmas. Note that, for this solution, $\xi_e \sim \mc{O}(k_\perp \rho_e) \ll 1$, as promised. The KAW dispersion relation for a bi-kappa ion-electron plasma is given by (\ref{eqn:KAW_bikappa}).

The equations governing the corresponding sub-ion-scale fluctuations in the electron density, the parallel flow velocity, and the magnetic-field strength are obtained from the gyrokinetic field equations (\ref{eqn:gkqn})--(\ref{eqn:gkprpamp}) after expanding in $k_\perp \rho_e \ll 1 \ll k_\perp \rho_i$. The resulting equations are equivalent to (I--C88)--(I--C90) with all the $\Gamma_{\ell m}(\alpha_i)$ coefficients set to zero, {\it viz.},
\begin{subequations}\label{eqn:KAWfluctuations}
\begin{align}
\frac{\dne}{\nem} &= - \frac{Z_i e\varphi}{\tprp{i}}  = - \frac{2}{\sqrt{\betaprp{i}}} \frac{\Phi}{\rho_i v_{\rm A}} , \label{eqn:KAWdn} \\*
\delta u_{\parallel e} &= \frac{c}{4\upi e\nem} \nabla^2_\perp A_\parallel = - \frac{\rho_i}{\sqrt{\betaprp{i}}} \nabla^2_\perp \Psi ,\label{eqn:KAWuprle} \\*
\delta u_{\parallel i} &= - \frac{c}{4\upi Z_i e\nip} \betaprl{i}\Delta_i \frac{A_\parallel}{\rho^2_i} = \frac{\rho_i}{\sqrt{\betaprp{i}}} \betaprl{i}\Delta_i \frac{\Psi}{\rho^2_i} ,\label{eqn:KAWuprli}\\
 \frac{\delta B_\parallel}{B_0} &= \frac{\betaprp{i}}{2} \biggl( 1 + \frac{Z_i}{\tauprp{i}} \biggr) \biggl( 1 -  \frac{\betaprp{e}}{2} \Delta_e \biggr)^{-1} \frac{Z_i e\varphi}{\tprp{i}}  \nonumber\\*
 \mbox{} &= \sqrt{\betaprp{i}} \biggl( 1 + \frac{Z_i}{\tauprp{i}} \biggr) \biggl( 1 -  \frac{\betaprp{e}}{2} \Delta_e \biggr)^{-1} \frac{\Phi}{\rho_i \valf} \label{eqn:KAWdBprl} ,
\end{align}
\end{subequations}
where we have introduced the stream and flux functions $\Phi$ and $\Psi$ (see (I--C54{\it a},{\it b})) via
\begin{equation}\label{eqn:PhiPsi}
\varphi = \frac{B_0}{c} \Phi \quad {\rm and} \quad A_\parallel = - \frac{B_0}{v_{\rm A}} \Psi .
\end{equation}
Equations (\ref{eqn:KAWfluctuations}{\it a},{\it b},{\it c},{\it d}) are to be compared with equations (221)--(223) of S09. They reflect the fact that, for $k_\perp \rho_i \gg 1$, the ion response is effectively Boltzmann (see (\ref{eqn:boltzmann})), with the gyrokinetic response $\widetilde{h}_i$ contributing nothing either to the fields or to the flows. Note that the parallel ion flow velocity (\ref{eqn:KAWuprli}) is $\sim$$(k_\perp \rho_i)^{-2} \ll 1$ smaller than the corresponding pressure-anisotropic terms in the parallel electron flow velocity (\ref{eqn:KAWuprle}), and thus contributes almost nothing to the parallel current.

There are several things to note about the KAW dispersion relation (\ref{eqn:kaw}). First, KAWs in a bi-Maxwellian plasma are subject to {\em both} the mirror {\em and} firehose instability thresholds, whose geometric mean appears as the final term in (\ref{eqn:kaw}), repeated here:
\[
\omega = \pm \frac{k_\parallel \valf k_\perp \rho_i}{ \sqrt{\betaprp{i} +  2 / (1 +  Z_i \tprl{e} / \tprp{i} ) - 2 \mc{K}} } \,
\sqrtexplained{
\underwrite{mirror factor}{\bigl( 1 - \betaprp{e} \Delta_e + \mc{K} \bigr)}  \underwrite{firehose factor}{\biggl( 1 + \frac{\betaprl{e}}{2} \Delta_e \biggr)}
}
\]
This makes sense, as Alfv\'{e}nic and compressive fluctuations are coupled in the KAW by finite-Larmor-radius effects. Indeed, the eigenfunctions corresponding to the frequencies (\ref{eqn:kaw}) are (cf.~(231) of S09)
\begin{align}\label{eqn:Theta}
\Theta^\pm_{\bs{k}} &= \sqrt{ \biggl(1 + \frac{Z_i\tprl{e}}{\tprp{i}} \biggr) \biggl[ 2 + \betaprp{i}  \biggl(1 + \frac{Z_i\tprl{e}}{\tprp{i}} \biggr) - \betaprl{e} \Delta^2_e \biggr]} ~\frac{(1 - \betaprp{e}\Delta_e + \mc{K})^{1/2}}{1-\betaprp{e}\Delta_e/2} \frac{\Phi_{\bs{k}}}{\rho_i} \nonumber\\*
\mbox{} &\mp \sqrt{ 1+ \frac{\betaprl{e}}{2} \Delta_e} ~k_\perp \Psi_{\bs{k}} .
\end{align}
We see that the factor $1 + \betaprl{e}\Delta_e/2$, related to the firehose threshold, is associated with the Alfv\'{e}nic fluctuation $\delta B_{\perp\bs{k}} \propto k_\perp \Psi_{\bs{k}}$; the factor $1 - \betaprp{e} \Delta_e + \mc{K}$, related to the mirror threshold, is associated with the compressive fluctuation $\delta B_{\parallel\bs{k}} \propto \Phi_{\bs{k}} / \rho_i$.\footnote{To obtain these thresholds, let $\alpha_i \gg 1$ and $\alpha_e \ll 1$ in the general firehose (\ref{eqn:generalfirehose}) and mirror (\ref{eqn:generalmirror}) stability criteria. The electron $\Gamma_{\ell m}$ factors all reduce to their long-wavelength counterparts (see (\ref{eqn:gammas}) and (\ref{eqn:gammaprl})), while $\Gamma^\parallel_{\ell m}(0,\alpha_i) = \Gamma^\perp_{\ell m}(0,\alpha_i) = 0$ for all $\ell$ and $m$. The result is $1 + (\betaprl{e}/2)\Delta_e \ge 0$ for firehose stability and $1 - \betaprp{e}\Delta_e \ge -\mc{K}$ for mirror stability. The firehose-unstable solution to (\ref{eqn:kaw}) that occurs for $\betaprl{e}\Delta_e < -2$ is likely related to the oblique ``electron firehose instability'' found by \citet{lh00}, which is non-resonant and purely growing. This section (\S\ref{sec:linKAW}) provides a physical explanation and approximate analytical description of this mode.}  The KAW eigenfunctions $\Theta^\pm_{\bs{k}}$ combine both effects.

Secondly, the ion pressure anisotropy does not appear in (\ref{eqn:kaw}). Physically, this is because the ion response is essentially Boltzmann (\ref{eqn:boltzmann}), which produces an isothermal pressure response
\begin{subequations}\label{eqn:KAWdp}
\begin{equation}\label{eqn:KAWdpi}
\dpprp{i} = \tprp{i} \dni \quad{\rm and}\quad \dpprl{i} = \tprl{i} \dni.
\end{equation}
By contrast, the electron pressure response is (see (I--2.45{\it a},{\it b}))
\begin{equation}\label{eqn:KAWdpe}
\dpprp{e} = \tprp{e} \dne - \pprp{e} \Delta_e \frac{\dBprl}{B_0} \quad{\rm and}\quad \dpprl{e} = \tprl{e} \dne ,
\end{equation}
\end{subequations}
so a magnetic-field-strength perturbation produces a perpendicular electron temperature perturbation proportional to the electron pressure anisotropy. This difference arises because, at scales satisfying $k_\perp \rho_i \gg 1$, the ions do not ``see'' the magnetic-field-strength fluctuation, which varies rapidly along the ion gyro-orbit and is thus ring-averaged away. In this situation, the ions have no reason to adjust their perpendicular pressure according to the changes in the magnetic-field strength. Put differently, because $\mu$ conservation for the ions holds with respect to the magnetic-field strength averaged over the particle orbit (see (\ref{eqn:mus})), fluctuations in field strength ($\dBprl$) as seen by the ions are reduced by a factor of $2J_1(a_i)/a_i \sim a^{-3/2}_i \ll 1$. For the  electrons, on the other hand, we have $2J_1(a_e)/a_e \simeq 1$, so $\dpprp{e}$ fully adjusts to the fluctuating field strength (see the second term in (\ref{eqn:KAWdpe}), as well as \S I--2.5.2). 

The ion pressure anisotropy is also absent from the firehose factor $1+(\betaprl{e}/2)\Delta_e$ in (\ref{eqn:kaw}) and (\ref{eqn:Theta}) for a similar reason: pressure-anisotropy corrections to the effective tension in the field lines stem from the $\eb\delta\eb (\pprp{s} - \pprl{s})$ term in the perturbed magnetized pressure tensor, which is only relevant if species $s$ can ``see'' the field fluctuation $\delta\eb$. This result explains why sub-ion-Larmor cascades of KAWs were observed in the hybrid-kinetic simulations of \citet{kss14} and \citet{kunz16}, despite the larger scales being driven mirror unstable by positive ion pressure anisotropy: in those calculations, the electrons were assumed to be pressure-isotropic and so KAWs were stable.

Thirdly, the KAW in the gyrokinetic limit satisfies perpendicular pressure balance:
\begin{equation}\label{eqn:KAW_pressure_balance}
\frac{B_0\dBprl}{4\upi} + \dpprp{e} + \dpprp{i} =  \pprp{i} \biggl( 1 - \frac{\betaprp{e}}{2} \Delta_e \biggr) \frac{2}{\betaprp{i}} \frac{\dBprl}{B_0} + \pprp{i} \biggl( 1 + \frac{Z_i}{\tauprp{i}} \biggr)   \frac{\dne}{\nem} = 0 ,
\end{equation}
which follows from combining (\ref{eqn:KAWdn},{\it d}) and (\ref{eqn:KAWdpi},{\it b}). This equation states that an increase in number density must be accompanied by a decrease in the magnetic-field strength, the amount of this decrease depending upon the factor $1-\betaprp{e}\Delta_e/2$. If $\Delta_e > 0$, then the magnetic-field lines must inflate further in order to maintain perpendicular pressure balance as large-pitch-angle particles are squeezed into the magnetic troughs. When the concentration of these particles leads to more perpendicular pressure than can be stably balanced by the magnetic pressure, the troughs must grow deeper to compensate. In the long-wavelength limit, the pressure-balanced slow mode then goes unstable to the mirror instability. In the short-wavelength limit, the KAW goes unstable for the same reason.

Finally, there is obviously something amiss about (\ref{eqn:KAWdBprl}) and (\ref{eqn:Theta}) when $\Delta_e = 2/\betaprp{e}$. In this case, $\dBprl \ne 0$ even though $\varphi = \dne = 0$. The perpendicular pressure balance (\ref{eqn:KAW_pressure_balance}) is then achieved by balancing the perturbed magnetic pressure $B_0 \dBprl / 4\upi$ with the adiabatic electron response $\dpprp{e} = \pprp{e} (\dtprp{e}/\tprp{e}) = - \pprp{e} \Delta_e \dBprl/B_0$. The KAW then consists only of magnetic-field fluctuations, with polarization
\begin{equation}\label{eqn:KAWmod_dBprl}
\frac{\dBprpk}{\dBprlk} = \pm \ez\btimes\frac{\bb{k}_\perp}{k_\perp} \biggl( 2 + \frac{2}{\betaprp{e}} \biggr)^{-1/2}
\end{equation}
and growth/decay rate
\begin{equation}\label{eqn:KAWmod_gamma}
\gamma = \pm k_\parallel v_A k_\perp d_i \biggl[\frac{\betaprl{e}}{2} \Delta_e \biggl( 1 + \frac{\betaprl{e}}{2} \Delta_e \biggr) \biggr]^{1/2} = \pm k_\parallel v_A k_\perp d_i \, \frac{(2+2/\betaprp{e})^{1/2}}{1+2/\betaprp{e}} .
\end{equation}
The instability is allowed because perpendicular pressure balance is maintained for any $\dBprl$ so long as $\Delta_e = 2/\betaprp{e}$.
 
Generally speaking, the location of the wavenumber transition from Alfv\'{e}n waves to KAWs during a turbulent cascade (at $k_\perp \rho_i \sim 1$ for a Maxwellian plasma) is a function of the electron pressure anisotropy. This dependence may be tested by looking for a shift in the ion-Larmor-scale spectral break in measurements of Alfv\'{e}nic turbulence in the non-Maxwellian solar wind and in simulations of gyrokinetic turbulence in bi-Maxwellian plasmas (using our theory). Furthermore, in a Maxwellian, high-$\beta$ plasma, the dispersion relation exhibits a sharp frequency jump at the Alfv\'{e}n-wave--KAW transition (see figure 8{\it c} of S09). This jump is accompanied by very strong ion Landau damping. It is easy to imagine from the above discussion that the electron pressure anisotropy, by affecting the rate of collisionless damping, would thus play an important role in determining the fraction of wave energy that is damped on the ions versus cascaded down to electron scales. This is manifest in numerical solutions to (\ref{eqn:gkdisprel}), which we present in \S\ref{sec:numerical}.

\subsubsection{Slow waves in the sub-ion-Larmor range: $k_\perp \rho_e \ll 1 \ll k_\perp \rho_i$, $\xi_i Z_{\rm M}(\xi_i) \sim \mc{O}(\alpha^{3/2}_i)$}\label{sec:KSW}

The approximate values of $\mc{A}$, $\mc{C}$, and $\mc{D}$ derived in \S\ref{sec:linKAW} made use of the asymptotic forms $\Gamma_0(\alpha_i) \sim 1/\sqrt{2\pi\alpha_i}$ and $\Gamma_1(\alpha_i) \sim 1/\sqrt{8\pi\alpha^3_i}$ for $\alpha_i \gg 1$. Terms proportional to these asymptotically small $\Gamma_m$ coefficients were dropped, assuming that all multiplicative factors of them were of order unity. However, solutions of the dispersion relation (\ref{eqn:gkdisprel}) exist with $\xi_i Z_{\rm M}(\xi_i) \sim \mc{O}(\alpha^{3/2}_i)$, and so some of these terms are non-negligible and must be retained. In this case, we find the leading-order expressions
\begin{gather}
\mc{A} \sim \frac{\xi_i Z_{\rm M}(\xi_i)}{\sqrt{2\pi\alpha_i}}  \frac{\tprp{i}}{\tprl{i}} , \\*
\mc{C} \sim \frac{\xi_i Z_{\rm M}(\xi_i)}{\sqrt{8\pi\alpha^3_i}} \frac{\tprp{i}}{\tprl{i}} - \biggl[ \Delta_e + \xi_e Z_{\rm M}(\xi_e) \frac{\tprp{e}}{\tprl{e}} \biggr] , \\*
\mc{D} \sim \frac{\xi_i Z_{\rm M}(\xi_i)}{\sqrt{2\pi\alpha^3_i}} \frac{\tprp{i}}{\tprl{i}} + \frac{2Z_i}{\tauprp{i}} \biggl[ \Delta_e + \xi_e Z_{\rm M}(\xi_e) \frac{\tprp{e}}{\tprl{e}} \biggr] .
\end{gather}
With $\mc{B} \simeq 1$ and $\mc{E} \simeq -1$, the dispersion relation (\ref{eqn:gkdisprel}) becomes
\begin{equation}\label{eqn:slowdisprel}
\biggl( \frac{\alpha_\ast}{\overline{\omega}^2} - 1 \biggr) \biggl( \frac{2}{\betaprp{i}} - \mc{D} \biggr) \simeq 1 .
\end{equation}
In general, each of the terms in this equation is $\mc{O}(1)$ and a numerical solution is required. However, when $\betaprp{s} \sim \betaprl{s} \sim \Delta^{-1}_s \gg 1$, a simple analytic solution can be obtained. In this limit, the dispersion relation (\ref{eqn:slowdisprel}) becomes $\mc{D}\simeq 1$. Writing $\xi_i = -\imag\widetilde{\xi}_i$ and using $Z_{\rm M}(\xi_i) \simeq 2\imag\sqrt{\upi}\exp(\widetilde{\xi}^2_i)$ for $\widetilde{\xi}_i \gg 1$, the solution is $\widetilde{\xi}_i \sim \sqrt{3\ln|k_\perp\rho_i|}$  up to logarithmically small corrections, or
\begin{equation}\label{eqn:KSW}
\gamma \sim - |k_\parallel| \valf \sqrt{3\betaprl{i} \ln|k_\perp\rho_i|} .
\end{equation}
This is an aperiodic, Barnes-damped slow mode in the sub-ion-Larmor range. Electron contributions to this solution are small, and the pressure anisotropy of either species makes no appearance to leading order. These features can be seen in the numerical solutions, which we now present.

\subsubsection{Numerical solutions for an electron-ion bi-Maxwellian plasma}\label{sec:numerical}

In this section, the dispersion relation (\ref{eqn:gkdisprel}) is solved numerically. We group the resulting solutions into ``Alfv\'{e}nic'' and ``compressive'' ones, based on whether their long-wavelength limit returns the Alfv\'{e}n-wave solution (\ref{eqn:gkaw}) or the compressive solution (\ref{eqn:compressive}), respectively. These two branches are coupled at $k_\perp \rho_i \gtrsim 1$ by finite-Larmor-radius effects.

%
%
\begin{figure}
\centering
\includegraphics[width=\linewidth,clip]{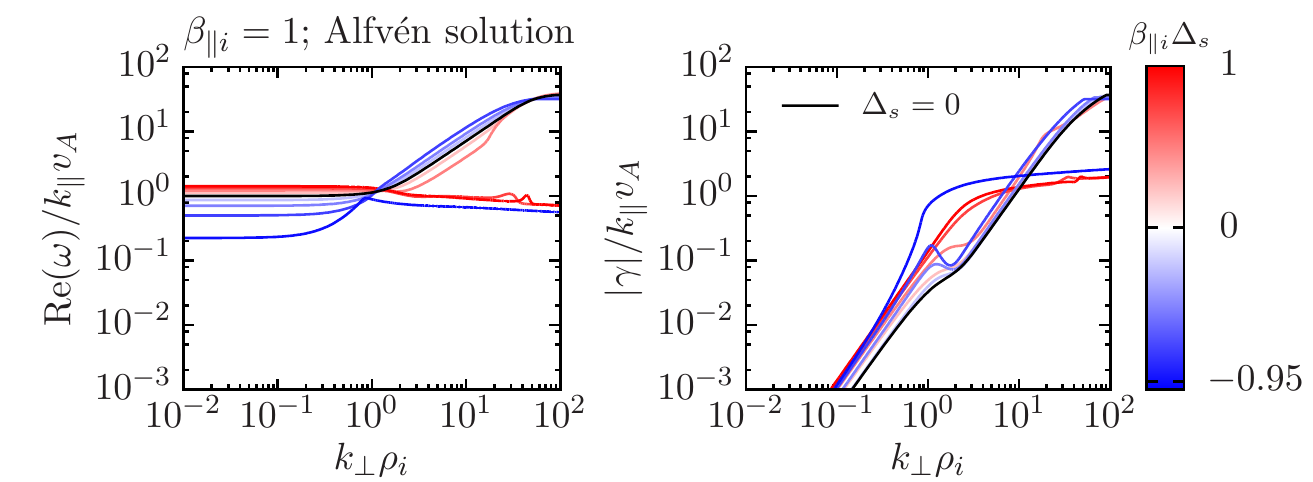}
\newline\vspace{0.1em}
\includegraphics[width=\linewidth,clip]{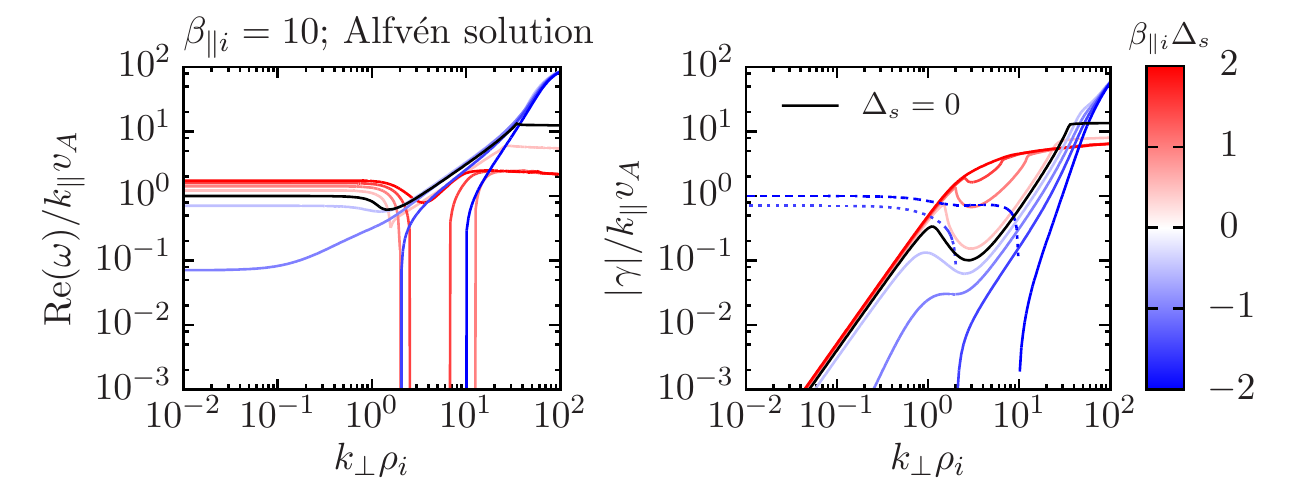}
\newline\vspace{0.1em}
\includegraphics[width=\linewidth,clip]{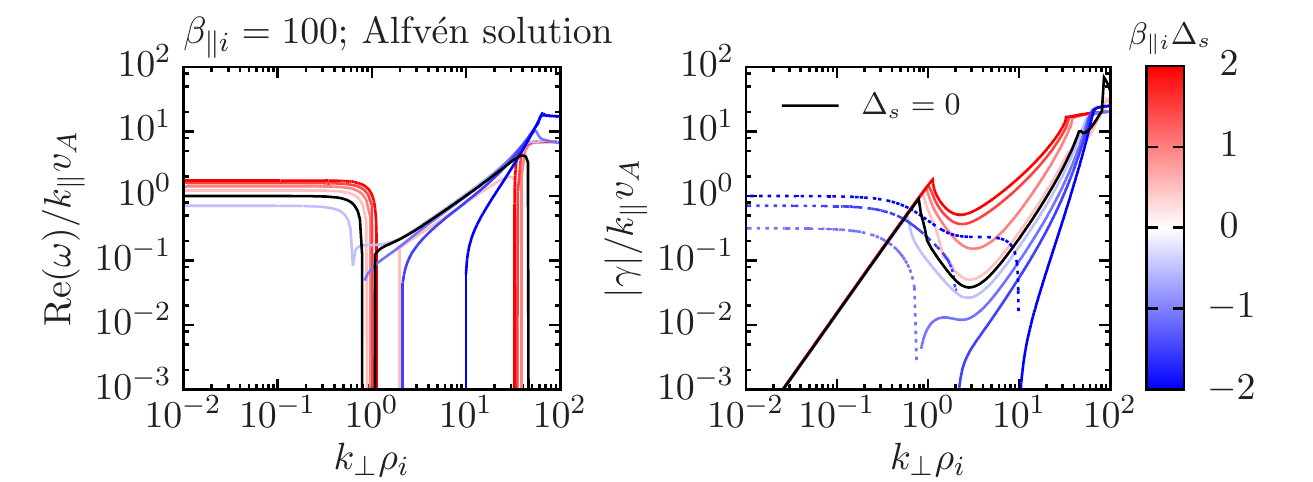}
\newline
\caption{Real and imaginary parts of the frequency $\omega = {\rm Re}(\omega) + \imag\gamma$ on the Alfv\'{e}nic branch as functions of $k_\perp\rho_i$ for $\betaprl{i} = \betaprl{e} = 1$, $10$, and $100$. (Recall that $\rho_i \doteq \vthprp{i}/\Omega_i \propto \sqrt{\betaprp{i}}$, and so the abscissa is a function of pressure anisotropy.) Equal ion and electron equilibrium pressure anisotropies $\Delta_s$ are equally spaced from negative (blue) to positive (red) values, the hues of the lines darkening towards larger absolute values (as indicated by the color bars). The solution for a Maxwellian equilibrium, $\Delta_s = 0$, is shown as the black line. Stable (unstable) modes are denoted by solid (dotted) lines. See \S\ref{sec:numerical} for discussion.}
\label{fig:alfven}
\end{figure}

Figure \ref{fig:alfven} displays the real part, ${\rm Re}(\omega)$, and the absolute value of the imaginary part, $|\gamma| \doteq |{\rm Im}(\omega)|$, of the complex frequency $\omega$ (normalized to $k_\parallel \valf$) describing waves on the Alfv\'{e}nic branch as a function of $k_\perp \rho_i$, for varying $\betaprl{i}$ and pressure anisotropy $\Delta_i$. We take $\betaprl{e} = \betaprl{i}$ and $\Delta_e = \Delta_i$ (i.e.~$\tprl{e} = \tprl{i}$), so that the equal-temperature ions and electrons both contribute to the total pressure anisotropy of the equilibrium plasma. The solid black lines correspond to $\Delta_s = 0$, i.e.~to a Maxwellian equilibrium, and match those presented in figure 4 of \citet{howes06} and figure 8 of S09. The red (blue) lines correspond to positive (negative) equilibrium pressure anisotropy, their hues darkening towards larger absolute values (as indicated by the accompanying color bars).

\begin{figure}
\centering
\includegraphics[width=\linewidth,clip]{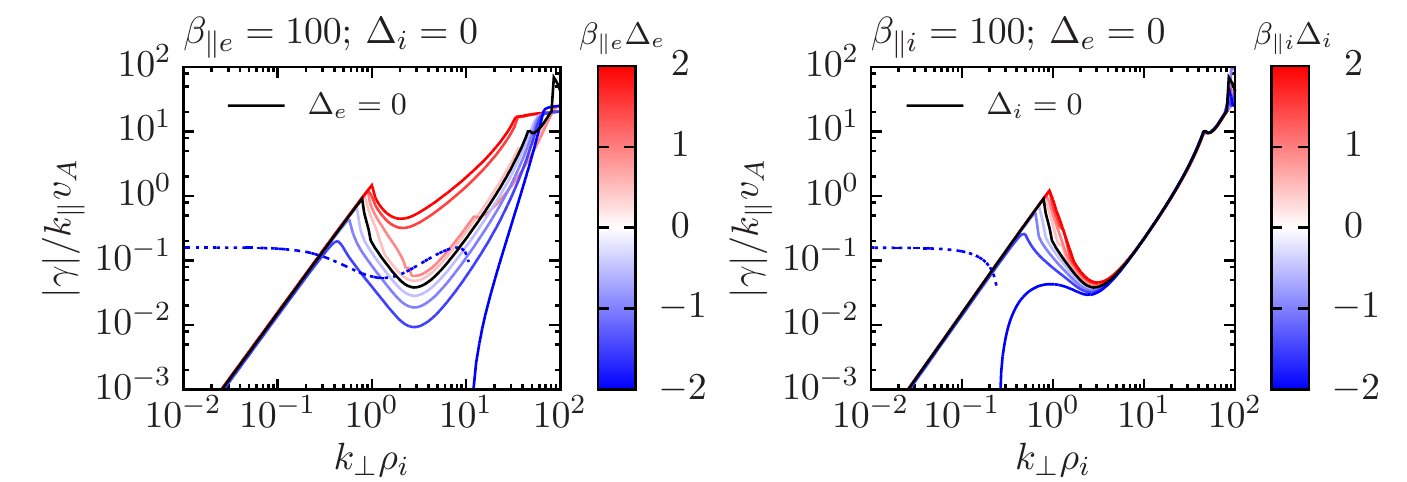}
\newline
\caption{Imaginary part of the frequency $\omega = {\rm Re}(\omega) + \imag\gamma$ on the Alfv\'{e}nic branch as a function of $k_\perp\rho_i$ for $\betaprl{i} = \betaprl{e} = 100$. Either (left) the equilibrium ion distribution is taken to be isotropic, $\Delta_i = 0$, or (right) the equilibrium electron distribution is taken to be isotropic, $\Delta_e = 0$. Stable (unstable) modes are denoted by solid (dotted) lines. Sub-ion-Larmor fluctuations are primarily affected by electron pressure anisotropy, while long-wavelength fluctuations and the transition at $k_\perp \rho_i \sim 1$ are affected by both. See \S\ref{sec:numerical} for discussion.}
\label{fig:alfven2}
\end{figure}

For all values of $\betaprl{i}$ and $\Delta_s$, the analytic solution (\ref{eqn:gkaw}) accurately describes the Alfv\'{e}n waves (and the firehose-unstable modes) at long wavelengths, with negative (positive) pressure anisotropies decreasing (increasing) the effective Alfv\'{e}n speed and, thus, the real part of the frequency. (Solutions for $\betaprl{i} \ll 1$ are not presented, as very extreme pressure anisotropies are required to appreciably modify the waves in this case.) The weak damping affecting these pressure-anisotropy-modified Alfv\'{e}n waves is obtained analytically in appendix \ref{app:highbeta} for $\betaprl{i} \gg 1$ with $k_\perp \rho_i \sim \mc{O}(\betaprl{i}^{-1/4})$, $\overline{\omega} \sim \mc{O}(1)$ (see (\ref{eqn:highbeta4})):
\begin{equation}\label{eqn:highbeta4_text}
\overline{\gamma} \simeq - \frac{9}{16} \frac{k^2_\perp \rho^2_i}{2} \sqrt{\frac{\betaprl{i}}{\pi}} ,
\end{equation}
which is independent (to leading order) of the pressure anisotropy (as it indeed is in the figure). For $\betaprl{i}^{-1/4} \ll k_\perp \rho_i \ll 1$ (see (\ref{eqn:alfweakly})), or $k_\perp \rho_i \sim \mc{O}(1)$, $\overline{\omega} \sim \mc{O}(\betaprl{i}^{-1/2})$ (see (\ref{eqn:alfweakly2})), we have
\begin{equation}\label{eqn:alfweakly2_text}
\overline{\gamma} \simeq - \frac{8}{9} \biggl( 1 + \frac{\betaprl{i}}{2} \Delta_i + \frac{\betaprl{e}}{2} \Delta_e \biggr) \biggl( \frac{k^2_\perp \rho^2_i}{2} \biggr)^{-1} \sqrt{\frac{\pi}{\betaprl{i}}} \frac{\tprp{i}}{\tprl{i}},
\end{equation}
which captures well the decrease of $|\overline{\gamma}|$ with $k_\perp \rho_i$ and its dependence on the pressure anisotropy that are seen in the $\betaprl{i} = 10$ and $100$ plots near $k_\perp \rho_i \sim 1$. For $k_\perp \rho_i \gg 1$, the damping rate increases again with $\alpha_i$ (see (\ref{eqn:kaw2})), this time with an explicit dependence on the electron pressure anisotropy:
\begin{equation}\label{eqn:kaw2_text}
\overline{\gamma} \simeq - \frac{k^2_\perp \rho^2_i}{2} \sqrt{\frac{\pi}{\betaprl{i}}} \biggl( 1 + \frac{\betaprl{e}}{2} \Delta_e \biggr) \frac{\tprp{e}^2}{\tprl{e}^2}\frac{\tprl{i}}{\tprp{i}} \biggl( \frac{Z_i}{\tauprl{i}} \frac{Z_i m_e}{m_i} \biggr)^{1/2} .
\end{equation}
Positive anisotropy $\Delta_e > 0$ increases the rate of collisionless damping by effectively increasing the Alfv\'{e}n speed and thereby bringing the wave frequency (see (\ref{eqn:kaw2})) closer to the ion-Landau resonance; this is manifest in the $\betaprl{i} = 10$ and $100$ panels of figure \ref{fig:alfven}. Negative electron pressure anisotropy, on the other hand, drives these KAWs firehose unstable for $\Delta_e < -2/\betaprl{e}$. Note that {\em there exist Alfv\'{e}nic fluctuations that are firehose unstable at long wavelengths, which become stable at short wavelengths} (the blue lines that are absent at low $k_\perp$ but appear at high $k_\perp$ in the ${\rm Re}(\omega)$ panel). This is because the ion contribution to the total pressure anisotropy is only relevant at long wavelengths. Both the analytic solutions (cf.~(\ref{eqn:gkaw}) and (\ref{eqn:kaw})) and the numerical solutions shown in figure \ref{fig:alfven2} demonstrate this point.

%
%

In \S\ref{sec:linKAW}, we speculated that the location of the wavenumber transition from Alfv\'{e}n waves to KAWs and the consequent ion-Larmor-scale spectral break that is routinely observed in kinetic turbulence \citep[e.g.][]{leamon98,leamon99,bale05,howes06,markovskii08,alexandrova09,chen14} should be a function of the pressure anisotropy. This conjecture is supported especially by the $\betaprl{i} = 10$ and $100$ plots, which display a sharp frequency jump near $k_\perp \rho_i \sim 1$ at a location dependent upon the pressure anisotropy. This dependence can be estimated by equating the asymptotic damping rates (\ref{eqn:highbeta4_text}) or (\ref{eqn:alfweakly2_text}) to the real frequency (\ref{eqn:gkaw}); this yields
\begin{equation}\label{eqn:breakpoint}
( k_\perp \rho_i )_{\rm break} \propto \biggl( 1 + \frac{\betaprl{i}}{2} \Delta_i + \frac{\betaprl{e}}{2} \Delta_e \biggr)^{1/4} .
\end{equation}
Positive (negative) anisotropies are thus predicted to shift the ion-Larmor-scale spectral break towards larger (smaller) perpendicular wavenumbers. This frequency jump is seen to be accompanied by very strong ion Landau damping ($|\overline{\gamma}| \sim 1$), with order-of-magnitude variations in the damping rate depending upon $\Delta_s$. Note further that the width of the ${\rm Re}(\omega) = 0$ gap at large $\betaprl{i}$ near $k_\perp \rho_i \sim 1$, evident in the $\betaprl{i} = 100$ plot, increases with $\Delta_s > 0$. (Figure \ref{fig:alfven2} shows that both the ion and electron pressure anisotropies affect the strength of the collisionless damping and the location of the break point.) 

\begin{figure}
\centering
\includegraphics[width=\linewidth,clip]{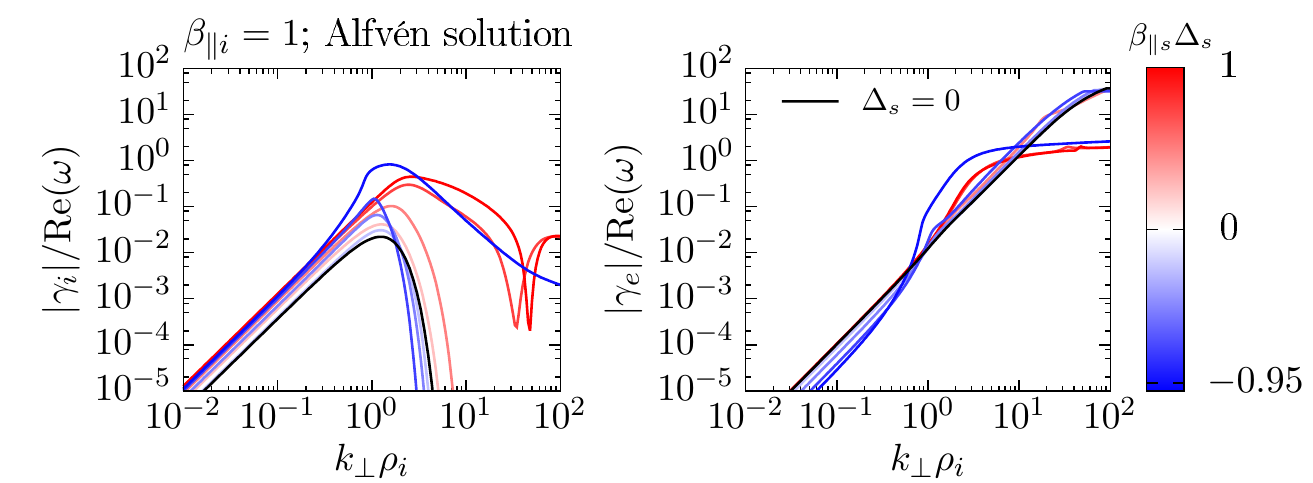}
\newline\vspace{0.1em}
\includegraphics[width=\linewidth,clip]{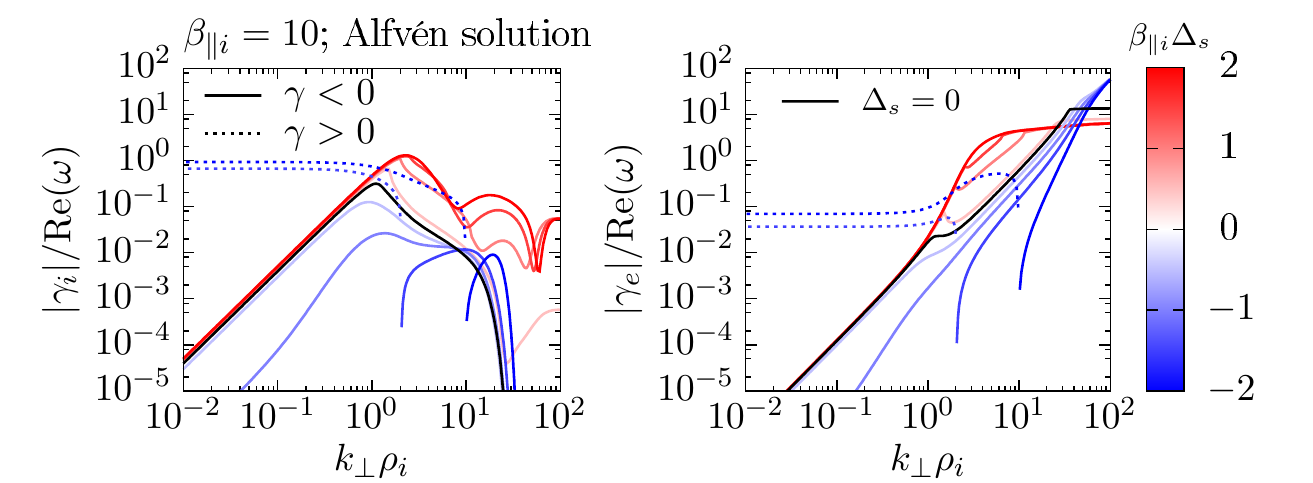}
\newline\vspace{0.1em}
\includegraphics[width=\linewidth,clip]{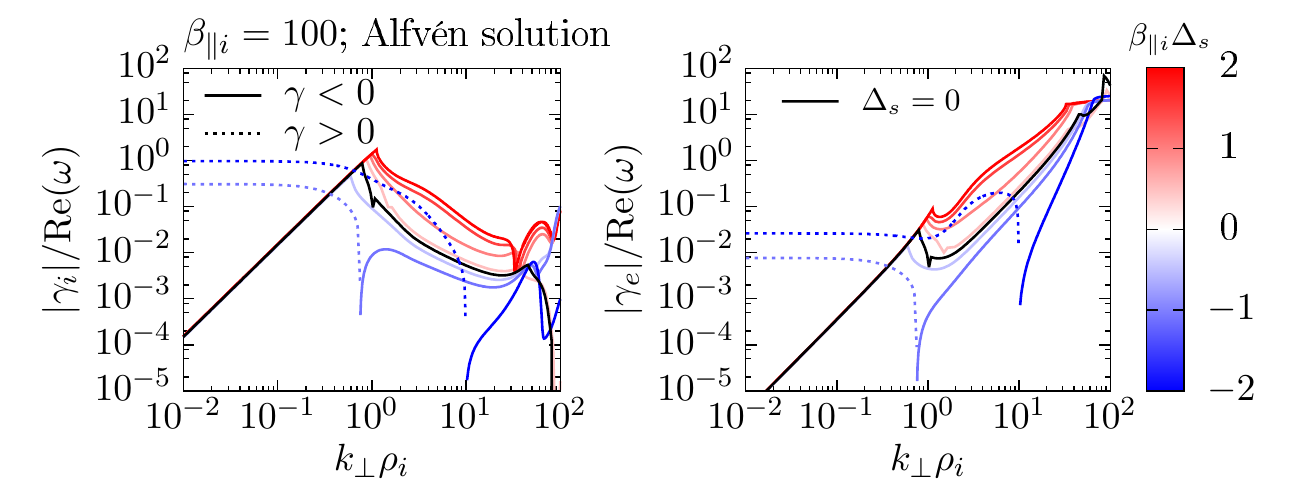}
\newline
\caption{Ratio of damping/growth rate $|\gamma|$ and ${\rm Re}(\omega)$ on the Alfv\'{e}nic branch, partitioned into contributions from damping on ions ($|\gamma_i|$; left panels) and electrons ($|\gamma_e|$; right panels), as a function of $k_\perp\rho_i$ for $\betaprl{i} = \betaprl{e} = 1$, $10$, and $100$. Equal ion and electron equilibrium pressure anisotropies $\Delta_s$ are varied from negative (blue) to positive (red) values, with $\Delta_s = 0$ indicated by the black line.  Stable (unstable) modes are denoted by solid (dotted) lines. See \S\ref{sec:numerical} for discussion.}
\label{fig:gammas}
\end{figure}

Because the amount of Landau damping at these scales is likely to be related to how much energy ultimately goes into heating the ions or the electrons, pressure anisotropy ought to be considered alongside $\beta_i$ and $T_{0i}/T_{0e}$ when assessing the efficiency of ion heating. This consideration may be important for developing a quantitative and predictive theory of collisionless radiatively inefficient accretion flows, for which the efficiency of ion heating is a key unknown \citep[e.g.][]{qg99,howes10} and in which pressure anisotropy is predicted to govern much of the nonlinear dynamics and thermodynamics \citep[e.g.][]{sharma06,sharma07,kunz16}. We investigate this possibility quantitatively using a simple cascade model in \S\ref{sec:heating}. Here we simply examine which species (ions or electrons) contributes the most to the Landau damping of the Alfv\'{e}nic fluctuations at different scales. In figure \ref{fig:gammas}, the ratio of the absolute value of the imaginary part and the real part of the frequency, separated into contributions from damping on the ions (left panels) and on the electrons (right panels), are shown versus $k_\perp \rho_i$, $\betaprl{i}$, and $\Delta_s$. (See \S\ref{sec:heating} and, in particular, (\ref{eqn:speciesgamma}) for how this partitioning of $\gamma$ into $\gamma_i$ and $\gamma_e$ is computed.) In all cases, the collisionless damping on the ions is stronger than that on the electrons for $k_\perp \rho_i \lesssim 2$, whereas the collisionless damping on the electrons at sub-ion-Larmor scales is dominant. While the latter is much stronger than the former, one would anticipate in a turbulent cascade that the energy carried by the sub-ion-Larmor fluctuations would be much smaller than that carried by the long-wavelength Alfv\'{e}nic fluctuations, and so the damping rates on both the ions and the electrons (and the impact of pressure anisotropy on them) are likely important for calculating ion versus electron heating (see \S\ref{sec:heating}).

%
%
%
%
%

%
%
\begin{figure}
\centering
\includegraphics[width=\linewidth,clip]{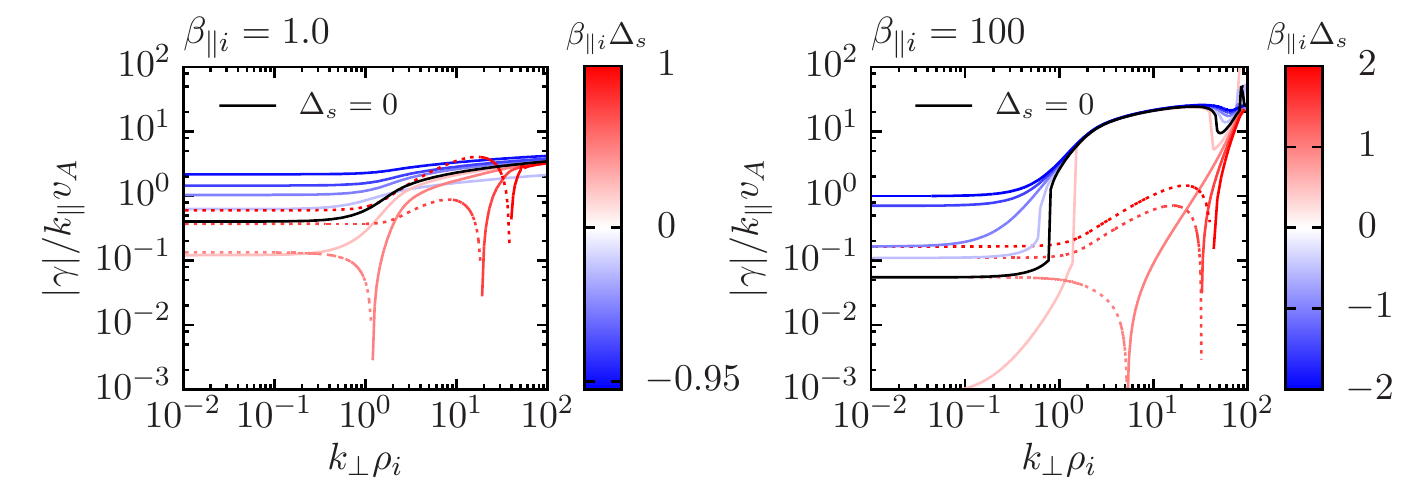}
\newline
\caption{Imaginary part of the frequency $\omega = {\rm Re}(\omega) + \imag\gamma$ on the compressive branch as a function of $k_\perp\rho_i$ for $\betaprl{i} = \betaprl{e} = 1$ and $100$. Equal ion and electron equilibrium pressure anisotropies $\Delta_s$ are varied from negative (blue) to positive (red) values, with $\Delta_s = 0$ indicated by the black line. Stable (unstable) modes are denoted by solid (dotted) lines. See \S\ref{sec:numerical} for discussion.}
\label{fig:mirror}
\end{figure}
%
%
\begin{figure}
\centering
\includegraphics[width=\linewidth,clip]{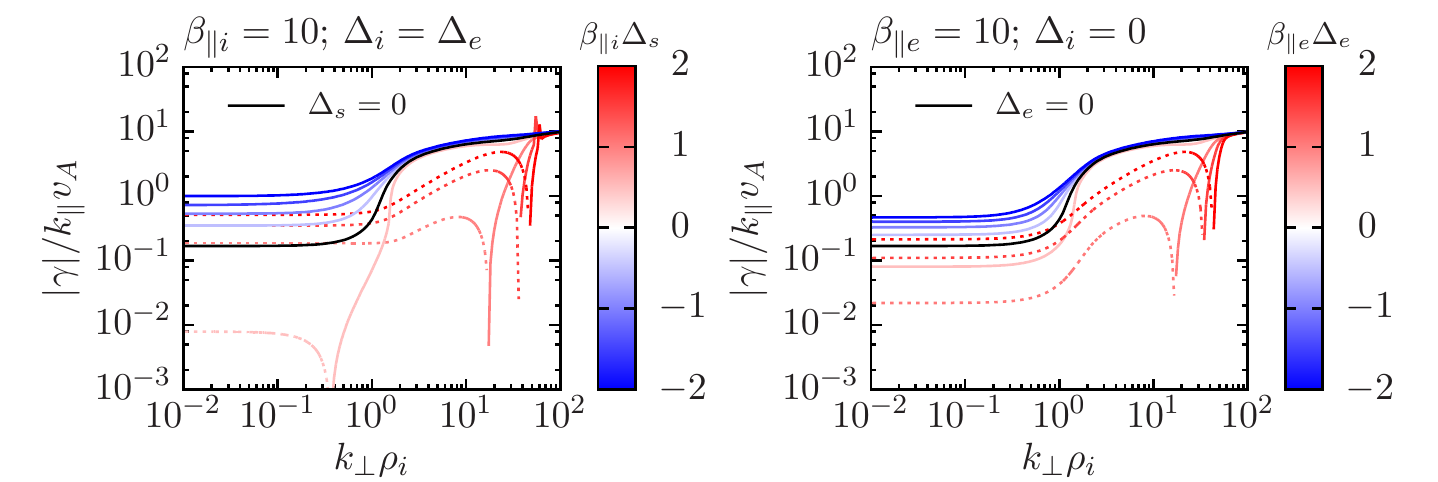}
\newline
\caption{Imaginary part of the frequency $\omega = {\rm Re}(\omega) + \imag\gamma$ on the compressive branch as a function of $k_\perp\rho_i$ for $\betaprl{i} = \betaprl{e} = 10$ and either (left) $\Delta_i = \Delta_e$ or (right) $\Delta_i = 0$. Pressure anisotropies are varied from negative (blue) to positive (red) values, the hues of the lines darkening towards larger absolute values (as indicated by the color bars). The solution for a Maxwellian equilibrium, $\Delta_s = 0$, is indicated by the black line. Stable (unstable) modes are denoted by solid (dotted) lines. Ion pressure anisotropy only affects the behaviour of the modes at long wavelengths, $k_\perp \rho_i \lesssim 1$. See \S\ref{sec:numerical} for discussion.}
\label{fig:mirrorduo}
\end{figure}

Figures \ref{fig:mirror} and \ref{fig:mirrorduo} display the imaginary part of the frequency describing waves on the compressive branch as a function of $k_\perp \rho_i$, for varying plasma $\betaprl{i}$ and pressure anisotropy $\Delta_s$. These modes always have ${\rm Re}(\omega) = 0$ and thus are aperiodic. As in figure \ref{fig:alfven}, the solid black lines correspond to $\Delta_s = 0$, while the red (blue) lines correspond to positive (negative) equilibrium pressure anisotropy. In figure \ref{fig:mirror}, we take $\betaprl{e} = \betaprl{i}$ and $\Delta_e = \Delta_i$, so that the equal-temperature ions and electrons both contribute to the total pressure anisotropy of the equilibrium plasma. By comparing solutions obtained using $\Delta_i = \Delta_e$ with those obtained using $\Delta_i = 0$, figure \ref{fig:mirrorduo} demonstrates that it is only the pressure anisotropy of the electrons that affects the sub-ion-Larmor range (as predicted in \S\ref{sec:KSW}).

At long wavelengths ($k_\perp \rho_i \ll 1$), the damping/growth rate follows closely the ``$+$'' branch obtained from the dispersion relation (\ref{eqn:compressive}), as it is the more weakly damped of the two branches.\footnote{In a $\betaprl{i} \sim 1$ plasma, large negative pressure anisotropy $\Delta_s \sim -1$ can cause the oscillatory ``$-$'' branch to be marginally less damped than the plotted aperiodic ``$+$'' branch. In a $\betaprl{i} \ll 1$ plasma, the damping rate of the ``$+$'' branch is logarithmically larger than that of the ``$-$'' branch (see \S I--4.4). Because the ``$-$'' branch is only marginally less damped than the ``$+$'' branch in these cases, we do not consider the long-wavelength ``$-$'' compressive branch in this paper, instead concentrating on the more generally longer-lived ``$+$'' compressive fluctuations. Order-unity pressure anisotropy does not change the fact that both branches are weakly damped when $\betaprl{i} \ll 1$, and so neither do we present the low-$\betaprl{i}$ limit. Unrealistic pressure anisotropies are required in a $\betaprl{i} \ll 1$ plasma to appreciably modify its wave properties.} This is the branch that goes over to the aperiodic Barnes-damped/mirror-unstable slow mode for $\betaprp{s}, \betaprl{s} \gg 1$ (see (\ref{eqn:slowmode})). As such, it is unstable if (\ref{eqn:iemirror2}) is satisfied; the figures show this. At $k_\perp \rho \sim 1$, the character of the mode changes abruptly to become either strongly damped at a rate independent of the pressure anisotropy (see (\ref{eqn:KSW}) for an approximation solution) or, if the pressure anisotropy is sufficiently positive, mirror-unstable at a rate given approximately by (\ref{eqn:kaw}). Note that, in the latter case, the compressive solution is coupled to the Alfv\'{e}nic solution, and the unstable mode is a KAW.

At electron-Larmor scales ($k_\perp \rho_e \gtrsim 1$), both branches of the dispersion relation -- Alfv\'{e}nic and compressive -- are strongly damped via Landau resonance on the electrons.

This completes our discussion of the linear gyrokinetic theory. We now make use of the physics contained in these results to explain how pressure anisotropy affects the nonlinear turbulent cascade of free energy to small scales in phase space.

\section{Generalised energy and the nonlinear kinetic cascade}\label{sec:invariant}

\subsection{KRMHD free-energy conservation for arbitrary $f_{0s}$}

In Paper I, we showed that the long-wavelength Alfv\'{e}nic and compressive fluctuations satisfy the following nonlinear conservation law:
\begin{equation}\label{eqn:WKRMHD}
\D{t}{W_{\rm KRMHD}} = - \int\rmd^3\bb{r} \sum_s \dupar \biggl( q_s \delta n_s E_\parallel - \dpprp{s} \eb\bcdot\grad \frac{\dBprl}{B_0} \biggr),
\end{equation}
where
\begin{align}\label{eqn:WK}
W_{\rm KRMHD} &\doteq  \int\rmd^3\bb{r} \,\Biggl\{ \sum_s \int\rmd^3\bb{v}\, \frac{\tprl{s} \delta\widetilde{f}^2_s}{2 f^\parallel_{0s}} + \frac{\rho_0 u^2_\perp}{2} \nonumber\\*
\mbox{} &\quad + \Biggl[ 1 + \sum_s \frac{\betaprl{s}}{2} \Biggl( \Delta_s - \frac{2\duparsq}{\vthprl{s}^2} \Biggr) \Biggr] \frac{\delta B^2_\perp}{8\upi} + \Biggl( 1 - \sum_s \betaprp{s} \Delta_{2s} \Biggr) \frac{\dBprl^2}{8\upi}  \Biggr\}
\end{align}
is the generalised free energy of KRMHD,
\begin{equation}
\delta\widetilde{f}_s = \delta f_s(v_\parallel, w_\perp) + \frac{w^2_\perp}{\vthprp{s}^2} \frac{\dBprl}{B_0} \mf{D}f_{0s}
\end{equation}
is the (long-wavelength) perturbed distribution function in the frame of the Alfv\'{e}nic fluctuations (see \S I--4.2 and (\ref{eqn:dftilde})), and $\Delta_{\ell s} \doteq \Delta_{\ell s}(0)$ (see (\ref{eqn:Delta_ells})). (Recall from (\ref{eqn:mus_longwavelength}) that $\bb{w}_\perp \doteq \bb{v} - \bb{u}_\perp - \bb{v}\bcdot\eb\eb$ is the perpendicular component of the particle velocity peculiar to the $\bb{E}\btimes\bb{B}$ flow.) The parallel electric field $E_\parallel$ on the right-hand side of (\ref{eqn:WKRMHD}) is given by (I--2.37) -- we have no need of restating its general form here; for a bi-Maxwellian equilibrium, it is given below by (\ref{eqn:eprl}).

In the presence of mean interspecies drifts $\dupar$, the right-hand side of (\ref{eqn:WKRMHD}) corresponds to the change in the free energy due to the minus work done by the fluctuating parallel electric and magnetic-mirror forces acting on the parallel-drifting particles. In the absence of these drifts, the right-hand side of (\ref{eqn:WKRMHD}) is zero. Then $W_{\rm KRMHD}$ is a quadratic invariant whose existence causes a turbulent cascade of generalised free energy in a pressure-anisotropic plasma to small scales in phase space across the inertial range. The invariant is comprised of three parts: two Alfv\'{e}nic invariants $W^+_{\rm AW}$ and $W^-_{\rm AW}$ ((I--3.10), arising from the second and third terms in (\ref{eqn:WK})) representing forward- and backward-propagating nonlinear Alfv\'{e}n waves, and a compressive invariant $W_{\rm compr}$ ((I--4.7), arising from the first and fourth terms in (\ref{eqn:WK})). In the pressure-isotropic case (eq.~(201) of S09), the first term in (\ref{eqn:WK}) is the entropy of the perturbed distribution function. For a single-ion-species bi-Maxwellian equilibrium, the compressive invariant $W_{\rm compr}$ splits into three independently cascading parts: $W^+_{\rm compr}$, $W^-_{\rm compr}$, and $W_{\widetilde{g}_i}$, the latter of which represents a purely kinetic cascade. All three cascade channels lead to small perpendicular spatial scales via passive mixing by the Alfv\'{e}nic turbulence and to small scales in $v_\parallel$ via the linear parallel phase mixing (\S I--4.3 through \S I--4.6; see however \citet{schekochihin16} for how nonlinear advection of the perturbed particle distribution by fluctuating flows might reduce the amount of parallel phase mixing). The rates of mixing are generally functions of the velocity-space anisotropy of the equilibrium function.

\subsection{Gyrokinetic free-energy conservation for arbitrary $f_{0s}$}

Our goal in this section is to derive the gyrokinetic generalisation of (\ref{eqn:WKRMHD}), valid at both long and short wavelengths and reducing to (\ref{eqn:WKRMHD}) in the former limit. The starting point is the gyrokinetic equation (\ref{eqn:gkequation2}), written in terms of the gyrokinetic response $\widetilde{h}_s$. Multiplying that equation by $\tprl{s} \widetilde{h}_s / f^\parallel_{0s}$ and integrating over the velocities and gyrocentre positions, we find that the nonlinear term conserves the variance of $\widetilde{h}_s$ and so
\begin{equation}\label{eqn:gkcons1}
\D{t}{} \int\rmd^3\bb{v} \int\rmd^3\bb{R}_s \, \frac{\tprl{s} \widetilde{h}^2_s}{2f^\parallel_{0s}} = \int\rmd^3\bb{v} \int\rmd^3\bb{R}_s \, q_s \biggl( \pD{t}{\langle\chi\rangle_{\gas}} + \dupar \pD{z}{\langle\chi\rangle_{\gas}} \biggr) \widetilde{h}_s .
\end{equation}
We now sum this equation over all species. The right-hand side becomes
\begin{align}\label{eqn:helpful1}
\sum_s q_s &\int\rmd^3\bb{v} \int\rmd^3\bb{R}_s \, \biggl( \pD{t}{\langle\chi\rangle_{\gas}} + \dupar \pD{z}{\langle\chi\rangle_{\gas}} \biggr) \widetilde{h}_s \nonumber\\*
\mbox{} &= \int\rmd^3\bb{r} \sum_s q_s \int\rmd^3\bb{v} \, \biggl\langle \biggl( \pD{t}{\chi} + \dupar \pD{z}{\chi} \biggr) \widetilde{h}_s \biggr\rangle_{\ggr} \nonumber\\*
\mbox{} &= \int\rmd^3\bb{r} \, \Biggl[ \pD{t}{\varphi} \sum_s q_s \int\rmd^3\bb{v} \, \langle\widetilde{h}_s\rangle_{\bs{r}} - \frac{1}{c} \pD{t}{\bb{A}} \bcdot \sum_s q_s \int\rmd^3\bb{v} \, \langle\bb{v}\widetilde{h}_s\rangle_{\bs{r}} \nonumber\\*
\mbox{} &\qquad + \sum_s q_s \dupar \int\rmd^3\bb{v}  \,\biggl\langle \pD{z}{\chi} \, \widetilde{h}_s \biggr\rangle_{\ggr} \Biggr] .
\end{align}
The first term on the right-hand side of (\ref{eqn:helpful1}) can be written in terms of the potentials ($\varphi, A_\parallel, \dBprl$) by using (\ref{eqn:htilde}) to unpack $\widetilde{h}_s$, employing the quasi-neutrality constraint (\ref{eqn:gkqn}), and performing the resulting integrals using the notation defined in appendix \ref{app:coeffs}. The second term on the right-hand side of (\ref{eqn:helpful1}) is most easily dealt with by using Faraday's law (\ref{eqn:EandB}) and Amp\`{e}re's law (\ref{eqn:ampere}) to write
\begin{align}\label{eqn:helpful2}
&- \D{t}{} \int\rmd^3\bb{r} \, \frac{|\delta\bb{B}|^2}{8\upi} = \int\rmd^3\bb{r} \, \bb{E}\bcdot\bb{j} = \int\rmd^3\bb{r} \, \Biggl( -\frac{1}{c} \pD{t}{\bb{A}} \bcdot \sum_s q_s \int\rmd^3\bb{v}\, \bb{v} \delta f_s \Biggr) \nonumber\\*
\mbox{} &\quad = \int\rmd^3\bb{r} \, \Biggl[ -\frac{1}{c} \pD{t}{\bb{A}} \bcdot \sum_s q_s \int\rmd^3\bb{v} \, \biggl\langle \bb{v} \biggl( \delta f_{1s,{\rm Boltz}} - \frac{q_s\langle\chi\rangle_{\gas} }{\tprp{s}} \mf{D}f_{0s}+ \widetilde{h}_s \biggr) \biggr\rangle_{\ggr} \Biggr] .
\end{align}
Then, substituting (\ref{eqn:boltzmann}) for $\delta f_{1s,{\rm Boltz}}$ and performing the resulting integrals (again, with the aid of appendix \ref{app:coeffs}), we may use (\ref{eqn:helpful2}) to write the second term on the right-hand side of (\ref{eqn:helpful1}) in terms of the potentials ($\varphi, A_\parallel, \dBprl$) and the rate of change of the magnetic energy. The third and final term on the right-hand side of (\ref{eqn:helpful1}) is markedly simplified by using (\ref{eqn:gyrocentredistribution}) to move from $\widetilde{h}_s$ to $g_s$:
\begin{align}\label{eqn:helpful3}
&\int\rmd^3\bb{r} \sum_s q_s \dupar \int\rmd^3\bb{v} \, \biggl\langle \pD{z}{\chi} \, \widetilde{h}_s \biggr\rangle_{\ggr} = \int\rmd^3\bb{r} \sum_s q_s \dupar \int\rmd^3\bb{v} \,\biggl\langle \pD{z}{\chi} \, g_s \biggr\rangle_{\ggr} \nonumber\\*
\mbox{}&\quad + \int\rmd^3\bb{r} \sum_s q_s \dupar \int\rmd^3\bb{v} \, \frac{q_s f^\parallel_{0s}}{\tprl{s}} \Biggl\langle \pD{z}{\chi} \Biggl\langle \chi + \frac{(v_\parallel-\dupar)A_\parallel}{c} \Biggr\rangle_{\ggs}  \Biggr\rangle_{\ggr} ,
\end{align}
from which we may remove the entire second line after integrating by parts with respect to $z$ (to eliminate the first term) and $v_\parallel$ (to eliminate the second). 

Assembling (\ref{eqn:helpful1})--(\ref{eqn:helpful3}) and cheerfully expending much algebraic effort, we find that (\ref{eqn:gkcons1}) summed over species is equivalent to the following conservation law:
\begin{equation}\label{eqn:WGK}
\D{t}{W_{\rm GK}} = \int\rmd^3\bb{r} \sum_s \dupar \int\rmd^3\bb{v} \, q_s g_s \Biggl[ \bigl\langle \eb \bigr\rangle_{\gas} \bcdot \grad \biggl\langle \varphi - \frac{\bb{v}_\perp\bcdot\bb{A}_\perp}{c} \biggr\rangle_{\ggs} + \frac{1}{c} \pD{t}{\langle A_\parallel\rangle_{\gas}} \Biggr] ,
\end{equation}
where
\begin{align}\label{eqn:W}
W_{\rm GK} &\doteq \int\rmd^3\bb{r} \, \Biggl\{ \sum_s \int\rmd^3\bb{v}\, \frac{\tprl{s}\delta\widetilde{f}^2_s}{2f^\parallel_{0s}} + \sum_s \frac{q^2_s\nsp }{2\tprp{s}} \, \varphi\Bigl( \kzero{s} - \widehat{\Gamma}^\perp_{00s} \Bigr) \varphi +  \frac{\delta B^2_\perp}{8\upi} \nonumber\\*
\mbox{} &\quad + \sum_s \frac{\betaprl{s}}{2} \frac{A_\parallel}{\sqrt{4\upi}\rho_s} \Biggl[ \frac{\tprp{s}}{\tprl{s}} \Bigl( 1-\widehat{\Gamma}_{00s} \Bigr) - \Bigl( \ktwo{s}-\widehat{\Gamma}^\perp_{02s} \Bigr) \Biggl( 1 + \frac{2\duparsq}{\vthprl{s}^2} \Biggr) \Biggr] \frac{A_\parallel}{\sqrt{4\upi}\rho_s} \nonumber\\*
\mbox{} &\quad + \frac{\dBprl}{\sqrt{8\upi}} \Biggl[ 1 - \sum_s \betaprp{s} \Biggl( \frac{\tprp{s}}{\tprl{s}} \frac{\widehat{\Gamma}^\parallel_{20s}}{2} - \frac{\widehat{\Gamma}^\perp_{20s}}{2} \Biggr) \Biggr] \frac{\dBprl}{\sqrt{8\upi}} - \sum_s q_s \nsp \dupar \frac{A_\parallel}{c} \widehat{\Gamma}^\perp_{11s} \frac{\dBprl}{B_0}  \Biggr\}
\end{align}
is the appropriately generalised gyrokinetic free energy. Here, $\delta\widetilde{f}_s$ is given by (\ref{eqn:dftilde}). The bi-linear differential operators in (\ref{eqn:W}), conspicuously ornamented with the symbol $\,\widehat{\mbox{\,\,}}\,$, are defined in appendix \ref{app:coeffs} by (\ref{eqn:Gammahat}). They result from integral combinations of various electromagnetic fields, gyro- and ring-averages of those fields, and various derivatives of the background distribution function: e.g.
\[
\int\rmd^3\bb{r} \int\rmd^3\bb{v}\, \varphi \langle\langle\varphi\rangle_{\gas}\rangle_{\bs{r}} \, f^\perp_{0s} \doteq \nsp \int\rmd^3\bb{r}\,\varphi \widehat{\Gamma}^\perp_{00s}\varphi \doteq \nsp\sum_{\bs{k}} \Gamma^\perp_{00s}(\alpha_s) |\varphik|^2,
\]
where the Fourier coefficient $\Gamma^\perp_{00}(\alpha_s)$ is given by (\ref{eqn:gammaprp00}). To leading order in $k^2_\perp \rho^2_s \ll 1$, the operators satisfy
\begin{gather}\label{eqn:longwavelengthG}
\Psi\Bigl( \kzero{s} - \widehat{\Gamma}^\perp_{00s} \Bigr) \Psi \approx \Psi\Bigl( 1 - \widehat{\Gamma}_{00s} \Bigr) \Psi \approx \Psi\Bigl( \ktwo{s} - \widehat{\Gamma}^\perp_{02s}  \Bigr) \Psi \approx  \sum_{\bs{k}} \alpha_s |\Psi_{\bs{k}}|^2 , \nonumber\\*
\widehat{\Gamma}^\parallel_{20s} \approx \frac{1}{\nsp} \int\rmd^3\bb{v}\, \frac{v^4_\perp}{\vthprp{s}^4} f^\parallel_{0s},\quad \widehat{\Gamma}^\perp_{20s} \approx 2,\quad \widehat{\Gamma}^\perp_{11s} \approx 1, 
\end{gather}
for any function $\Psi(\bb{r})$. Substituting these long-wavelength expressions into (\ref{eqn:W}), eliminating its final term by using $\sum_s q_s \nsp \dupar = 0$, and manipulating (\ref{eqn:uexb}) and (\ref{eqn:dBprp}) to write $\sum_{\bs{k}} k^2_\perp|\varphik|^2$ and $\sum_{\bs{k}} k^2_\perp |A_{\parallel\bs{k}}|^2$ in terms of $u^2_\perp$ and $\delta B^2_\perp$, respectively, we find that the gyrokinetic invariant reduces to its KRMHD counterpart (\ref{eqn:WK}), as it should. More generally, equation (\ref{eqn:W}) may be equivalently written in Fourier space using Parseval's theorem as $W_{\rm GK} = \sum_{\bs{k}} W_{{\rm GK},\bs{k}}$, where
\begin{align}\label{eqn:Wk}
W_{{\rm GK},\bs{k}} &= \sum_s \int\rmd^3\bb{v} \, \frac{\tprl{s} |\delta\widetilde{f}_{s\bs{k}}|^2}{2 f^\parallel_{0s}} + \sum_s \frac{\kzero{s}-\Gamma^\perp_{00s}(\alpha_s)}{\alpha_s} \frac{m_s \nsp|\bb{u}_{\perp\bs{k}}|^2}{2}   \nonumber\\*
\mbox{} &+ \Biggl\{ 1 + \sum_s \frac{\betaprl{s}}{2} \Biggl[ \frac{\tprp{s}}{\tprl{s}} \frac{1-\Gamma_{00s}(\alpha_s)}{\alpha_s} - \frac{\ktwo{s}-\Gamma^\perp_{02s}(\alpha_s)}{\alpha_s} \Biggl( 1 + \frac{2\duparsq}{\vthprl{s}^2} \Biggr) \Biggr] \Biggr\} \frac{|\dBprpk|^2}{8\upi} \nonumber\\*
\mbox{} &+ \Biggl\{ 1 - \sum_s \betaprp{s} \Biggl[ \frac{\tprp{s}}{\tprl{s}} \frac{\Gamma^\parallel_{20s}(0,\alpha_s)}{2} - \frac{\Gamma^\perp_{20s}(\alpha_s)}{2} \Biggr] \Biggr\} \frac{|\dBprlk|^2}{8\upi} \nonumber\\*
\mbox{} &- \sum_s q_s \nsp \dupar \Gamma^\perp_{11s}(\alpha_s) \frac{\Aprlk^\star\dBprlk}{cB_0} ,
\end{align}
where $\star$ denotes the complex conjugate.

Each of the terms in (\ref{eqn:W}) (or (\ref{eqn:Wk})) deserves some discussion. The first term ($\propto$$\delta\widetilde{f}^2_s / f^\parallel_{0s}$) is due to the piece of the distribution function that represents changes in the kinetic energy of the particles due to interactions with the compressive fluctuations. In it are contributions from Landau-resonant particles, whose energy is changed by the parallel-electric and magnetic-mirror forces in such a way as to enable \citet{landau46} and \citet{barnes66} damping of compressive fluctuations. In the pressure-isotropic case, this term is simply the perturbed entropy of the system in the frame of the Alfv\'{e}nic fluctuations (S09). The second term ($\propto$$\varphi^2$) represents the energy associated with the $\bb{E}\btimes\bb{B}$ motion. At long wavelengths, it is equal to $\rho_0 u^2_\perp / 2$ (see (\ref{eqn:uexb})). The next two terms represent the energetic cost of bending the magnetic-field lines (note that $k^2_\perp|\Aprlk|^2 \propto |\dBprpk|^2$), with an increase or decrease in this cost dependent upon the pressure anisotropy of the mean distribution function and the presence of interspecies drifts. The $\dBprl^2$ term signals a change in the energetic cost of compressing the magnetic-field lines ($\dBprl\ne 0$) because of background pressure anisotropy. The final term ($\propto$$A_\parallel\dBprl$) has no long-wavelength limit. It is related to conservation of helicity of the perturbed magnetic field,
\begin{align}
\int\rmd^3\bb{r}\, \bb{A}\bcdot\delta\bb{B} &= \int\rmd^3\bb{r} \, \bigl[ \bb{A}_\perp\bcdot \bigl(\grad_\perp\btimes A_\parallel \ez \bigr) + A_\parallel \dBprl \bigr] \nonumber\\*
\mbox{} &= \int\rmd^3\bb{r} \, \bigl[ A_\parallel \ez \bcdot \bigl( \grad_\perp\btimes\bb{A}_\perp \bigr) + A_\parallel \dBprl \bigr] = 2 \int\rmd^3\bb{r}\, A_\parallel \dBprl ,
\end{align}
which is broken by parallel electric fields.

These parallel electric fields are on the right-hand side of (\ref{eqn:WGK}), which is comprised of the fluctuating parallel force on gyrocentres multiplied by the number density of gyrocentres and the parallel interspecies drifts (recall that $g_s$ is the gyrocentre distribution function, (\ref{eqn:gyrocentredistribution})). In the long-wavelength limit, the right-hand side of (\ref{eqn:WGK}) is precisely the same as the right-hand side of (\ref{eqn:W}) -- the work done on the plasma by the fluctuating parallel-electric and magnetic-mirror forces acting on the equilibrium drifts. The only difference in the gyrokinetic limit is that these (ring-averaged) parallel forces effectively act on the guiding centres instead of on the particles.

As explained in Paper I, because we have ordered collisions out of our equations, the long-wavelength invariant given by (\ref{eqn:WK}) is just one of an infinite number of invariants of the system. The same holds true for the more general invariant $W_{\rm GK}$ (\ref{eqn:W}) derived here. The invariant $W_{\rm GK}$ does, however, have special significance for (at least) two reasons.

First, $W_{\rm GK}$ properly reduces to the gyrokinetic free-energy invariant for a (collisional) Maxwellian plasma (cf.~(74) of S09):
\begin{subequations}
\begin{align}
W_{\rm GK} &\rightarrow \int\rmd^3\bb{r} \, \Biggl[ \sum_s \int\rmd^3\bb{v} \, \frac{T_{0s} \delta\widetilde{f}^2_s}{2 f_{0s}} + \sum_s \frac{q^2_s \nsp}{2T_{0s}} \, \varphi \Bigl( 1 - \widehat{\Gamma}_{00s} \Bigr) \varphi + \frac{|\delta\bb{B}|^2}{8\upi} \Biggr] , \\*
\mbox{} &= \int\rmd^3\bb{r} \,\Biggl( \sum_s \int\rmd^3\bb{v} \, \frac{T_{0s} \delta f^2_s}{2f_{0s}} + \frac{|\delta\bb{B}|^2}{8\upi} \Biggr) ,
\end{align}
\end{subequations}
which is the gyrokinetic version of a kinetic invariant variously referred to as the generalised grand canonical potential \citep{hallatschek04} or free energy \citep{fowler68,brizard94,scott10} because of its similarity to the Helmholtz free energy. It is the only quadratic invariant of Maxwellian gyrokinetics in three dimensions (S09).

Secondly, $W_{\rm GK}$ affords a thermodynamic interpretation of the effect of the mirror and firehose (in)stability parameters on nonlinear fluctuations. When the plasma is linearly stable, $W_{\rm GK}$ is a positive-definite quantity that measures the amount of (generalised) free energy carried by the fluctuations. As the stability thresholds are approached, it becomes energetically `cheaper' to bend ($\dBprp$) or compress/rarefy ($\dBprl$) the magnetic-field lines, depending upon the sign of the pressure anisotropy; negative anisotropy effectively weakens the restoring magnetic-tension force, while positive anisotropy reduces the magnetic pressure response. When the plasma is unstable, $W_{\rm GK}$ is minimised by growing fluctuations.\footnote{A perceptive reader might notice that the factor multiplying $\dBprl^2$ in the free-energy invariant (\ref{eqn:W}), which reduces to $1 - \sum_s \betaprp{s} \Delta_{2s}$ in the long-wavelength limit and $1-\betaprp{e} \Delta_{2e}$ at sub-ion-Larmor scales, is not the exact mirror stability parameter (cf.~(I--B14) for arbitrary $f_{0s}$ at long wavelengths and (\ref{eqn:generalmirror}) more generally; at sub-ion-Larmor scales, the exact mirror stability parameter for a bi-Maxwellian equilibrium is $1-\betaprp{e} \Delta_{e} + \mc{K}$ -- see (\ref{eqn:kaw})). As discussed in the final paragraph of \S I--4.1, this factor does not capture the stabilizing influence due to the interaction of linearly resonant particles with the parallel electric field (i.e.~Landau damping). This physics is instead contained inside the first term in the invariant, proportional to $\delta\widetilde{f}^2_s$, and can be ignored at very high $\betaprp{s}, \betaprl{s}$ or for cold electrons.} See \S I--3.3 and \S I--4.1 and for further discussion.

With a general invariant $W_{\rm GK}$ (\ref{eqn:W}) in hand, valid across all scales (as long as they satisfy the gyrokinetic ordering), we now ask how the phase-space cascade begun in the inertial range (i.e.~$k_\perp \rho_i \ll 1$) proceeds through the sub-ion-Larmor kinetic range (i.e.~$k_\perp \rho_i \gg 1$). In the next section, we show that the power arriving from the inertial range is redistributed at $k_\perp \rho_i \sim 1$ into two independent cascades: a KAW cascade and an ion-entropy cascade. Just {\em how} this power is redistributed is a function of the background pressure anisotropy, the plasma beta parameter, the ion-to-electron temperature ratio, and, in particular, the background interspecies drifts, which are responsible for the source term on the right-hand side of (\ref{eqn:WGK}). This source term represents the work done by fluctuations as they extract free energy from these flows, a process that requires the electric and magnetic-mirror forces involved to be coherent on time scales long enough to affect said flows (which are taken to be $\sim$$\vthprl{i}$). For $k_\perp \rho_i \gg 1$, the fluctuations have frequencies too large and amplitudes too small for this source term to matter, and $W_{\rm GK}$ is approximately conserved. In fact, for $\betaprp{s} \sim \betaprl{s} \sim 1$, all terms proportional to  $\dupar$ vanish to leading order in $k_\perp \rho_i$ from the dynamical equations,\footnote{This is because all appearances of $\dupar$ in (\ref{eqn:W}) are accompanied by $A_\parallel$, which is $\sim$$\varphi / k_\perp \rho_i \ll \varphi$ for KAWs and therefore small deep in the sub-ion-Larmor range.} and the KAW and ion-entropy cascades proceed independently, without regard for the presence of background interspecies drifts. For that reason, we keep the following exposition as simple as possible by focusing exclusively on the single-ion-species, bi-Maxwellian case. Nothing of great import appears to be gainable by carrying around sums over ion species and various $\cell{s}$ and $\kell{s}$ coefficients.

\subsection{Gyrokinetic free energy in the sub-ion-Larmor range}\label{sec:KAWfreeenergy}

In the wavenumber range $k_\perp \rho_e \ll 1 \ll k_\perp \rho_i$, the $\widehat{\Gamma}_{\ell m}$ operators in the gyrokinetic free energy (\ref{eqn:W}) take on the long-wavelength limits (\ref{eqn:longwavelengthG}) for the electrons and vanish to leading order for the ions. Equation (\ref{eqn:W}) then becomes
\begin{align}\label{eqn:WGKsubion1}
W_{\rm GK} &\rightarrow \int\rmd^3\bb{r} \, \Biggl\{  \int\rmd^3\bb{v}\, \frac{\tprl{i}\widetilde{h}^2_i}{2f_{0i}} + \int\rmd^3\bb{v}\, \frac{\tprl{e}\delta\widetilde{f}^2_e}{2f_{0e}} + \frac{\pprp{i}}{2} \biggl( \frac{Z_i e\varphi}{\tprp{i}}\biggr)^2  \nonumber\\*
\mbox{} &\quad  + \biggl( 1 + \frac{\betaprl{e}}{2} \Delta_e \biggr) \frac{\delta B^2_\perp}{8\upi} + \bigl( 1 - \betaprp{e} \Delta_e \bigr) \frac{\dBprl^2}{8\upi} \Biggr\} .
\end{align}
To simplify this expression further, we employ the mass-ratio expansion implicit in the $k_\perp \rho_e \ll 1 \ll k_\perp \rho_i$ ordering to obtain the leading-order electron kinetic response (see (I--C73))
\begin{equation}\label{eqn:dfe}
\delta\widetilde{f}_e = g_e + \frac{v^2_\perp}{\vthprl{e}^2} \frac{\dBprl}{B_0} f^\parallel_{0e} = \biggl( \frac{\dne}{\nem} + \Delta_e \frac{\dBprl}{B_0} \biggr) f_{0e} ,
\end{equation}
as well as the reduced quasi-neutrality constraint $\dne/\nem = -Z_i e\varphi/\tprp{i}$ (see (\ref{eqn:KAWdn})). Substituting these approximations into (\ref{eqn:WGKsubion1}) gives
\begin{align}
W_{\rm GK} &\rightarrow \int\rmd^3\bb{r} \, \Biggl\{ \int\rmd^3\bb{v}\, \frac{\tprl{i}\widetilde{h}^2_i}{2f_{0i}} + \frac{\pprl{e}}{2} \biggl( -\frac{Z_i e\varphi}{\tprp{i}} + \Delta_e \frac{\dBprl}{B_0} \biggr)^2 + \frac{\pprp{i}}{2} \biggl( \frac{Z_i e\varphi}{\tprp{i}}\biggr)^2 \nonumber\\*
\mbox{} &\quad + \biggl( 1 + \frac{\betaprl{e}}{2} \Delta_e \biggr) \frac{\delta B^2_\perp}{8\upi} + \bigl( 1 - \betaprp{e} \Delta_e \bigr) \frac{\dBprl^2}{8\upi} \Biggr\} \label{eqn:WGKsubion2}\\*
\mbox{} &\doteq W_{\widetilde{h}_i} + W_{\rm KAW} .\label{eqn:WhWKAW}
\end{align}
The first term, $W_{\widetilde{h}_i}$, is proportional to the total variance of $\widetilde{h}_i$; its cascade to small scales in phase space is discussed in \S\ref{sec:entropy}. The remaining terms in (\ref{eqn:WGKsubion2}) constitute the independently cascaded KAW energy $W_{\rm KAW}$. We shall show in the next two subsections that these two parts, $W_{\widetilde{h}_i}$ and $W_{\rm KAW}$, are independently conserved. How the total generalised gyrokinetic free energy $W_{\rm GK}$ is partitioned between them is determined at $k_\perp \rho_i \sim 1$, which ultimately sets the amount of ion versus electron heating.

\subsubsection{ERMHD and the KAW cascade}\label{sec:KAWcascade}

Using (\ref{eqn:KAWdBprl}) to relate the magnetic-field-strength fluctuation $\dBprl$ in the sub-ion-Larmor range to the electrostatic potential $\varphi$, writing the energy in the perpendicular magnetic field $\delta B^2_\perp$ as $|\grad_\perp A_\parallel|^2$, and re-introducing the stream and flux functions via (\ref{eqn:PhiPsi}), the KAW invariant that appears in (\ref{eqn:WhWKAW}) may be written as
\begin{subequations}\label{eqn:WKAW}
\begin{align}
W_{\rm KAW} &= \int\rmd^3\bb{r} \, \frac{m_i \nip}{2} \biggl\{ \biggl( 1+ \frac{\betaprl{e}}{2} \Delta_e \biggr) | \grad_\perp \Psi |^2  \nonumber\\*
\mbox{} &+ \frac{1 - \betaprp{e}\Delta_e + \mc{K}}{(1-\betaprp{e}\Delta_e/2)^2} \biggl(1+\frac{Z_i\tprl{e}}{\tprp{i}}\biggr) \biggl[ 2 + \betaprp{i} \biggl( 1 + \frac{Z_i\tprl{e}}{\tprp{i}} \biggr) - \betaprl{e} \Delta^2_e \biggr] \frac{\Phi^2}{\rho^2_i} \biggr\} \\*
\mbox{} &= \int\rmd^3\bb{r}\, \frac{m_i \nip}{4} \bigl( \, |\Theta^+|^2 + |\Theta^-|^2 \,\bigr) ,
\end{align}
\end{subequations}
which is the sum of the energies of the ``$+$'' and ``$-$'' linear KAW eigenmodes found in equation (\ref{eqn:Theta}).\footnote{Equation (246) of S09 specifies $W_{\rm KAW}$ for the case of a Maxwellian equilibrium, which may be compared with our equation (\ref{eqn:WKAW}). There are two typographical errors in their formula: the $\Phi^2/\rho^2_i$ term is missing a multiplicative factor of $2$, and the pre-factor of $1/2$ on the final line of that formula ought to be $1/4$. Neither error affects their subsequent analysis.} This decomposition of the KAW invariant into a sum of energies of the $\Theta^+$ and $\Theta^-$ fluctuations underlies the scaling theory of KAW turbulence presented in S09. That theory is based on the reasoning (i) that $\Theta^+$ and $\Theta^-$ are nonlinearly coupled and thus should have similar scaling with $k_\perp$, (ii) that they obey a constant-flux cascade to small spatial scales through local interactions, and (iii) that critical balance holds for the KAW fluctuations. As long as the plasma remains mirror- and firehose-stable, the resulting scaling laws are not expected to change (although the overall fluctuation amplitudes will; see \S\ref{sec:scalings}). 

We now show that the form of $W_{\rm KAW}$ in (\ref{eqn:WKAW}) can be similarly obtained from the equations describing nonlinear KAWs -- so-called {\em electron reduced magnetohydrodynamics} (ERMHD) -- and thus is the ``fluid energy'' conserved during the nonlinear cascade of KAWs to small spatial scales.

In appendix C.8 of Paper I, we derived, via a mass-ratio expansion, nonlinear equations describing the electron kinetics for $k_\perp \rho_e \sim k_\perp \rho_i (m_e/m_i)^{1/2} \ll 1$. For the purposes of the present paper, the two most important electron equations are those specifying the parallel electric field (see (I--C72))
\begin{equation}\label{eqn:eprl}
E_\parallel = - \eb\bcdot\grad\varphi -\frac{1}{c} \pD{t}{A_\parallel}  = - \eb\bcdot\grad \frac{\tprl{e}}{e} \biggl( \frac{\dne}{\nem} + \Delta_e \frac{\dBprl}{B_0} \biggr) 
\end{equation}
and what amounts to a reduced electron continuity equation (see (I--C78) and the accompanying discussion in \S I--C.8.3):
\begin{equation}\label{eqn:econt}
\D{t}{} \biggl( \frac{\dne}{\nem} - \frac{\dBprl}{B_0} \biggr) + \eb\bcdot\grad u^{(1)}_{\parallel e} + \frac{c\tprp{e}}{eB_0} \biggl\{ \frac{\dne}{\nem} , \frac{\dBprl}{B_0} \biggr\} = 0,
\end{equation}
where
\begin{equation}
\D{t}{} \doteq \pD{t}{} + \bb{u}_\perp\bcdot\grad = \pD{t}{} + \frac{c}{B_0} \bigl\{ \varphi , \dots \bigr\}
\end{equation}
is the Lagrangian time derivative measured in a frame transported at the $\bb{E}\btimes\bb{B}$ drift velocity, $\bb{u}_\perp$ (defined by (\ref{eqn:uexb})), and
\begin{equation}\label{eqn:uprle1}
u^{(1)}_{\parallel e} \doteq \frac{1}{\nem} \int\rmd^3\bb{v} \, v_\parallel g^{(1)}_e = \frac{c}{4\upi e\nem} \biggl( 1 + \frac{\betaprl{e}}{2} \Delta_e \biggr) \nabla^2_\perp A_\parallel
\end{equation}
is the parallel electron flow velocity associated with the $\mc{O}(\sqrt{m_e/m_i})$ piece of $g_e$, denoted $g^{(1)}_e$ in Paper I (see \S I--C.8).\footnote{There is a error in (I--C76), which defines $n_{0e} u^{(1)}_{\parallel e}$ as the first parallel-velocity moment of $g^{(1)}_e$ and equates it to $\nem \delta u_{\parallel e}$. A comparison between (\ref{eqn:uprle1}) and (\ref{eqn:KAWuprle}) reveals that $u^{(1)}_{\parallel e}$ is, in fact, {\em not} equal to the perturbed parallel electron flow velocity $\delta u_{\parallel e}$, as was asserted by (I--C76). This error is due to a notational inconsistency in \S I--C.8 concerning electron parallel flows, one that does not affect any of the other equations or conclusions in that paper or in this one.}

Using (\ref{eqn:KAWdn}) and (\ref{eqn:KAWdBprl}) for the electron density fluctuation and the magnetic-field-strength fluctuation in the sub-ion-Larmor scale range, and introducing $\Phi$ and $\Psi$ via (\ref{eqn:PhiPsi}), we find that equations (\ref{eqn:eprl}) and (\ref{eqn:econt}) become, respectively,
\begin{equation}\label{eqn:KAWPsi}
\pD{t}{\Psi} = \valf \Biggl\{ 1 + \frac{Z_i\tprl{e}}{\tprp{i}} \Biggl[ 1 -   \frac{\betaprp{e}}{2} \Delta_{e} \biggl( \frac{1+\tauprp{i}/Z_i}{1-\betaprp{e}\Delta_e/2} \biggr) \Biggr] \Biggr\} \eb\bcdot\grad \Phi ,
\end{equation}
\begin{equation}\label{eqn:KAWPhi}
 \pD{t}{\Phi} \Biggl[ 2 + \betaprp{i} \biggl( \frac{1+Z_i/\tauprp{i}}{1-\betaprp{e}\Delta_e/2}\biggr) \Biggr] = - \valf  \biggl( 1 + \frac{\betaprl{e}}{2} \Delta_e \biggr) \eb\bcdot\grad  \rho^2_i \nabla^2_\perp  \Psi .
\end{equation}
These equations generalize the linear theory of KAWs (\S\ref{sec:linKAW}) to the nonlinear regime (cf.~equations (226)--(227) of S09, to which (\ref{eqn:KAWPsi})--(\ref{eqn:KAWPhi}) reduce in the pressure-isotropic limit). Introducing the perturbed magnetic-field vector
\begin{equation}
\frac{\delta\bb{B}}{B_0} = \frac{1}{\valf} \ez\btimes\grad_\perp \Psi + \ez \frac{\dBprl}{B_0} 
\end{equation}
with $\dBprl$ given by (\ref{eqn:KAWdBprl}), equations (\ref{eqn:KAWPsi}) and (\ref{eqn:KAWPhi}) can be recast as two coupled evolution equations for the perpendicular and parallel components of the perturbed magnetic field, respectively.

We now show that (\ref{eqn:KAWPsi}) and (\ref{eqn:KAWPhi}) conserve the energy $W_{\rm KAW}$ given by (\ref{eqn:WKAW}). First, apply $\grad_\perp$ to (\ref{eqn:KAWPsi}) and dot the result with $( 1 + \betaprl{e}\Delta_e/2) \grad_\perp \Psi$; then integrate over space and use integration by parts to obtain
\begin{align}\label{eqn:KAWproof1}
&\D{t}{} \int\rmd^3\bb{r} \, \frac{1}{2} \biggl( 1 + \frac{\betaprl{e}}{2} \Delta_e \biggr) |\grad_\perp\Psi|^2 = - \int\rmd^3\bb{r} \,  \valf \biggl( 1 + \frac{\betaprl{e}}{2} \Delta_e \biggr)   \nonumber\\*
\mbox{} &\quad\times \Biggl\{ 1 + \frac{Z_i\tprl{e}}{\tprp{i}} \Biggl[ 1 -   \frac{\betaprp{e}}{2} \Delta_{e} \biggl( \frac{1+\tauprp{i}/Z_i}{1-\betaprp{e}\Delta_e/2}\biggr) \Biggr] \Biggr\} \bigl( \nabla^2_\perp\Psi \bigr)  \eb\bcdot\grad \Phi .
\end{align}
Next, we multiply (\ref{eqn:KAWPhi}) by $(1+Z_i\tprl{e}/\tprp{i}) (1-\betaprp{e}\Delta_e/2) (\Phi/\rho^2_i)$, integrate over space, and use integration by parts to find
\begin{align}\label{eqn:KAWproof2}
&\D{t}{} \int\rmd^3\bb{r} \, \frac{1}{2} \biggl( 1 + \frac{Z_i\tprl{e}}{\tprp{i}} \biggr) \biggl[ 2 + \betaprp{i} \biggl( 1 + \frac{Z_i\tprl{e}}{\tprp{i}} \biggr) - \betaprl{e} \Delta^2_e \biggr]  \frac{\Phi^2}{\rho^2_i} \nonumber\\*
\mbox{} &\quad =  \int\rmd^3\bb{r} \, \valf \biggl( 1 + \frac{\betaprl{e}}{2}\Delta_e \biggr) \biggl( 1 + \frac{Z_i\tprl{e}}{\tprp{i}} \biggr)  \biggl(1 - \frac{\betaprp{e}}{2} \Delta_e \biggr)  ( \eb\bcdot\grad \Phi ) \nabla^2_\perp\Psi .
\end{align}
Further multiplying (\ref{eqn:KAWproof2}) by a factor of $ (1-\betaprp{e}\Delta_e + \mc{K})/(1-\betaprp{e}\Delta_e/2)^2$ and adding it to (\ref{eqn:KAWproof1}) cancels their right-hand sides and gives
\begin{equation}\label{eqn:KAWconservation}
\D{t}{W_{\rm KAW}} = 0,
\end{equation}
with the KAW invariant $W_{\rm KAW}$ identical to that given by (\ref{eqn:WKAW}).

It is clear from the ERMHD equations (\ref{eqn:KAWPsi}) and (\ref{eqn:KAWPhi}), as well as from the form of the quadratic quantity they conserve (\ref{eqn:WKAW}), that electron pressure anisotropy influences the ability of the plasma to support certain magnetic-field perturbations. As the firehose threshold is approached from the stable side (i.e.~$1 + \betaprl{e}\Delta_e/2 \rightarrow 0^+$), the energetic cost for a KAW to bend the magnetic-field lines is reduced. Likewise, as the mirror threshold is approached from the stable side (i.e.~$1 - \betaprp{e} \Delta_e + \mc{K} \rightarrow 0^+$), the energetic cost for a KAW to compress/rarefy the magnetic-field lines is reduced. Beyond these thresholds, it becomes energetically profitable to grow these fluctuations. (Again, note that these instabilities themselves fall outside of the gyrokinetic ordering employed here.) Equation (\ref{eqn:KAWconservation}) thus establishes a thermodynamic connection between the linear and nonlinear stability of KAWs.

Before proceeding to investigate the ion-entropy cascade, in which $W_{\widetilde{h}_i}$ is independently conserved, we briefly remark on the case of $\Delta_e = 2/\betaprp{e}$, for which the nonlinear ERMHD equations (\ref{eqn:KAWPsi}) and (\ref{eqn:KAWPhi}) are ill-posed. Instead, the appropriate nonlinear equations are
\begin{equation}
\frac{1}{v_A d_i} \pD{t}{A_\parallel} = \frac{\betaprl{e}}{2} \Delta_e \,\eb\bcdot\grad \dBprl = \frac{1}{1+2/\betaprp{e}} \,\eb\bcdot\grad \dBprl ,
\end{equation}
\begin{equation}
\frac{1}{v_A d_i} \pD{t}{\dBprl} = \biggl( 1 + \frac{\betaprl{e}}{2} \Delta_e \biggr) \eb\bcdot\grad \nabla^2_\perp A_\parallel = \frac{2+2/\betaprp{e}}{1+2/\betaprp{e}} \,\eb\bcdot\grad\nabla^2_\perp A_\parallel ,
\end{equation}
which of course return the eigenvectors (\ref{eqn:KAWmod_dBprl}) and growth/decay rates (\ref{eqn:KAWmod_gamma}) in the linear regime. The corresponding conserved energy is
\begin{equation}
W_{\rm KAW} = \int\rmd^3\bb{r}  \, \Biggl[ \biggl(2+ \frac{2}{\betaprp{e}} \biggr) \frac{\delta B^2_\perp}{8\upi} - \frac{\dBprl^2}{8\upi} \Biggr] \biggl( 1 + \frac{2}{\betaprp{e}} \biggr)^{-1} ,
\end{equation}
which, as with (\ref{eqn:WKAW}), may be written as the sum of the energies of the ``$+$'' and ``$-$'' linear KAW eigenmodes found in equation (\ref{eqn:KAWmod_dBprl}). Note that, if $\dBprl^2$ grows or decays, so too must $\delta B^2_\perp$ grow or decay proportionally to keep $W_{\rm KAW}$ constant. As explained after (\ref{eqn:KAWmod_gamma}), the resulting instability occurs because perpendicular pressure balance is maintained for any $\dBprl$ so long as $\Delta_e = 2/\betaprp{e}$.

\subsubsection{Ion-entropy cascade}\label{sec:entropy}

The other piece of the gyrokinetic invariant, $W_{\widetilde{h}_i}$ (see (\ref{eqn:WhWKAW})), represents the cascade of ion-entropy fluctuations to small scales in phase space (velocity and position) via nonlinear phase mixing in the gyrocentre space (see \S 7.9 of S09 for further details). It may be obtained directly from the gyrokinetic equation (\ref{eqn:gkequation2}) as follows.

In the sub-ion-Larmor range, the ring-averaged gyrokinetic potential $\langle\chi\rangle_{\gai}$ is dominated by the contribution from the electrostatic potential:
\begin{align}
\frac{Z_i e}{\tprp{i}} \langle\chik \rangle_{\gai} &= \frac{2}{\sqrt{\betaprp{i}}} \biggl\{ \biggl[ J_0(a_i) + \frac{v^2_\perp}{\vasq} \frac{J_1(a_i)}{a_i} \frac{1+Z_i/\tauprp{i}}{1-\betaprp{e}\Delta_e/2} \biggr] \frac{\Phik}{\rho_i \valf}  + J_0(a_i) \frac{v_\parallel}{\valf} \frac{\Psik}{\rho_i \valf} \biggr\} \nonumber\\*
\mbox{} &\simeq \frac{2}{\sqrt{\betaprp{i}}} \sqrt{\frac{2}{\pi a_i}} \cos\Bigl(a_i - \frac{\pi}{4}\Bigr) \frac{\Phik}{\rho_i \valf} \quad {\rm for}~a_i \doteq k_\perp \rho_i \frac{v_\perp}{\vthprp{i}} \gg 1 ,
\end{align}
because of the KAW scaling $\Psik \sim \Phik / k_\perp \rho_i$ (see (\ref{eqn:Theta})).\footnote{Note that having $\beta_i \sim a_i$ can interfere with this large-$a_i$ expansion by allowing larger $A_\parallel$ and $\dBprl$, as can having an electron pressure anisotropy within $\sim$$a^{-2}_i$ of the mirror or firehose instability threshold when $\beta_i \sim \beta_e \sim 1$.} The gyrokinetic equation (\ref{eqn:gkequation2}) in the sub-ion-Larmor range is then
\begin{equation}\label{eqn:KAWgkequation}
\pD{t}{\widetilde{h}_i} + v_\parallel \pD{z}{\widetilde{h}_i} + \{ \langle\Phi\rangle_{\gai} , \widetilde{h}_i \} = \frac{Z_i e B_0}{c\tprl{i}} \pD{t}{\langle \Phi \rangle_{\gai}}  f_{0i} .
\end{equation}
This equation states that $\widetilde{h}_i$ is linearly phase-mixed by parallel-streaming ions, nonlinearly phase-mixed in the gyrocentre space by ring-averaged $\bb{E}\btimes\bb{B}$ flows (i.e.~KAW turbulence), and sourced by wave-particle interactions, which become asymptotically small as $k_\perp \rho_i \gg 1$. Multiplying (\ref{eqn:KAWgkequation}) by $\tprl{i}\widetilde{h}_i/f_{0i}$ and integrating over the phase space, we obtain
\begin{equation}\label{eqn:Wh}
\D{t}{W_{\widetilde{h}_i}} \doteq \D{t}{} \int\rmd^3\bb{v} \int\rmd^3\bb{R}_i \, \frac{\tprl{i}\widetilde{h}^2_i}{2 f_{0i}} = \frac{Z_i e B_0}{c} \int\rmd^3\bb{v} \int\rmd^3\bb{R}_i \, \pD{t}{\langle\Phi\rangle_{\gai}} \widetilde{h}_i .
\end{equation}
Deep in the sub-ion-Larmor range, the right-hand side of (\ref{eqn:Wh}) representing the collisionless damping at the ion gyroscale is small and the total variance of $\widetilde{h}_i$ is conserved (as was shown also by equation (\ref{eqn:WhWKAW})). The ion-entropy fluctuations are transferred across this scale range by means of a cascade, in which particles with like gyrocentre positions but with different $v_\perp$ (and so different gyroradii) sample spatially decorrelated electromagnetic fluctuations and thus $\ez\btimes\grad\langle\Phi\rangle_{\gai}$ drift with different velocities.

Equation (\ref{eqn:Wh}) makes clear that neither ion nor electron pressure anisotropy (in a bi-Maxwellian plasma) affects the ion-entropy cascade itself. Instead, by shifting the precise $k$-space location of the frequency break that occurs near $k_\perp \rho_i \sim 1$ (see \S\ref{sec:numerical}), pressure anisotropy could adjust the values of $\Phi$ and $\widetilde{h}_i$ at the end of the inertial range, where the ion-entropy cascade is sourced. This may, in turn, alter the fraction of free energy that enters the entropy cascade rather than the KAW cascade. Since the entropy cascade is what ultimately carries the fraction of the energy diverted from the inertial-range electromagnetic fluctuations by the collisionless damping (wave--particle interaction) to collisional scales, the ion and electron pressure anisotropies are likely to affect the ratio of ion to electron heating.

\subsection{Turbulence scaling relations}\label{sec:scalings}

Our final stop on our tour of the nonlinear kinetic cascade of the generalised energy concerns the $k$-space scaling relations: in the phase-space cascades of Alfv\'{e}n waves, compressive fluctuations, KAWs, and ion entropy, how do the fluctuation amplitudes depend upon scale? For each of these cascades, we follow the arguments presented in S09, which have their origin in \citet{kolmogorov41} and \citet{obukhov41}. That is, we assume (a) that the flux of energy through any scale $\lambda$ is independent of $\lambda$, (b) that the interactions leading to this flux are local in scale, and, crucially, (c) that the linear and nonlinear time scales are related scale by scale by the {\em critical balance} \citep{gs95,cl04}. The latter amounts to the assumption that the turbulence is strong and fluctuations on a particular scale nonlinearly interact and cascade on a time comparable to the wave-crossing time at that scale. The resulting energy spectra are identical in their slopes to those found in S09 and differ only in their normalization. This is to be expected: the background temperature anisotropies considered in this theory are spatially constant over the turbulence length scales of interest, and so all scales in a given cascade are affected by these anisotropies in the same way. To go beyond this -- that is, to consider the feedback of fluctuation-driven (rather than background), scale-dependent temperature anisotropies on the properties of the turbulent cascade -- would require an altogether different ordering than was set out in \S\ref{sec:gkordering} (see \citealt{squire16,squire17} for progress in this direction). Nevertheless, in what follows, we provide scaling relations for fluctuation-driven temperature anisotropies under the assumption (implied by the gyrokinetic ordering (\ref{eqn:ordering})) that they do not back-react on the fluctuations that drive them.

\subsubsection{Inertial-range scaling relations}

We first treat the long-wavelength inertial range $k_\perp \rho_i \ll 1$ (Paper I) by considering Alfv\'{e}nic fluctuations with perpendicular scale $\lambda$ and parallel scale $\ell_\lambda \gg \lambda$. The critical balance condition at scale $\lambda$ is given by
\begin{equation}
\omega_\lambda \sim \frac{\valfeff}{\ell_\lambda} \sim \frac{u_{\perp\lambda}}{\lambda} ,
\end{equation}
where $\omega_\lambda$ is the typical frequency of the fluctuations at scale $\lambda$ and $\valfeff$ includes the effect of temperature anisotropy on the propagation speed of linear Alfv\'{e}n waves (see (\ref{eqn:alfvenwave})). The cascade time is then $\tau_\lambda \sim \ell_\lambda / \valfeff \sim \lambda / u_{\perp\lambda}$, whence
\begin{equation}\label{eqn:AWscalings}
\frac{u^2_{\perp\lambda}}{\tau_\lambda} \sim \varepsilon = {\rm const} \quad \Longrightarrow \quad u_{\perp\lambda} \sim (\varepsilon \lambda)^{1/3} \quad{\rm and}\quad \ell_\lambda \sim \ell^{1/3}_0 \lambda^{2/3},
\end{equation}
where $\varepsilon$ is the (spatially constant) average energy per unit time per unit volume that the system dissipates and $\ell_0 \doteq \valfeff^3/\varepsilon$. These scaling relations state that the kinetic-energy spectrum is $\sim$$k^{-5/3}_\perp$ and that the spatial anisotropy of the turbulent fluctuations increases at smaller scales. For a given rate of energy injection, background temperature anisotropies do not affect the fluctuation amplitude at each scale, but do affect the spatial anisotropy at each scale by requiring larger (smaller) $\ell_\lambda$ for positive (negative) temperature anisotropies in order to maintain critical balance.

The compressive fluctuations are passively advected by the Alfv\'{e}nic component of the turbulence (\citealt{lg01}; S09), a feature that carries over unaltered to the pressure-anisotropic case (\S I--2.6). If these compressive fluctuations are weakly damped, they share the Alfv\'{e}nic $k^{-5/3}_\perp$ spectrum. In particular, this is true for the fluctuating pressure anisotropies produced by these modes, which may be obtained from the KRMHD equations (I--2.41), (I--2.42), (I--2.49), and (I--2.52). The fluctuating electron pressure anisotropy is given by
\begin{equation}
\frac{\dpprp{e}-\dpprl{e}}{\pprl{e}} = \Delta_e \biggl( \frac{\dne}{\nem} - \frac{\tprp{e}}{\tprl{e}} \frac{\dBprl}{B_0} \biggr) ,
\end{equation}
which indicates that no scale-dependent electron pressure anisotropy will be generated by the compressive fluctuations if no electron pressure anisotropy exists in the background state. Since electron pressure anisotropy affects the sub-ion-scale dynamics, this is worth noting. On the other hand, the fluctuating ion perpendicular pressure satisfies perpendicular pressure balance and is thus related to the density and magnetic-field-strength fluctuations as follows (see (I--2.52) with (I--2.46) substituted for $g_i$):
\begin{equation}
\frac{\dpprp{i}}{\pprp{i}} =  - \frac{Z_i}{\tauprp{i}} \frac{\dne}{\nem} -  \frac{2}{\betaprp{i}} \biggl( 1 -  \frac{\betaprp{e}}{2} \Delta_e \biggr) \frac{\dBprl}{B_0} .
\end{equation}
The fluctuating ion parallel pressure cannot be written in terms of $\dne$ and $\dBprl$ without further assumptions; its evolution is given by
\begin{equation}
\D{t}{} \biggl( \frac{\dpprl{i}}{\pprl{i}} - \frac{\dBprl}{B_0} \biggr) + \eb\bcdot\grad \frac{Q_{\parallel i}}{\pprl{i}} = 0
\end{equation}
where $Q_{\parallel i} \doteq \int\rmd^3\bb{v} \, m_i v^3_\parallel \delta f_i$ is the parallel flow of parallel ion heat. Whether the parallel ion heat flow is negligible ($\omega \gg |k_\parallel| c_s$) or dominant ($\omega \ll |k_\parallel| c_s$), scale-dependent ion pressure anisotropy will be generated by the fluctuations even when $\Delta_i = 0$. 

If the strength of the collisionless damping of these fluctuations determines whether a cascade of compressive fluctuations can occur, then the background pressure anisotropy will affect the amount of $\dBprl$ and $\dne$, and consequently of $\dpprp{s}$ and $\dpprl{s}$, in this cascade (see \S I--4.4.2 and equation I--4.32{\it a},{\it b}); for example, positive $\Delta_s$ reduces the rate of Barnes damping of magnetic-field-strength fluctuations in a high-$\beta$ plasma (see (\ref{eqn:slowmode})). If compressive fluctuations are strongly damped, or if no compressive fluctuations are excited at long wavelengths in the first place, then the temperature anisotropy would instead be driven adiabatically by the Alfv\'{e}nic fluctuations \citep[e.g.][]{squire16,squire17}:
\begin{equation}\label{eqn:AlfDelta}
\biggl( \frac{\dpprp{s}-\dpprl{s}}{\pprl{s}} \biggr)_{\!\lambda} \propto \biggl| \frac{\delta\bb{B}_{\perp\lambda}}{B_0} \biggr|^2 \propto \lambda^{2/3},
\end{equation}
which translates into a $k^{-7/3}_\perp$ power spectrum of pressure anisotropy. To the extent one can differentiate between $-5/3$ and $-7/3$, the source of the scale-dependent pressure anisotropy could thus be inferred. While it would be interesting to investigate how such scale-dependent pressure anisotropies affect the Alfv\'{e}n-wave cascade, this physics, as well as the scaling relation (\ref{eqn:AlfDelta}), lie outside of the gyrokinetic ordering (\ref{eqn:ordering}).

Before moving on to the kinetic range, we note that modern theories of strong inertial-range MHD turbulence have appeared to converge on a perpendicular spectral slope of $-3/2$, rather than the Goldreich-Sridhar $-5/3$ (see \citealt{schekochihin17rev} for a biased but up-to-date review), the former scaling law having its origins in the phenomenon of dynamical alignment proposed by \citet{boldyrev06}. It is possible to argue that the discrepancy between $-5/3$ and $-3/2$ is fundamentally an intermittency effect \citep{chandran15,ms17}; the specific scaling laws (\ref{eqn:AWscalings}) and (\ref{eqn:AlfDelta}) do not account for intermittency. In any case, the point of this exposition is not tied to whether the turbulence follows one power law or the other, but rather is that the Alfv\'{e}nic and compressive fluctuations in the inertial range can independently drive scale-dependent pressure anisotropies, whose scaling laws say something about the dominant physics and the possible role of pressure anisotropy in influencing the efficacy of collisionless damping.

\subsubsection{Kinetic-range scaling laws}

In the sub-ion-Larmor kinetic range $k_\perp \rho_i \gg 1$, a nonlinear cascade of KAWs can be set up. To obtain scalings associated with it, we follow S09 in assuming that the KAW fluctuations satisfy $\Theta^+_\lambda \sim \Theta^-_\lambda$ (see (\ref{eqn:Theta})), or
\begin{equation}\label{eqn:KAWbalance}
\frac{1-\betaprp{e}\Delta_e/2}{(1-\betaprp{e}\Delta_e + \mc{K})^{1/2}} \,\Psi_\lambda \sim \sqrt{\frac{1+\betaprp{i}-2\mc{K}}{1+ \betaprl{e}\Delta_e/2}} \, \frac{\lambda}{\rho_i} \,\Phi_\lambda
\end{equation}
(for the purposes of these scaling arguments, we set $Z_i \tprl{e}/\tprp{i} \sim 1$, but retain the $\beta_i$ and $\beta_e\Delta_e$ dependences). Using (\ref{eqn:KAWbalance}), the constant-flux assumption becomes
\begin{equation}\label{eqn:KAWflux}
\frac{(\Psi_\lambda/\lambda)^2}{\tau_{{\rm KAW}\lambda}} \sim \frac{1+\betaprp{i}-2\mc{K}}{1+ \betaprl{e}\Delta_e/2} \frac{1-\betaprp{e}\Delta_e + \mc{K}}{(1-\betaprp{e}\Delta_e/2)^2}  \, \frac{(\Phi_\lambda/\rho_i)^2}{\tau_{{\rm KAW}\lambda}} \sim \varepsilon_{\rm KAW} = {\rm const},
\end{equation}
where $\tau_{{\rm KAW}\lambda}$ is the cascade time and $\varepsilon_{{\rm KAW}}$ is the energy flux of KAWs. The latter is the fraction of the total energy flux $\varepsilon$ that is converted into the KAW cascade at the ion gyroscale (which itself is likely a function of the background pressure anisotropy). The cascade time follows from examining the nonlinear terms proportional to $\valf (\dBprp/B_0)\bcdot\grad_\perp \sim \Psi_\lambda / \lambda^2$ in (\ref{eqn:KAWPsi}) and (\ref{eqn:KAWPhi}). Assuming the turbulence is critically balanced, this time is comparable to the inverse of the KAW frequency (\ref{eqn:kaw}):\footnote{There is a typographical error in equation (239) of S09, which is the isotropic-Maxwellian version of (\ref{eqn:tauKAW}); its right-hand side should be inverted.}
\begin{equation}\label{eqn:tauKAW}
\tau_{{\rm KAW}\lambda} \sim  \frac{1-\betaprp{e}\Delta_e/2}{1-\betaprp{e}\Delta_e + \mc{K}} \frac{\lambda^2}{\Phi_\lambda} \sim \sqrt{\frac{1+\betaprp{i}-2\mc{K}}{(1-\betaprp{e}\Delta_e + \mc{K})(1+\betaprl{e}\Delta_e/2)}} \, \frac{\lambda}{\rho_i} \frac{\ell_\lambda}{\valf} .
\end{equation}
Combining (\ref{eqn:KAWflux}) and (\ref{eqn:tauKAW}) leads to the following scaling relations for KAW turbulence:
\begin{gather}
\Phi_\lambda \sim  \biggl( \frac{\varepsilon_{\rm KAW}}{\varepsilon} \biggr)^{1/3} \ell^{-1/3}_0 \rho^{2/3}_i \lambda^{2/3} \,  \frac{\valfeff(1-\betaprp{e}\Delta_e/2)}{(1-\betaprp{e}\Delta_e + \mc{K})^{2/3}} \biggl( \frac{1+\betaprl{e}\Delta_e/2}{1+\betaprp{i}-2\mc{K}} \biggr)^{1/3}  , \\*
\Psi_\lambda \sim  \biggl( \frac{\varepsilon_{\rm KAW}}{\varepsilon} \biggr)^{1/3} \ell^{-1/3}_0 \rho^{-1/3}_i \lambda^{5/3} \, \frac{\valfeff}{(1-\betaprp{e}\Delta_e + \mc{K})^{1/6}} \biggl( \frac{1+\betaprp{i}-2\mc{K}}{1+\betaprl{e}\Delta_e/2} \biggr)^{1/6} , \label{eqn:KAWPsik} \\*
\ell_\lambda \sim \biggl( \frac{\varepsilon}{\varepsilon_{\rm KAW}} \biggr)^{1/3}  \ell^{1/3}_0 \rho^{1/3}_i \lambda^{1/3}  \,\frac{\valf}{\valfeff} \biggl[ \frac{(1-\betaprp{e}\Delta_e + \mc{K})(1+\betaprl{e}\Delta_e/2)}{1+\betaprp{i}-2\mc{K}} \biggr]^{1/6} . \label{eqn:KAWell}
\end{gather}
The corresponding $k^{-7/3}_\perp$ spectrum of magnetic energy is identical to that found in the isotropic-Maxwellian case, but with fluctuation amplitudes that depend on the electron anisotropy and its distance from the firehose and mirror instability thresholds. KAWs also generate scale-dependent ion and electron pressure anisotropy, which may be calculated from the relations (\ref{eqn:KAWdn},{\it d}) and (\ref{eqn:KAWdpi},{\it b}):
\begin{equation}
\biggl(\frac{\dpprp{i}-\dpprl{i}}{\pprl{i}} \biggr)_{\!\lambda} = \Delta_i \frac{\delta n_{i\lambda}}{\nip} = -\Delta_i \frac{2}{\sqrt{\betaprp{i}}} \frac{\Phi_\lambda}{\rho_i \valf} \propto \Delta_i \,\lambda^{2/3} ,
\end{equation}
\begin{align}
\biggl(\frac{\dpprp{e}-\dpprl{e}}{\pprl{e}} \biggr)_{\!\lambda} &= \Delta_e \biggl( \frac{\delta n_{e\lambda}}{\nem} - \frac{\tprp{e}}{\tprl{e}} \frac{\delta B_{\parallel\lambda}}{B_0} \biggr) \nonumber\\*
\mbox{} &= - \Delta_e \frac{2}{\sqrt{\betaprp{i}}} \biggl( 1 + \frac{\tprp{e}}{\tprl{e}} \frac{\betaprp{i}}{2} \frac{1+Z_i/\tauprp{i}}{1-\betaprp{e}\Delta_e/2} \biggr) \frac{\Phi_\lambda}{\rho_i \valf} \propto \Delta_e \,\lambda^{2/3} .
\end{align}
Interestingly, KAWs do not produce scale-dependent ion (electron) pressure anisotropy if no ion (electron) anisotropy exists in the background state. This is because the KAW pressure response is isothermal when $\Delta_{i,e} = 0$ (see (\ref{eqn:KAWdp})). 

Once again, these specific scaling laws are probably not accurate, and the coefficients accompanying them are probably not measurable. (For example, \citet{bp12} appeal to intermittency corrections to predict a steeper spectrum of $-8/3$, in closer agreement with solar-wind observations that yield slopes between $-3.1$ and $-2.5$ with a mode of $-2.8$ \citep[e.g.][]{sahraoui13}.) But here we have a clear demonstration of three salient features: (1) that background pressure anisotropy is required to excite scale-dependent pressure anisotropy in the KAW range (at least within the gyrokinetic theory); (2) that the relative amplitudes of the magnetic-field-strength and density fluctuations depend on the electron pressure anisotropy and not the ion one; and (3) that the nonlinear KAW cascade time at any given scale becomes larger the closer the plasma is to the mirror/firehose thresholds (if critical balance continues to hold). 

This last point becomes particularly important for the ion-entropy cascade, in which $W_{\widetilde{h}_i}$ is conserved. Following similar reasoning, a constant flux of entropy implies
\begin{equation}\label{eqn:entropyflux}
\frac{\vthprp{i}^6\vthprl{i}^2}{\nip^2} \frac{\widetilde{h}^2_{i\lambda}}{\tau_{h\lambda}} \sim \varepsilon_{h} = {\rm const} ,
\end{equation}
where $\varepsilon_h$ is the entropy flux proportional to the fraction of the total energy flux $\varepsilon$ that is converted into the entropy cascade at the ion gyroscale (again, which itself is a function of the background pressure anisotropy). Following the arguments presented in \S 7.9.2 of S09, the nonlinear time is obtained by weighting $\tau_{{\rm KAW}\lambda}$ (see (\ref{eqn:tauKAW})) by a factor of $(\rho_i/\lambda)^{1/2}$, due to the ring averaging in the gyrokinetic nonlinearity (the third term in (\ref{eqn:KAWgkequation})). An additional factor of $(\rho_i/\lambda)^{1/2}$ accounts for the random-walk accumulation of changes to $\widetilde{h}_{i\lambda}$ during one cascade time -- an argument analogous to that used to describe weak turbulence of Alfv\'{e}n waves (\citealt{kraichnan65}; see equation (255) of S09). Thus,
\begin{equation}\label{eqn:tauh}
\tau_{h\lambda} \sim \frac{\rho_i}{\lambda} \,\tau_{{\rm KAW}\lambda} \sim \biggl( \frac{\varepsilon}{\varepsilon_{\rm KAW}} \biggr)^{1/3} \frac{\ell^{1/3}_0 \rho^{1/3}_i \lambda^{1/3}}{\valfeff} \biggl[ \frac{1+\betaprp{i}-2\mc{K}}{(1+\betaprl{e}\Delta_e/2)(1-\betaprp{e}\Delta_e+\mc{K})} \biggr]^{1/3} .
\end{equation}
Substituting this into (\ref{eqn:entropyflux}) provides the scaling
\begin{equation}
\widetilde{h}_{i\lambda} \sim \frac{\nip}{\vthprp{i}^3 } \biggl( \frac{\varepsilon_{h}}{\varepsilon} \biggr)^{1/2} \biggl( \frac{\varepsilon}{\varepsilon_{\rm KAW}} \biggr)^{1/6} \frac{\ell^{-1/3}_0 \rho^{1/6}_i \lambda^{1/6}}{\sqrt{\betaprl{i}}} \frac{\valfeff}{\valf} \biggl[ \frac{1+\betaprp{i}-2\mc{K}}{(1+\betaprl{e}\Delta_e/2)(1-\betaprp{e}\Delta_e+\mc{K})} \biggr]^{1/6} ,
\end{equation}
which corresponds to a $k^{-4/3}_\perp$ entropy spectrum. 

Again, the point is not so much the $-4/3$ ion-entropy spectrum, which is unchanged from the prediction of isotropic-pressure gyrokinetics (S09), but that the time over which $\widetilde{h}_i$ is decorrelated due to mixing by ring-averaged $\bb{E}\btimes\bb{B}$ flows is a function of the electron pressure anisotropy. One instance in which this dependence may become important concerns the ratio of parallel (i.e.~linear) versus perpendicular (i.e.~nonlinear) phase mixing. If we assume that the parallel decorrelation scale of $\widetilde{h}_i$ is inherited from the KAW scaling (\ref{eqn:KAWell}), then after one cascade time $\tau_{h\lambda}$ (see (\ref{eqn:tauh})) $\widetilde{h}_i$ becomes decorrelated on the parallel velocity scales
\begin{equation}\label{eqn:hprlmixing}
\frac{\delta v_\parallel}{\vthprl{i}} \sim \frac{\ell_\lambda}{\vthprl{i}\tau_{h\lambda}} \sim \sqrt{\frac{(1+\betaprl{e}\Delta_e/2)(1-\betaprp{e}\Delta_e+\mc{K})}{\betaprl{i}(1+\betaprp{i}-2\mc{K})}} .
\end{equation}
In S09, it was argued that within the ion-entropy cascade the parallel phase mixing can be ignored relative to the perpendicular phase mixing. This was done by comparing the estimate $\delta v_\parallel / v_{{\rm th}i} \sim 1/\sqrt{\beta_i(1+\beta_i)} \sim 1$ after one cascade time (i.e.~equation (\ref{eqn:hprlmixing}) with $\Delta_e = 0$) to the smallest perpendicular-velocity-space interval between spatially decorrelated electromagnetic fluctuations, $\delta v_\perp / v_{{\rm th}i} \sim (k_\perp \rho_i)^{-1}$ with $k_\perp \rho_i$ given by its value at the collisional cutoff. Recent gyrokinetic simulations of KAW turbulence that demonstrate predominantly perpendicular collisional heating of the ions \citep{navarro16} support this argument. But in a pressure-anisotropic plasma, the entropy cascade slows down near the firehose and mirror thresholds (see (\ref{eqn:tauh})). The parallel phase mixing could then compete with the perpendicular phase mixing, since there is more time for small scales in $v_\parallel$ to be secularly produced and ultimately activate collisional dissipation. This idea can be tested using high-resolution gyrokinetic simulations of KAW turbulence in a  pressure-anisotropic plasma.

\subsection{Ion versus electron heating}\label{sec:heating}

Throughout this paper, we have commented a number of times on the likelihood that pressure anisotropy in the mean distribution function influences the partitioning of free energy amongst the various cascade channels and, consequently, the differential heating of ions and electrons. In this section, we seek to quantify this idea, if only very roughly.

As a first attempt, we leverage the fact that the energy spectrum of magnetic-field fluctuations observed in several gyrokinetic \citep{howes08,told15} and kinetic \citep{franci15a,franci15b,franci16,franci17,cerri16,cerri17a,cerri17b,groselj17,arzamasskiy18} simulations of Alfv\'{e}nic turbulence is continuous across the $k_\perp \rho_i \sim 1$ spectral transition and impose continuity of $\delta B_{\perp \lambda}$ across $\lambda \sim \rho_i$. Setting the inertial-range scaling $\delta B_{\perp\lambda}/B_0 \sim \ell^{-1/3}_0 \lambda^{1/3}$ (from (\ref{eqn:AWscalings})) similar to the kinetic-range scaling $\delta B_{\perp\lambda}/B_0 \sim \Psi_\lambda / \lambda v_A$ (with $\Psi_\lambda$ given by (\ref{eqn:KAWPsik})) at $\lambda \sim \rho_i$ implies that the fraction of the total energy flux that is converted into the KAW cascade at the ion gyroscale satisfies
\begin{equation}\label{eqn:epsKAW}
\left( \frac{\varepsilon_{\rm KAW}}{\varepsilon} \right)^2 \sim \frac{(1-\betaprp{e}\Delta_e+\mc{K})(1+\betaprl{e}\Delta_e/2)}{(1+\betaprp{i}-2\mc{K})(1+\betaprl{i}\Delta_i/2+\betaprl{e}\Delta_e/2)^3} .
\end{equation}
In a pressure-isotropic plasma, this equation states that less energy goes into the KAW cascade as $\beta_i$ is increased: $\varepsilon_{\rm KAW}/\varepsilon \sim (1+\betaprp{i})^{-1/2}$. In other words, the electron heating decreases (relative to that of the ions) as $\beta_i$ increases, a qualitative prediction that is made quantitative in some gyrokinetic cascade models (e.g.~\citealt{howes10}; Y.~Kawazura, M.~Barnes \& A.~Schekochihin, in prep.). In a pressure-anisotropic plasma, however, this ratio can vary widely, with the KAW channel being shut off as the firehose or mirror thresholds are approached (either because $1-\betaprp{e}\Delta_e + \mc{K} = 0$ and $1+\betaprl{e}\Delta_e/2 = 0$ preclude the existence of stable KAWs, or because the inertial-range cascade of Alfv\'{e}nic fluctuations is interrupted at the long-wavelength firehose threshold, $1+ \sum_s \betaprl{s} \Delta_s/2 = 0$).  Given the importance of ion versus electron heating in many weakly collisional astrophysical plasmas (perhaps most notably radiatively inefficient accretion flows; e.g.~\citet{qg99}), it is worthwhile to investigate in more detail how pressure anisotropy might affect this thermodynamics.

To do so, we follow the model developed by \citet{howes10}. The evolution of the energy of critically balanced perpendicular magnetic-field fluctuations with perpendicular wavenumber $k_\perp$ is modeled by the following continuity equation:
\begin{equation}
\D{t}{b^2_k} = -k_\perp \pD{k_\perp}{\varepsilon(k_\perp)} + S\, \delta(k_\perp - k_0) -  2 \gamma b^2_k ,
\end{equation}
where $b^2_k \doteq \delta B^2_\perp(k_\perp)/4\upi m_i\nip$ is the turbulent magnetic energy (in velocity units) at scale $k_\perp$, $\varepsilon(k_\perp) = C^{-3/2}_1 k_\perp \overline{\omega} b^3_k$ is the energy cascade rate, $S$ is the energy injection rate at the driving scale $k_0$, and $\gamma$ is the linear kinetic damping rate. The steady-state solution for the energy cascade rate is given by equation (2) of \citet{howes10}:
\begin{equation}\label{eqn:epsheating}
\varepsilon(k_\perp) = \varepsilon_0 \exp \left[ -2 C^{3/2}_1 C_2 \int^{k_\perp}_{k_0} \frac{\rmd k'_\perp}{k'_\perp} \, \frac{\overline{\gamma}(k'_\perp)}{\overline{\omega}(k'_\perp)} \right] ,
\end{equation}
where $\varepsilon_0$ is the rate of energy input at $k_0$ and $C_1$ and $C_2$ are order-unity dimensionless Kolmogorov constants. (In the calculation below, we follow \citet{howes10} and set $C_1 = 1.9632$ and $C_2 = 1.0906$, ignoring the possibility that these coefficients vary with the plasma parameters.) The dimensionless gyrokinetic frequencies $\overline{\omega} \doteq \omega / |k_\parallel| v_A$ and damping rates $\overline{\gamma} \doteq \gamma / |k_\parallel|v_A$ are obtained by solving the dispersion relation (\ref{eqn:gkdisprel}). The spectrum of heating on species $s$ is then
\begin{equation}\label{eqn:Qheating}
Q_s(k_\perp) = 2 C^{3/2}_1 C_2 \,\frac{\overline{\gamma}_s(k_\perp)}{\overline{\omega}(k_\perp)} \frac{\varepsilon(k_\perp)}{k_\perp} .
\end{equation}
The damping rate of species $s$ due to waves with perpendicular wavenumber $k_\perp$, $\overline{\gamma}_s(k_\perp)$, is calculated via
\begin{equation}\label{eqn:speciesgamma}
\frac{\overline{\gamma}_s}{\overline{\omega}} = \frac{\bb{E}^\ast \bcdot \bb{\chi}^a_s \bcdot\bb{E}}{4W_{\rm EM}} ,
\end{equation}
where $\bb{\chi}^a_s$ is the anti-Hermitian part of the linear susceptibility tensor for species $s$ evaluated at the real component of the normal-mode frequency $\overline{\omega}$, $\bb{E}$ and $\bb{E}^\ast$ are the electric field associated with the normal mode and its complex conjugate, and $W_{\rm EM}$ is the electromagnetic wave energy (see, e.g., \citet{stix92}, \S11.8; \citet{quataert98}, eqn.~(4); or \citet{klein17}, eqn.~(3.1)). (Note that the total damping rate $\gamma = \gamma_i + \gamma_e$.) To obtain the total species-dependent heating rate due to the kinetic dissipation of the turbulent cascade, $Q_s$, we integrate (\ref{eqn:Qheating}) over $k_\perp$. Then, $Q_i/Q_e$ may be obtained as a function of the plasma parameters $(\betaprl{i}, \tauprp{i}, \Delta_i, \Delta_e)$. 

Before proceeding with this programme, it is important to note three things. First, compressive fluctuations are neglected. If a significant fraction of the turbulent energy resides in the compressive cascade channel, then its dissipation and the resulting plasma heating must be accounted for. This may be particularly important near the mirror threshold, where the collisionless damping rate of compressive fluctuations decreases (see figure \ref{fig:mirror}). Secondly, the existence of ${\rm Re}(\omega)=0$ Alfv\'{e}nic fluctuations near $k_\perp \rho_i \sim 1$ at high $\beta_i$ (see figure \ref{fig:alfven}) makes the exact implementation of (\ref{eqn:Qheating}) problematic. For these fluctuations, the nonlinear cascade frequency, which is matched by critical balance to the linear wave frequency, vanishes, and the cascade model used here technically breaks down. To circumvent this difficulty, we follow \citet{howes10} and replace ${\rm Re}(\omega)=0$ by a small number; this has the effect of allowing the cascade to continue to smaller scales even when the nonlinear transfer is formally arrested by the ${\rm Re}(\omega)=0$ region. As long as this region is relatively narrow (and it need not be -- see, e.g., the $\betaprl{i}\Delta_i = 0.5$ curve in the $\betaprl{i} = 100$ panel of figure \ref{fig:alfven}), one can argue that non-local effects on the energy transfer should enable the cascade to proceed through this gap. This ought to be checked in future numerical studies of kinetic turbulence at high $\beta_i$. (Consideration of refined cascade models that include non-local transfers, such as the one developed by \citet{howes11}, could also serve as a useful future study.) Thirdly, the amount of linear Landau/Barnes damping within a given cascade channel is not necessarily related to the amount of particle heating. There are (at least) five reasons for this: (i) the steady-state dissipation rate in a turbulent kinetic system may not equal the Landau damping rate (for relatively simple examples, see \citet{plunk13} and \citet{kanekar15}); (ii) nonlinear advection of the perturbed particle distribution by fluctuating flows may greatly reduce the amount of parallel phase mixing (a stochastic version of the plasma echo; see \citet{schekochihin16}); (iii) even if collisionless damping occurs at the rates predicted by linear theory, true irreversible thermodynamic heating can only be accomplished through collisions \citep{howes06}; (iv) the nonlinear energy transfers may be nonlocal and/or the mechanism by which free energy is carried to collisional scales in velocity space may be nonlinear \citep[e.g.][]{told15}; and (v) turbulent heating of particles in realistic space and astrophysical plasmas may occur via other mechanisms (e.g.~cyclotron heating, stochastic heating, magnetic reconnection, kinetic instabilities) that are outside of the gyrokinetic ordering \citep[e.g.][]{hi02,dmitruk04,chandran10,cranmer14,sharma07,sn15}.

\begin{figure}
\centering
\includegraphics[width=\linewidth,clip]{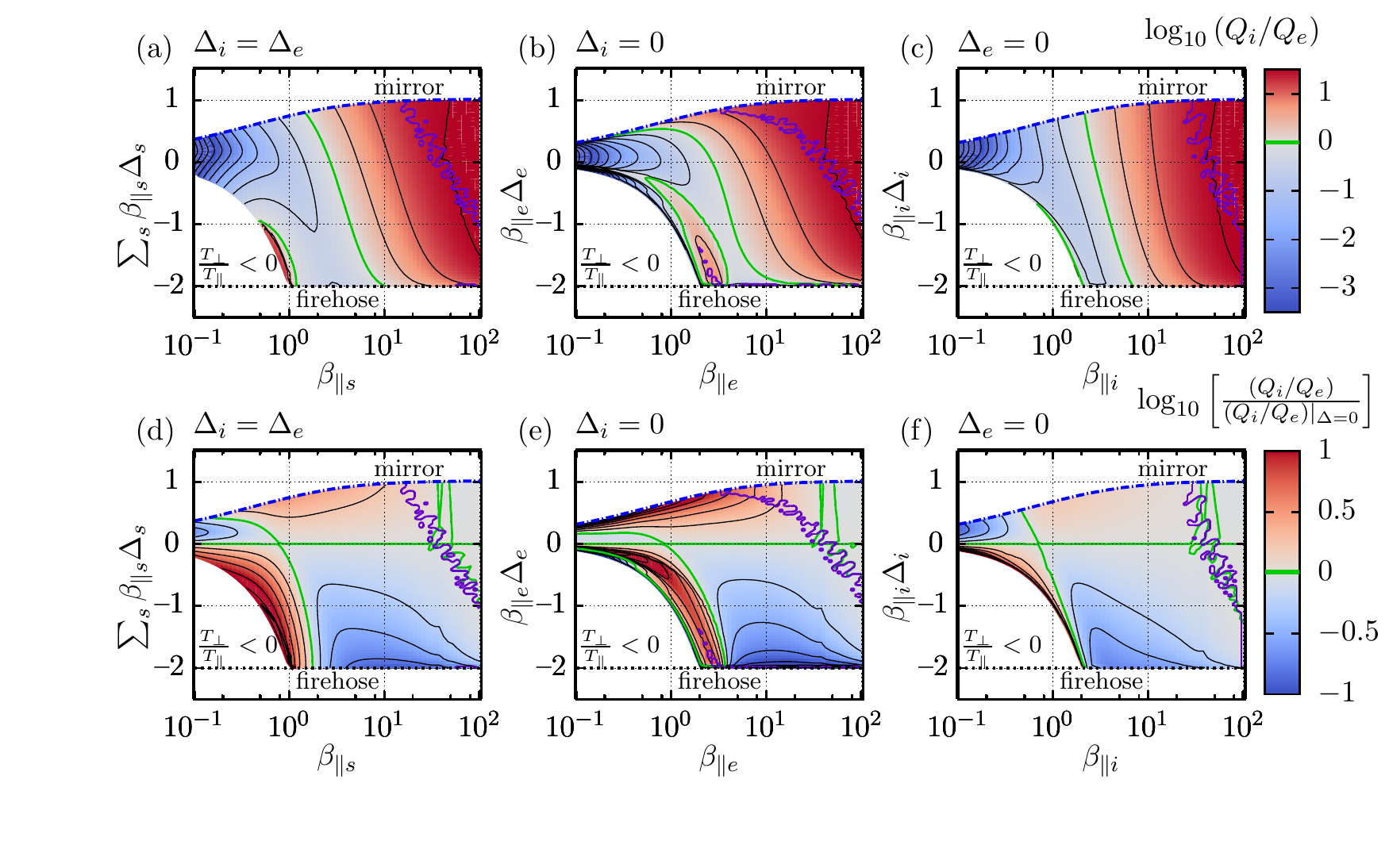}
\caption{Ratio of ion and electron heating rates, as calculated using (\ref{eqn:Qheating}) for (left) $\Delta_i = \Delta_e$, (middle) $\Delta_i =0$, and (right) $\Delta_e=0$ as a function of $\betaprl{s}$ and $\sum_s\betaprl{s} \Delta_s$. The top row, panels (a)--(c), illustrates the ratio $Q_i/Q_e$; the bottom row, panels (d)--(e), illustrates this ratio normalized to $Q_i/Q_e$ evaluated at $\Delta_s=0$.  The purple contour serves as the boundary above which a finite extent of the Alfv\'{e}n dispersion relation contains ${\rm Re}(\omega)=0$ modes. See \S\ref{sec:heating} for details.}
\label{fig:heating}
\end{figure}

These many caveats declared, we present in Figure \ref{fig:heating} our calculation of $Q_i/Q_e$ versus $\betaprl{s}$ and $\Delta_s$. We have elected to use equal ion and electron parallel temperatures, since variations in $T_i/T_e$ do not cause large changes in heating ratio \citep[cf.][]{howes10}. The heating rate is calculated for three cases, with $\Delta_i = \Delta_e$, $\Delta_i=0$, and $\Delta_e=0$ for $\betaprl{s}$ ranging from $0.1$ to $100$. We limit the value of $\Delta_s$ so that the system is stable to both the mirror and firehose instabilities and so that $T_\perp / T_\parallel > 0$. Both the ratio $Q_i/Q_e$ and this ratio normalized to $Q_i/Q_e$ evaluated in a pressure-isotropic plasma (i.e.~$\Delta_s = 0$, the case treated by \citet{howes10}) are shown. The region where ${\rm Re}(\omega) = 0$ modes arise on the Alfv\'{e}n dispersion surface is bounded by a purple contour.

We see the general trend of an increase in $Q_i/Q_e$ with increasing $\betaprl{s}$, with ion (electron) heating enhanced as the mirror (firehose) threshold is approached. These deviations from the $\Delta_s = 0$ result can be up to an order of magnitude, with the most significant alterations being due to near-threshold electron pressure anisotropy for $\betaprl{s} \lesssim 10$. While the precise quantitative predictions should perhaps be taken with a grain of salt (given the caveats stipulated earlier in this section), it is clear that the pressure anisotropy in the mean distribution function has an influence on the partitioning of free energy amongst the ion and electron cascade channels. Future studies of ion versus electron heating in space and astrophysical plasmas should take into consideration the velocity-space anisotropy of the mean distribution function.

%
%
\section{Summary}

In this paper, we have provided a theoretical framework for describing electromagnetic plasma turbulence in a multi-species, magnetized, pressure-anisotropic plasma characteristic of those found in many space and astrophysical systems. Using linear and nonlinear gyrokinetic theory in concert with the concept of a turbulent cascade of free energy to small scales in phase space, we have arrived at the following principal conclusions:
\begin{enumerate}
\item Because of their impact on collisionless damping (see (\ref{eqn:slowmode})) and the rigidity of magnetic-field lines (see (\ref{eqn:alfvenwave})), positive (negative) background pressure anisotropies should shift the ion-Larmor-scale spectral break towards larger (smaller) perpendicular wavenumbers (see (\ref{eqn:breakpoint})). This prediction may be tested by comparing turbulent spectra measured in different parts of the $\beta_i \gtrsim 1$ solar wind that exhibit disparate velocity-space anisotropies.
\item The strong ion Landau damping that occurs at the ion-Larmor-scale spectral break for $\beta_i \gtrsim 1$ exhibits order-of-magnitude variations in the damping rate depending upon the pressure anisotropy. Since the amount of Landau damping at this scale is related to how much energy ultimately goes into heating the ions or the electrons, pressure anisotropy ought to be considered alongside $\beta_i$ and $T_{0i}/T_{0e}$ when assessing the efficiency of ion heating. This is particularly important in the context of radiatively inefficient accretion flows onto supermassive black holes \citep[e.g.][]{qg99}, in which the temperature of the electrons is expected to be smaller than the poorly radiating ions and in which pressure anisotropy is thought to play a key role in both the turbulent transport of angular momentum \citep{quataert02,sharma06,riquelme12,chandra15,hoshino15,kunz16,foucart16} and the non-adiabatic heating of particles \citep{sharma07,sn15,sironi15}. A generalized version of the \citet{howes10} cascade model for assessing ion versus electron heating suggests that ion (electron) heating is enhanced as the mirror (firehose) threshold is approached, with order-of-magnitude variations in $Q_i/Q_e$ between these thresholds (see figure \ref{fig:heating}).
\item Kinetic Alfv\'{e}n waves are subject to both the firehose and mirror instability criteria (see (\ref{eqn:kaw})), which, in the sub-ion-Larmor range, are functions of the pressure anisotropy of the electrons only. That the background pressure anisotropy of the ions only affects fluctuations in the inertial range is a consequence of the isothermal pressure response of the ions (see (\ref{eqn:KAWdpi})), which results from the attenuating effect of ring averaging on an otherwise adiabatic response to the fluctuating magnetic-field strength. As a result, Alfv\'{e}nic fluctuations can be firehose unstable at long ($k_\perp \rho_i \ll 1$) wavelengths, and yet remain stable at short ($k_\perp \rho_i \gg 1$) wavelengths. This explains why hybrid-kinetic simulations of firehose and mirror instabilities by \citet{kss14} and of magnetorotational turbulence by \citet{kunz16}, which used an isothermal and isotropic equation of state for the electrons, demonstrated a KAW cascade at sub-ion-Larmor scales, even though the larger scales were firehose/mirror unstable. In the case where $\betaprl{e} \Delta_e < -2$, the KAW is driven unstable to what appears to be the zero-frequency, oblique ``electron firehose instability'' found by \citet{lh00}, for which our work provides a physical explanation and approximate analytical description (\S\ref{sec:linKAW}).
\item The gyrokinetic system (\ref{eqn:gkequation2}), (\ref{eqn:gkqn})--(\ref{eqn:gkprpamp}) satisfies the conservation law (\ref{eqn:WGK}) for the generalized free energy $W_{\rm GK}$ (\ref{eqn:W}). In the absence of background interspecies drifts, the conservation of $W_{\rm GK}$ causes a turbulent cascade to small scales in phase space across both the inertial and kinetic range. At long wavelengths, $W_{\rm GK}$ is comprised of two Alfv\'{e}nic invariants and a compressive invariant, the latter of which is related to the entropy of the perturbed distribution function (Paper I). At short wavelengths, the cascade of $W_{\rm GK}$ splits into an ion-entropy cascade and a KAW cascade, which is governed by the nonlinear equations of ERMHD (modified to include pressure anisotropic species). These cascades are independent of one another so long as $k_\perp \rho_i \gg 1$ and $\betaprp{s} \sim \betaprl{s} \sim 1$, with the ion-entropy cascade taking the collisionlessly damped free energy at the end of the inertial range onwards to collisional scales and the KAW cascade carrying the remainder down to electron scales. In both regimes, as the firehose or mirror instability thresholds are approached, the dynamics of the plasma are modified so as to reduce the energetic cost of bending magnetic-field lines or of compressing them. At $k_\perp \rho_i \sim 1$, the partitioning of energy into the KAW and ion-entropy cascades is influenced by the strength of the collisionless damping (itself a function of $\betaprl{i}$, $\tauprp{i}$, $\Delta_i$, and $\Delta_e$) and the work done by the fluctuating electric and magnetic-mirror forces on the interspecies drifts.
\item Background pressure anisotropy, so long as it remains within the firehose/mirror stability boundaries, does not affect the spectral scaling laws predicted from pressure-isotropic gyrokinetics for strong Alfv\'{e}nic turbulence in the inertial range (to which the compressive fluctuations are slaved), KAW turbulence in the sub-ion-Larmor range, and the accompanying phase-space cascade of ion entropy to collisional scales (see \S\ref{sec:scalings}). However, it does place constraints on whether {\em scale-dependent} pressure anisotropy can be generated and, if so, sustained. We have shown that no scale-dependent electron pressure anisotropy will be generated by compressive fluctuations in either the inertial or kinetic range if no electron pressure anisotropy exists in the background state. By contrast, scale-dependent ion pressure anisotropy can be generated in the inertial range even when $\Delta_i = 0$, either linearly by the compressive fluctuations (if they are not strongly Landau damped) or nonlinearly by the Alfv\'{e}nic fluctuations (at second order in the fluctuation amplitude). While the latter source falls out of the range of validity of the gyrokinetic theory that we have developed, it is nevertheless interesting to note that each of these sources will likely produce ion pressure anisotropy with different scaling relations, and that distinguishing between these scaling relations in, say, solar-wind data could help identify the main drivers of fluctuating pressure anisotropy and assess the role of pressure anisotropy in influencing the efficacy of collisionless damping.
\end{enumerate}

In closing, we return to one of the main conclusions of Paper I: that, despite the many quantitative changes arising from the non-Maxwellian nature of the background distribution function, most of the salient qualitative features of the gyrokinetic theory of astrophysical turbulence are robust. This is because, as long as the plasma remains stable, the nature of the nonlinear wave-wave and wave-particle couplings that are central to the theory are insensitive to deviations from velocity-space isotropy, even if the linear physics is not (see \citealt{kh15} for an alternate route to the same conclusion). That being said, the study of kinetic turbulence, both theoretical and observational, is becoming increasingly quantitative, and so it is perhaps worth restating Paper I's coda: ``In light of the ever-increasing scrutiny placed upon theories of Alfv\'{e}nic turbulence by the wealth of data from the solar wind, as well as the astrophysical importance of knowing the proportion of turbulent energy that is distributed between the ion and electron populations, such details matter.'' A numerical programme to quantify the broader impact of these details using the theoretical framework developed in this paper is now underway.

\vspace{0.1in}
Support for M.W.K.~was provided by US DOE Contract DE-AC02-09CH11466 and NASA grant NNX16AK09G. Support for I.G.A.~was provided by the Princeton Center for Theoretical Science and the Framework grant for Strategic Energy Research (Dnr.~2014-5392) from Vetenskapsr\aa det. A.A.S.~was supported in part by UK STFC Consolidated Grant ST/N000919/1 and ERSPC Grant EP/M022331/1. K.G.K.~was supported by NASA grant NNX16AM23G. The authors thank J.~Burby, F.~Rincon, J.~Squire, J.~TenBarge, D.~Verscharen and especially S.~Cowley for useful conversations; G.~Howes for providing the code for the cascade model in \citet{howes10}; and the referees for a careful reading of the manuscript. The completion of this work was facilitated by the generous hospitality and material support provided by the Wolfgang Pauli Institute in Vienna. 

\appendix

\section{Definitions of $C^\perp_{\ell s}$, $C^\parallel_{\ell s}$, $\Gamma^\perp_{\ell m}$, and $\Gamma^\parallel_{\ell m}$ coefficients}\label{app:coeffs}

This paper is replete with velocity-space integrals of $f^\perp_{0s}$ and $f^\parallel_{0s}$ (see (\ref{eqn:fprlfprp}) for their definitions), which we have allowed to masquerade as deceptively benign coefficients. The first set of these are the $C^\perp_{\ell s}$ coefficients:
\begin{subequations}
\begin{align}
\kzero{s} & \doteq \frac{1}{\nsp} \int\rmd^3\bb{v} \, f^\perp_{0s} , \\*
\kone{s} & \doteq \frac{1}{\nsp} \int\rmd^3\bb{v} \, \frac{v_\parallel}{\vthprl{s}} f^\perp_{0s} \times \Biggl( \frac{\dupar}{\vthprl{s}} \Biggr)^{-1} , \\*
\ktwo{s} & \doteq \frac{1}{\nsp} \int\rmd^3\bb{v} \, \frac{v^2_\parallel}{\vthprl{s}^2} f^\perp_{0s} \times \Biggl( \frac{1}{2} + \frac{\duparsq}{\vthprl{s}^2} \Biggr)^{-1} ,
\end{align}
\end{subequations}
normalized so that $C^\perp_{\ell s} = 1$ for a parallel-drifting bi-Maxwellian distribution function. The next set of coefficients were borne out of the linear theory:
\begin{equation}\label{eqn:scoeff}
\cell{s}(\xi_s) = \frac{1}{\nsp} \int\rmd^3\bb{v} \, \frac{1}{\ell !} \biggl( \frac{v_\perp}{\vthprp{s}} \biggr)^{2\ell} \frac{ v_\parallel - \dupar}{v_\parallel - \omega / k_\parallel} f^\parallel_{0s}
\end{equation}
for integer $\ell$, where $\xi_s \doteq ( \omega - k_\parallel \dupar ) / k_\parallel \vthprl{s}$ is the dimensionless Doppler-shifted phase velocity of the (linear) fluctuations. The functions defined by (\ref{eqn:scoeff}) are generalisations of the plasma dispersion function for non-Maxwellian distributions: e.g.
\begin{equation}
\cell{s}(\xi_s) = 1 + \xi_s Z_{\rm M} (\xi_s) ~ \textrm{for a bi-Maxwellian, where~} 
Z_{\rm M}(\xi) \doteq \frac{1}{\sqrt{\upi}} \int^\infty_{-\infty} \rmd x \, \frac{e^{-x^2}}{x-\xi}
\end{equation}
is the plasma dispersion function \citep{fc61}. 

Accounting for finite-Larmor-radius effects is most easily achieved in the Fourier domain, and the $C_{\ell s}$ coefficients can be profitably generalised by including various combinations of the $n$th-order Bessel function ${\rm J}_n(a_s)$, where $a_s \doteq k_\perp v_\parallel /\Omega_s$. Suitably normalised, they are
\begin{subequations}\label{eqn:gammas}
\begin{align}
\Gamma_{00} (\alpha_s) & \doteq \frac{1}{\nsp} \int \rmd^3\bb{v} \, \bigl[ {\rm J}_0(a_s) \bigr]^2 \, f_{0s} = 1 - \alpha_s + \dots \label{eqn:gamma00} \\*
\Gamma^\perp_{00} (\alpha_s) & \doteq \frac{1}{\nsp} \int \rmd^3\bb{v} \, \bigl[ {\rm J}_0(a_s) \bigr]^2 \, f^\perp_{0s} = \kzero{s} - \alpha_s + \dots  \label{eqn:gammaprp00}\\*
\Gamma^\perp_{01} (\alpha_s) & \doteq \frac{1}{\nsp} \int \rmd^3\bb{v} \, \bigl[ {\rm J}_0(a_s) \bigr]^2 \biggl(\frac{v_\parallel}{\vthprl{s}}\biggr)  f^\perp_{0s} \times \Biggl( \frac{\dupar}{\vthprl{s}} \Biggr)^{-1}= \kone{s} - \alpha_s + \dots \\*
\Gamma^\perp_{02} (\alpha_s) & \doteq \frac{1}{\nsp} \int \rmd^3\bb{v} \, \bigl[ {\rm J}_0(a_s) \bigr]^2 \biggl(\frac{v_\parallel}{\vthprl{s}}\biggr)^2 f^\perp_{0s} \times \Biggl( \frac{1}{2} + \frac{\duparsq}{\vthprl{s}^2} \Biggr)^{-1} = \ktwo{s} - \alpha_s + \dots \label{eqn:gammaprp02} \\*
\Gamma^\perp_{10} (\alpha_s) & \doteq \frac{1}{\nsp} \int \rmd^3\bb{v} \, \frac{v^2_\perp}{\vthprp{s}^2} \frac{2{\rm J}_0(a_s) {\rm J}_1(a_s)}{a_s} \, f^\perp_{0s} = 1 - \frac{3}{2} \alpha_s + \dots \\*
\Gamma^\perp_{11} (\alpha_s) & \doteq \frac{1}{\nsp} \int \rmd^3\bb{v} \, \frac{v^2_\perp}{\vthprp{s}^2} \frac{2{\rm J}_0(a_s) {\rm J}_1(a_s)}{a_s} \biggl(\frac{v_\parallel}{\vthprl{s}}\biggr) f^\perp_{0s} \times \Biggl( \frac{\dupar}{\vthprl{s}} \Biggr)^{-1} = 1 - \frac{3}{2} \alpha_s C_{11s} + \dots \label{eqn:gammaprp11} \\*
\Gamma^\perp_{20} (\alpha_s) & \doteq \frac{1}{\nsp} \int \rmd^3\bb{v} \biggl[ \frac{2v^2_\perp}{\vthprp{s}^2} \frac{{\rm J}_1(a_s)}{a_s} \biggr]^2 f^\perp_{0s} = 2 \biggl( 1 - \frac{3}{2} \alpha_s C_{20s} + \dots \biggr), \label{eqn:gammaprp20}
\end{align}
\end{subequations}
\begin{subequations}\label{eqn:gammaprl}
\begin{align}
\Gamma^\parallel_{00}(\xi_s, \alpha_s) &\doteq \frac{1}{\nsp} \int\rmd^3\bb{v} \, \bigl[ {\rm J}_0(a_s) \bigr]^2 \,\frac{v_\parallel - \dupar}{v_\parallel - \omega / k_\parallel} f^\parallel_{0s} = \czero{s}(\xi_s) - \alpha_s \cone{s}(\xi_s) + \dots ,\\*
\Gamma^\parallel_{01}(\xi_s,\alpha_s) &\doteq \frac{1}{\nsp} \int\rmd^3\bb{v} \, \bigl[ {\rm J}_0(a_s) \bigr]^2 \biggl( \frac{v_\parallel}{\vthprl{s}} \biggr) \frac{v_\parallel - \dupar }{v_\parallel - \omega / k_\parallel} f^\parallel_{0s}  \times \Biggl( \frac{\dupar}{\vthprl{s}} \Biggr)^{-1} \nonumber\\*
\mbox{} &= \frac{\omega}{k_\parallel\dupar} \Gamma^\parallel_{00} (\xi_s, \alpha_s ) ,\\*
\Gamma^\parallel_{02}(\xi_s,\alpha_s) &\doteq \frac{1}{\nsp} \int\rmd^3\bb{v} \, \bigl[ {\rm J}_0(a_s) \bigr]^2 \biggl( \frac{v_\parallel}{\vthprl{s}} \biggr)^2 \frac{ v_\parallel - \dupar }{v_\parallel - \omega / k_\parallel} f^\parallel_{0s} \times \Biggl( \frac{1}{2} + \frac{\duparsq}{\vthprl{s}^2} \Biggr)^{-1} \nonumber\\*
\mbox{} &=\Biggl[ \Gamma_{00}(\alpha_s) + \frac{2\duparsq}{\vthprl{s}^2} \frac{\omega^2}{k^2_\parallel\duparsq} \Gamma^\parallel_{00} (\xi_s, \alpha_s ) \Biggr]  \Biggl( 1 + \frac{2\duparsq}{\vthprl{s}^2} \Biggr)^{-1} ,\\*
\Gamma^\parallel_{10}(\xi_s,\alpha_s) &\doteq \frac{1}{\nsp} \int\rmd^3\bb{v} \, \frac{v^2_\perp}{\vthprp{s}^2} \frac{2{\rm J}_0(a_s){\rm J}_1(a_s)}{a_s}  \frac{v_\parallel - \dupar}{v_\parallel - \omega / k_\parallel} f^\parallel_{0s} \nonumber\\*
\mbox{} &= \cone{s}(\xi_s) - \frac{3}{2} \alpha_s \ctwo{s}(\xi_s) + \dots ,\\*
\Gamma^\parallel_{11}(\xi_s,\alpha_s) &\doteq \frac{1}{\nsp} \int\rmd^3\bb{v} \, \frac{v^2_\perp}{\vthprp{s}^2} \frac{2{\rm J}_0(a_s){\rm J}_1(a_s)}{a_s} \biggl( \frac{v_\parallel}{\vthprl{s}} \biggr)\frac{v_\parallel - \dupar}{v_\parallel - \omega / k_\parallel} f^\parallel_{0s}  \times \Biggl( \frac{\dupar}{\vthprl{s}} \Biggr)^{-1} \nonumber\\*
\mbox{} &= \frac{\omega}{k_\parallel\dupar} \Gamma^\parallel_{10}(\xi_s,\alpha_s) ,\\*
\Gamma^\parallel_{20}(\xi_s,\alpha_s) &\doteq \frac{1}{\nsp} \int\rmd^3\bb{v} \,\biggl[ \frac{2v^2_\perp}{\vthprp{s}^2} \frac{{\rm J}_1(\alpha_s)}{a_s} \biggr]^2 \frac{v_\parallel - \dupar}{v_\parallel - \omega / k_\parallel} f^\parallel_{0s} \nonumber\\*
\mbox{} &= 2 \biggl[ \ctwo{s}(\xi_s) - \frac{3}{2} \alpha_s \cthree{s}(\xi_s) + \dots \biggr] , \label{eqn:gammaprl20}
\end{align}
\end{subequations}
where $\alpha_s \doteq (k_\perp \rho_s)^2/2$, and
\begin{equation}\label{eqn:c11c20}
C_{11s} \doteq \frac{1}{\nsp} \int \rmd^3\bb{v} \, \frac{v^2_\perp}{\vthprp{s}^2} \frac{v_\parallel}{\dupar} f_{0s}  \quad {\rm and} \quad C_{20s} \doteq \frac{1}{\nsp} \int \rmd^3\bb{v} \, \frac{1}{2} \frac{v^4_\perp}{\vthprp{s}^4} f_{0s},
\end{equation}
both of which equate to unity for a drifting bi-Maxwellian distribution (\ref{eqn:biMax}). To facilitate comparison with the long-wavelength results of Paper I, the final equalities in (\ref{eqn:gammas}) and (\ref{eqn:gammaprl}) provide their leading-order expansions in $\alpha_s \ll 1$. It is helpful to note the numbering scheme used for the $\Gamma_{\ell m}$ subscripts, which reflects the number of powers $\ell$ of $v^2_\perp$ and $m$ of $v_\parallel$ in the integrand.

In \S\ref{sec:invariant}, we promoted several of these Fourier-space $\Gamma_{\ell m}(\alpha_s)$ integrals to real-space bi-linear operators by dressing them with hats. Their action on an arbitrary function $\Psi(\bb{r}) = \sum_{\bs{k}} \Psi_{\bs{k}} \exp(\imag\bb{k}\bcdot\bb{r})$ is best expressed in Fourier space:
\begin{subequations}\label{eqn:Gammahat}
\begin{align}
\int\rmd^3\bb{r}\, \Psi(\bb{r}) \widehat{\Gamma}_{00} \Psi(\bb{r}) &= \sum_{\bs{k}} \Gamma_{00}(\alpha_s) |\Psi_{\bs{k}} |^2 , \\*
\int\rmd^3\bb{r}\, \Psi(\bb{r}) \widehat{\Gamma}^\perp_{00} \Psi(\bb{r}) &= \sum_{\bs{k}} \Gamma^\perp_{00}(\alpha_s) |\Psi_{\bs{k}} |^2 , \\*
\int\rmd^3\bb{r}\,\Psi(\bb{r})  \widehat{\Gamma}^\perp_{02} \Psi(\bb{r}) &= \sum_{\bs{k}} \Gamma^\perp_{02}(\alpha_s) |\Psi_{\bs{k}} |^2 , \\*
\int\rmd^3\bb{r}\,\Psi(\bb{r})  \widehat{\Gamma}^\perp_{11} \Psi(\bb{r}) &= \sum_{\bs{k}} \Gamma^\perp_{11}(\alpha_s) |\Psi_{\bs{k}} |^2 , \\*
\int\rmd^3\bb{r}\,\Psi(\bb{r})  \widehat{\Gamma}^\parallel_{20} \Psi(\bb{r}) &= \sum_{\bs{k}} \Gamma^\parallel_{20}(0,\alpha_s) |\Psi_{\bs{k}} |^2 , \\*
\int\rmd^3\bb{r}\, \Psi(\bb{r}) \widehat{\Gamma}^\perp_{20} \Psi(\bb{r}) &= \sum_{\bs{k}} \Gamma^\perp_{20}(\alpha_s) |\Psi_{\bs{k}} |^2 
\end{align}
\end{subequations}
with the Fourier-space $\Gamma_{\ell m}(\alpha_s)$ integrals being given by (\ref{eqn:gammas}{\it a,b,d,f,g}) and (\ref{eqn:gammaprl20}). 

This completes our catalogue of integrals.

\section{Limiting cases of the bi-Maxwellian linear gyrokinetic dispersion relation}\label{app:biMaxlimits}

In this appendix, the transition between the long-wavelength solutions of \S\ref{sec:gkkrmhd} and the short-wavelength solutions of \S\ref{sec:linKAW} is treated in the analytically tractable limits of high and low $\betaprl{i}$. This follows the procedure in appendix of D \citet{howes06} of identifying the analytically solvable cases.

\subsection{High-$\betaprl{i}$ limit: $\betaprl{i} \gg 1$, $k_\perp \rho_i \sim 1$}\label{app:highbeta}

For $\betaprl{i} \gg 1$, we have $\xi_i = \overline{\omega}/\betaprl{i}^{1/2} \ll 1$ and $\xi_e = (m_e/m_i)^{1/2} ( \tprl{i} / \tprl{e} )^{1/2} \xi_i \ll 1$, and we can use the small-argument expansion of the plasma dispersion function, $Z_{\rm M}(\xi_s) \simeq \imag\sqrt{\upi}$. This requires $\tauprl{i} \ll (m_i / m_e) \betaprl{i}$, which is not particularly restrictive. We also take $\alpha_e \ll 1$ because $m_e/m_i \ll 1$, as well as order the pressure anisotropy $\Delta_s \sim 1/\betaprl{s}$. Retaining $k_\perp \rho_i \sim 1$, the coefficients of the gyrokinetic dispersion relation (\ref{eqn:howescoeffs}) become
\begin{subequations}\label{eqn:highbetacoeffs}
\begin{align}
\mc{A} &\simeq 1 + \Gamma_0(\alpha_i) \Delta_i + \frac{\tauprp{i}}{Z_i} \frac{\tprp{e}}{\tprl{e}} + \imag\sqrt{\upi} \,\xi_i \, \frac{\tprp{i}}{\tprl{i}} \Biggl[ \Gamma_0(\alpha_i) +  \biggl( \frac{\tauprl{i}}{Z_i} \biggr)^{3/2} \biggl( \frac{Z_i m_e}{m_i} \biggr)^{1/2} \Biggr] , \\*
\mc{B} &\simeq 1 - \Gamma_0(\alpha_i) ,\\*
\mc{C} &\simeq \Gamma_1(\alpha_i) \Delta_i - \Delta_e + \imag\sqrt{\upi} \,\xi_i \, \frac{\tprp{i}}{\tprl{i}} \Biggl[ \Gamma_1(\alpha_i) -  \frac{\tauprl{i}}{\tauprp{i}} \biggl( \frac{\tauprl{i}}{Z_i}\frac{Z_i m_e}{m_i} \biggr)^{1/2} \Biggr] , \\*
\mc{D} &\simeq 2 \biggl[ \Gamma_1(\alpha_i) \Delta_i + \frac{Z_i}{\tauprp{i}} \Delta_e \biggr] + 2\imag\sqrt{\upi} \,\xi_i \, \frac{\tprp{i}}{\tprl{i}} \Biggl[ \Gamma_1(\alpha_i) + \frac{\tauprl{i}^2}{\tauprp{i}^2} \biggl( \frac{Z_i}{\tauprl{i}} \frac{Z_i m_e}{m_i} \biggr)^{1/2} \Biggr] \nonumber\\*
\mbox{} &\doteq \frac{2}{\betaprp{i}} \bigl[ 1 - \mc{F}(\alpha_i) \bigr] + 2\imag\sqrt{\upi} \,\xi_i \mc{G}(\alpha_i) ,\label{eqn:Dcoeff}\\*
\mc{E} &\simeq \Gamma_1(\alpha_i) - 1,
\end{align}
\end{subequations}
where we have dropped all terms of first order and higher in $Z_i m_e/m_i$. As in \S\ref{sec:linKAW}, we must be careful to retain the final term in the definition of $\alpha_\ast$ (\ref{eqn:alphatilde}), despite its dependence on the higher-order $1-\Gamma_0(\alpha_e)$ factor. The auxiliary functions $\mc{F}(\alpha_i)$ and $\mc{G}(\alpha_i)$, which are defined implicitly by (\ref{eqn:Dcoeff}), will become useful below. 

We proceed by taking two instructive limits.

\subsubsection{The limit $k_\perp \rho_i \sim \mc{O}(\betaprl{i}^{-1/4})$, $\overline{\omega} \sim \mc{O}(1)$}

In this ordering, we have $\alpha_i \sim \xi_i \sim \mc{O}(\betaprl{i}^{-1/2})$, and so we may expand $\Gamma_0(\alpha_i) \simeq 1 - \alpha_i$ and $\Gamma_1(\alpha_i) \simeq 1 - (3/2)\alpha_i$. We find from (\ref{eqn:highbetacoeffs}) that $\mc{A} \sim \mc{O}(1)$ and $\mc{B}$, $\mc{C}$, $\mc{D}$, and $\mc{E} \sim \mc{O}(\betaprl{i}^{-1/2})$. Then, the dispersion relation (\ref{eqn:gkdisprel}) becomes
\begin{equation}
- \biggl( \frac{\alpha_\ast}{\overline{\omega}^2} - \mc{B} \biggr) \mc{D} = \mc{E}^2 ,
\end{equation}
where, to leading order, we have $\mc{B} \simeq \alpha_i$, $\mc{E} \simeq -(3/2)\alpha_i$, $\mc{D} \simeq 2\imag\overline{\omega}(\upi/\betaprl{i})^{1/2}$, and $\alpha_\ast \simeq \alpha_i [ 1 + (\betaprl{i}/2) \Delta_i + (\betaprl{e}/2) \Delta_e]$. This is a quadratic equation for $\overline{\omega}$, whose solutions are
\begin{equation}\label{caseAiomega}
\overline{\omega} = -\imag \frac{9}{16} \sqrt{\frac{\betaprl{i}}{\upi}} \alpha_i \pm \sqrt{ 1 + \frac{\betaprl{i}}{2} \Delta_i + \frac{\betaprl{e}}{2} \Delta_e - \Biggr( \frac{9}{16} \sqrt{\frac{\betaprl{i}}{\upi}} \alpha_i \Biggr)^2 } .
\end{equation}
In the subsidiary limit, $k_\perp \rho_i \ll 1/\betaprl{i}^{1/4}$, we recover, as expected, the Alfv\'{e}n wave, now with weak collisionless damping (cf.~(\ref{eqn:gkaw})):
\begin{equation}\label{eqn:highbeta4}
\overline{\omega} = \pm \sqrt{ 1 + \frac{\betaprl{i}}{2} \Delta_i + \frac{\betaprl{e}}{2} \Delta_e } - \imag\frac{9}{16} \frac{k^2_\perp \rho^2_i}{2} \sqrt{\frac{\betaprl{i}}{\upi} } .
\end{equation}
In the intermediate asymptotic limit $\betaprl{i}^{-1/4} \ll k_\perp \rho_i \ll 1$, we have two solutions:
\begin{subequations}
\begin{align} \label{eqn:alfweakly} 
\overline{\omega} &= - \imag \frac{8}{9}\biggl( 1 + \frac{\betaprl{i}}{2} \Delta_i + \frac{\betaprl{e}}{2} \Delta_e \biggr) \biggl( \frac{k^2_\perp\rho^2_i}{2} \biggr)^{-1} \sqrt{\frac{\upi}{\betaprl{i}}} \qquad \textrm{(weakly damped)};\\*
\overline{\omega} &= - \imag \frac{9}{8} \frac{k^2_\perp\rho^2_i}{2} \sqrt{\frac{\betaprl{i}}{\upi}} \qquad \textrm{(strongly damped)} .
\end{align}
\end{subequations}

\subsubsection{The limit $k_\perp \rho_i \sim \mc{O}(1)$, $\overline{\omega} \sim \mc{O}(\betaprl{i}^{-1/2})$}

In this ordering, $\alpha_i \sim \mc{O}(1)$ and $\xi_i \sim \mc{O}(\betaprl{i}^{-1})$. Then $\mc{A}$, $\mc{B}$, $\mc{E} \sim \mc{O}(1)$, and $\mc{C}$, $\mc{D} \sim \mc{O}(\betaprl{i}^{-1})$. The dispersion relation (\ref{eqn:gkdisprel}) becomes
\begin{equation}
\frac{\alpha_\ast}{\overline{\omega}^2} \biggl( \frac{2}{\betaprp{i}} - \mc{D} \biggr) = \mc{E}^2 .
\end{equation}
Since $\mc{D} \simeq (2/\betaprp{i}) [ 1 - \mc{F}(\alpha_i) ] + 2\imag \sqrt{\upi} \,\xi_i \mc{G}(\alpha_i)$, this is again a quadratic equation for $\overline{\omega}$, with solutions given by
\begin{equation}\label{eqn:caseAiiomega}
\overline{\omega} = -\imag\sqrt{\frac{\upi}{\betaprl{i}}} \frac{\alpha_\ast \, \mc{G}(\alpha_i)}{\bigl[ \Gamma_1(\alpha_i) - 1 \bigr]^2 }  \pm \sqrt{ \frac{2}{\betaprp{i}} \frac{ \alpha_\ast \, \mc{F}(\alpha_i) }{\bigl[ \Gamma_1(\alpha_i) - 1 \bigr]^2} - \Biggl\{ \sqrt{\frac{\upi}{\betaprl{i}}}  \frac{\alpha_\ast \, \mc{G}(\alpha_i)}{\bigl[ \Gamma_1(\alpha_i) - 1 \bigr]^2} \Biggr\}^2 } .
\end{equation}

In the long-wavelength limit, $k_\perp \rho_i \ll 1$, these become
\begin{subequations}
\begin{align}
\overline{\omega} &= - \frac{\imag}{\sqrt{\upi\betaprl{i}}} \frac{\tprl{i}^2}{\tprp{i}^2} \Bigl( 1 - \betaprp{i} \Delta_i - \betaprp{e} \Delta_e \Bigr) ,\\*
\overline{\omega} &= -\imag \frac{8}{9} \biggl( 1 + \frac{\betaprl{i}}{2} \Delta_i + \frac{\betaprl{e}}{2} \Delta_e \biggr) \biggl( \frac{k^2_\perp \rho^2_i}{2} \biggr)^{-1} \sqrt{\frac{\upi}{\betaprl{i}}} \frac{\tprp{i}}{\tprl{i}} .\label{eqn:alfweakly2}
\end{align}
\end{subequations}
The first solution is the Barnes-damped (or mirror-unstable) slow wave (cf.~I--4.33); the second solution matches the weakly damped Alfv\'{e}n wave in the intermediate limit (see (\ref{eqn:alfweakly})). 

In the subsidiary short-wavelength limit, $k_\perp \rho_i \gg 1$, we have $\Gamma_1(\alpha_i) \rightarrow 0$, $\mc{F}(\alpha_i) \rightarrow 1 - \betaprp{e} \Delta_e$, and $\mc{G}(\alpha_i) \rightarrow (\tprp{e}/\tprl{e})(\tauprl{i}/\tauprp{i})(Z_i/\tauprl{i})^{1/2}(Z_i m_e / m_i)^{1/2}$. Then (\ref{eqn:caseAiiomega}) reproduces the $\betaprl{i} \gg 1$ limit of the KAW dispersion relation (cf.~\ref{eqn:kaw}):
\begin{align}\label{eqn:kaw2}
\overline{\omega} &= \pm \frac{k_\perp \rho_i}{\sqrt{\betaprp{i}}} \biggl( 1 + \frac{\betaprl{e}}{2} \Delta_e \biggr)^{1/2} \Bigl( 1 - \betaprp{e} \Delta_e \Bigr)^{1/2} \nonumber\\*
\mbox{} & -\imag \frac{k^2_\perp\rho^2_i}{2} \sqrt{\frac{\upi}{\betaprl{i}}} \biggl( 1 + \frac{\betaprl{e}}{2} \Delta_e \biggr) \frac{\tprp{e}}{\tprl{e}} \frac{\tauprl{i}}{\tauprp{i}} \biggl(\frac{Z_i}{\tauprl{i}} \frac{Z_i m_e}{m_i} \biggr)^{1/2}  .
\end{align}
\subsection{Low-$\betaprl{i}$ limit: $\betaprl{i} \ll 1$, $k_\perp \rho_i \sim 1$}

For $\betaprl{i} \ll 1$ and $\Delta_s \sim 1$, the coefficients $\mc{A}$, $\mc{B}$, $\mc{C}$, $\mc{D}$, $\mc{E} \sim \mc{O}(1)$. Then the gyrokinetic dispersion relation (\ref{eqn:gkdisprel}) reduces to
\begin{equation}\label{eqn:caseBdisprel}
\biggl( \frac{\alpha_\ast\mc{A}}{\overline{\omega}^2} - \mc{A}\mc{B} + \mc{B}^2 \biggr) \frac{2\mc{A}}{\betaprp{i}} = 0 .
\end{equation}
The long-wavelength limit of the second factor ($\mc{A} = 0$) gives the Landau-damped ion acoustic wave (see (\ref{eqn:ionacoustic})). For the first factor, we order $\overline{\omega}\sim\mc{O}(1)$ and consider two interesting limits: 

\subsubsection{The limit $(Z_i m_e/m_i) (\tauprp{i}/Z_i) \ll \betaprl{i} \ll 1$}

In this limit, $\xi_i = \overline{\omega} / \sqrt{\betaprl{i}} \gg 1$ and $\xi_e = (Z_i m_e / m_i)^{1/2} ( \tauprp{i} / Z_i)^{1/2} \ll 1$ (slow ions, fast electrons). Expanding the ion and electron plasma dispersion functions in large and small arguments, respectively, we get
\begin{align}
\mc{A} &\simeq 1 - \Gamma_0(\alpha_i) + \frac{\tauprp{i}}{Z_i} \bigl[ 1 + \Gamma_0(\alpha_e) \Delta_e \bigr] \nonumber\\*
\mbox{} &+ \imag \overline{\omega} \sqrt{\frac{\upi}{\betaprl{i}}} \frac{\tprp{i}}{\tprl{i}} \Biggl[ \Gamma_0(\alpha_i) \exp\biggl( - \frac{\overline{\omega}^2}{\betaprl{i}} \biggr) + \biggl( \frac{\tauprl{i}}{Z_i} \biggr)^{3/2} \biggl( \frac{Z_i m_e}{m_i} \biggr)^{1/2} \Gamma_0(\alpha_e) \Biggr] .
\end{align}
The dispersion relation (\ref{eqn:caseBdisprel}) then becomes
\begin{equation}
\frac{\tauprp{i}}{Z_i} \frac{\tprp{e}}{\tprl{e}} \mc{B} \Gamma_0(\alpha_e) \overline{\omega}^2  - \alpha_\ast \biggl\{ 1 - \Gamma_0(\alpha_i) + \frac{\tauprp{i}}{Z_i} \bigl[ 1 + \Delta_e \Gamma_0(\alpha_e) \bigr] \biggr\} = - \imag \bigl(  \mc{B}\overline{\omega}^2 - \alpha_\ast \bigr) {\rm Im}(\mc{A}) .
\end{equation}
This equation may be iteratively solved to find
\begin{subequations}
\begin{align}
{\rm Re}(\overline{\omega}) &= \pm \sqrt{ \frac{\alpha_\ast \bigl\{ 1 - \Gamma_0(\alpha_i) + (\tauprp{i}/Z_i) \bigl[ 1 + \Delta_e \Gamma_0(\alpha_e)\bigr]\bigr\}}{(\tauprp{i}/Z_i)(\tprp{e}/\tprl{e})\mc{B}\Gamma_0(\alpha_e)}} , \label{eqn:caseBiomega}\\*
\overline{\gamma} &= - \frac{\alpha_\ast}{2 \bigl[ (\tauprl{i}/Z_i)\Gamma_0(\alpha_e) \bigr]^2} \sqrt{\frac{\upi}{\betaprl{i}}} \frac{\tprl{i}}{\tprp{i}} \nonumber\\*
\mbox{} &\quad\times \Biggl[ \Gamma_0(\alpha_i) \exp\biggl( - \frac{\overline{\omega}^2}{\betaprl{i}} \biggr) + \biggl( \frac{\tauprl{i}}{Z_i} \biggr)^{3/2} \biggl( \frac{Z_i m_e}{m_i} \biggr)^{1/2} \Gamma_0(\alpha_e) \Biggr] ,
\end{align}
\end{subequations}
where $\overline{\gamma} \doteq {\rm Im}(\omega)/k_\parallel\valf$. In the limit $\alpha_i \ll 1$, (\ref{eqn:caseBiomega}) reduces to the Alfv\'{e}n wave solution (\ref{eqn:gkaw}).

\subsubsection{The limit $\betaprl{i} \sim Z_i m_e/m_i \ll 1$, $\tauprp{i}/Z_i \gg 1$}

In this limit, $\xi_i \sim (m_i/m_e)^{1/2} \gg 1$, and so $\xi_e \sim (\tauprp{i}/Z_i)^{1/2} \gg 1$ (cold ions and electrons). Expanding all plasma dispersion functions in their large arguments, the coefficient
\begin{align}
\mc{A} &\simeq \mc{B} - \frac{\Gamma_0(\alpha_e)}{2\overline{\omega}^2} \frac{m_i}{Z_i m_e} \betaprp{i} + \imag \overline{\omega} \sqrt{\frac{\upi}{\betaprl{i}}}  \frac{\tprp{i}}{\tprl{i}} \nonumber\\*
\mbox{} &\times \Biggl[  \Gamma_0(\alpha_i) \exp\biggl(-\frac{\overline{\omega}^2}{\betaprl{i}} \biggr) + \biggl( \frac{\tauprl{i}}{Z_i} \biggr)^{3/2} \biggl( \frac{Z_i m_e}{m_i}\biggr)^{1/2} \Gamma_0(\alpha_e) \exp\biggl( - \frac{\tprl{i}}{\tprl{e}} \frac{m_e}{m_i} \frac{\overline{\omega}^2}{\betaprl{i}} \biggr) \Biggr] .
\end{align}
The dispersion relation (\ref{eqn:caseBdisprel}) then becomes
\begin{equation}
\frac{\alpha_\ast\Gamma_0(\alpha_e)}{2\overline{\omega}^2} \frac{m_i}{Z_i m_e} \betaprp{i} -\mc{B} \biggl[ \alpha_\ast + \frac{\Gamma_0(\alpha_e)}{2} \frac{m_i}{Z_i m_e} \betaprp{i} \biggr] = - \imag \bigl( \mc{B} \overline{\omega}^2 - \alpha_\ast \bigr) {\rm Im}(\mc{A}) ,
\end{equation}
which may be iteratively solved to find\footnote{There is a type-setting error in equation (D25) of \citet{howes06}, which is the Maxwellian counterpart of (\ref{eqn:howesfix}). The two terms proportional to $\Gamma_0(\alpha_i)$ and $\Gamma_0(\alpha_e)$ in their formula, which are type set as separated by brackets, should instead be added together as in (\ref{eqn:howesfix}).}
\begin{subequations}
\begin{align}
\overline{\omega} &= \pm \sqrt{ \frac{\alpha_\ast \Gamma_0(\alpha_e) (m_i / Z_i m_e) \betaprp{i}}{\bigl[ 2 \alpha_\ast + \Gamma_0(\alpha_e) (m_i/Z_i m_e) \betaprp{i} \bigr] \mc{B}} } , \\*
\overline{\gamma} &= -\frac{2 \alpha^3_\ast \Gamma_0(\alpha_e) (m_i / Z_i m_e) \betaprp{i}}{\bigl[ 2 \alpha_\ast + \Gamma_0(\alpha_e) (m_i/Z_i m_e) \betaprp{i} \bigr]^3 \mc{B}^2} \sqrt{\frac{\upi}{\betaprl{i}}}  \frac{\tprp{i}}{\tprl{i}} \nonumber\\*
\mbox{}&\times \Biggl[  \Gamma_0(\alpha_i) \exp\biggl(-\frac{\overline{\omega}^2}{\betaprl{i}} \biggr) + \biggl( \frac{\tauprl{i}}{Z_i} \biggr)^{3/2} \biggl( \frac{Z_i m_e}{m_i}\biggr)^{1/2} \Gamma_0(\alpha_e) \exp\biggl( - \frac{\tprl{i}}{\tprl{e}} \frac{m_e}{m_i} \frac{\overline{\omega}^2}{\betaprl{i}} \biggr) \Biggr] .\label{eqn:howesfix}
\end{align}
\end{subequations}
\subsection{Gyrokinetic dispersion relation for an electron-ion bi-kappa plasma}\label{sec:bikappa}

A bi-kappa distribution function is often used to describe the non-thermal electron population in the solar wind and, in particular, its suprathermal ($T_e \sim 60~{\rm eV}$) halo \citep[see, e.g.][]{vasyliunas68,mpl97a,mpr97b,maksimovic05}. In this section, we specialize the linear gyrokinetic theory of \S\S\ref{sec:gklin} and \ref{sec:gkfieldeqns} to a mean distribution function equal to
\begin{equation}
f_{\textrm{bi-}\kappa,s} ( v_\parallel , v_\perp ) \doteq \frac{\nsp}{\sqrt{\upi\kappa} \theta_{\parallel s}} \frac{1}{\upi\kappa\theta^2_{\perp s}} \frac{\Gamma(\kappa+1)}{\Gamma(\kappa-1/2)} \Biggl[ 1 + \frac{(v_\parallel - \dupar)^2}{\kappa\theta^2_{\parallel s}} + \frac{v^2_\perp}{\kappa\theta^2_{\perp s}} \Biggr]^{-(\kappa+1)} ,
\end{equation}
where $\Gamma$ is the Gamma function, $\kappa > 3/2$ is the spectral index, and
\begin{equation}
\theta_{\parallel s} \doteq \vthprl{s} \sqrt{1 - \frac{3}{2\kappa}} \quad {\rm and} \quad \theta_{\perp s} \doteq \vthprp{s} \sqrt{1 - \frac{3}{2\kappa}}
\end{equation}
are the effective parallel and perpendicular thermal speeds, respectively. At low and thermal energies, the bi-kappa distribution approaches a Maxwellian distribution, whereas at high energies it exhibits a non-thermal tail that can be described as a decreasing power law. The $C_{\ell s}$ coefficients are evaluated for $f_{\textrm{bi-}\kappa,s}$ in appendix D of Paper I.

The changes to the results presented in \S\S\ref{sec:lineargk_biMax}, \ref{sec:KAWfreeenergy}, \ref{sec:scalings} made by instead using a bi-kappa background distribution function are only quantitative. For example, the dispersion relation for the KAW (\ref{eqn:kaw}) becomes
\begin{equation}\label{eqn:KAW_bikappa}
\omega = \pm \frac{k_\parallel \valf k_\perp \rho_i}{\sqrt{\betaprp{i} + 2 / ( 1 + Z_i \tprl{e} / \tprp{i} ) - 2\mc{K}_\kappa}} \biggl( 1 + \frac{\betaprl{e}}{2} \Delta_e \biggr)^{1/2} \biggl( 1 - \betaprp{e} \Delta_e + \frac{\mc{K}}{C_\kappa} \biggr)^{1/2} ,
\end{equation}
where
\begin{equation}
\mc{K}_\kappa \doteq \mc{K} \, \frac{(2C_\kappa-1) \tprp{e}/\tprl{e} - 1}{\tprp{e}/\tprl{e} - 1} \quad {\rm and} \quad C_\kappa \doteq \biggl( 1 - \frac{1}{2\kappa} \biggr) \biggl( 1 - \frac{3}{2\kappa} \biggr)^{-1} .
\end{equation}
The modification is due to the effect of high-parallel-energy particles in the tail of the $\kappa$ distribution on the mirror stability threshold: a smaller fraction of particles are Landau resonant and thus the stabilizing influence of the parallel electric field on the mirror instability is reduced.

\bibliographystyle{jpp}

\end{document}